\documentclass[useAMS,usenatbib]{mn2e} 

\usepackage[T1]{fontenc}
\usepackage{ae,aecompl}

\usepackage{graphicx}	
\usepackage{amsmath}	
\usepackage{amssymb}	
\usepackage{epsfig}     
\usepackage{pifont}     
\usepackage{grffile}    

\newcommand{\tick}{\ding{52}}  
\def\spose#1{\hbox to 0pt{#1\hss}}
\def\ltsimm{\mathrel{\spose{\lower 3pt\hbox{$\sim$}}
        \raise 2.0pt\hbox{$<$}}}
\def\gtsimm{\mathrel{\spose{\lower 3pt\hbox{$\sim$}}
        \raise 2.0pt\hbox{$>$}}}

\title[3D adiabatic shock-cloud simulations]{The turbulent destruction of clouds
  - III. Three dimensional adiabatic shock-cloud simulations}

\author[J.~M.~Pittard \& E.~R.~Parkin]
{J.~M.~Pittard\thanks{E-mail: jmp@ast.leeds.ac.uk (JMP)} \& E.~R.~Parkin\\
School of Physics and Astronomy, University of
       Leeds, Woodhouse Lane, Leeds LS2 9JT, UK
}

\date{Accepted 4$^{th}$ Jan 2016. Received 7$^{th}$ Dec 2015; in
  original form 19$^{th}$ Oct 2015}

\pubyear{2016}

\begin{document}
\label{firstpage}
\pagerange{\pageref{firstpage}--\pageref{lastpage}}
\maketitle

\begin{abstract} 
  We present 3D hydrodynamic simulations of the adiabatic interaction
  of a shock with a dense, spherical cloud. We compare how the nature
  of the interaction changes with the Mach number of the shock, $M$,
  and the density contrast of the cloud, $\chi$. We examine the
  differences with 2D axisymmetric calculations, perform detailed
  resolution tests, and compare ``inviscid'' results to those obtained
  with the inclusion of a $k$-$\epsilon$ subgrid turbulence
  model. Resolutions of 32-64 cells per cloud radius are the minimum
  necessary to capture the dominant dynamical processes in 3D
  simulations, while the 3D inviscid and $k$-$\epsilon$ simulations
  typically show very good agreement. Clouds accelerate and mix up to
  5 times faster when they are poorly resolved. The interaction
  proceeds very similarly in 2D and 3D - although non-azimuthal modes
  lead to different behaviour, there is very little effect on key
  global quantities such as the lifetime of the cloud and its
  acceleration. In particular, we do not find significant differences
  in the hollowing or ``voiding'' of the cloud between 2D and 3D
  simulations with $M=10$ and $\chi=10$, which contradicts previous
  work in the literature.
\end{abstract}




\section{Introduction}
\label{sec:intro}
The interaction of fast, rarefied gas with denser ``clouds'' is a
common occurence in astrophysics and much effort has been invested to
understand this process. Clouds struck by shocks or winds can be
destroyed, ``mass-loading'' the flow and affecting its nature.  Such
interactions have implications for our understanding of the nature of
the interstellar medium \citep[ISM; see,
e.g.,][]{Elmegreen:2004,Scalo:2004}, for the evolution of diffuse
sources such as supernova remnants
\citep{McKee:1977,Chieze:1981,Cowie:1981,White:1991,Arthur:1996,Dyson:2002,Pittard:2003},
and for galaxy formation and evolution \citep[e.g.,][]{Sales:2010}.

Shock-cloud interactions in supernova remnants (SNRs) are some of the
best observed and studied cases to date. Some SNRs display large-scale
distortions which are associated with their interaction with nearby
molecular clouds \citep[see, e.g., a recent review
by][]{Slane:2015}. Examples include the Cygnus Loop
\citep{Graham:1995,Levenson:1999}, Tycho
\citep{Katsuda:2010,Williams:2013}, and SN~1006
\citep{Miceli:2014,Winkler:2014}. 
SNR-cloud interactions are also revealed by the presence of OH
(1720\,MHz) masers \citep[e.g.,][]{Brogan:2013}, an enhanced
$^{12}$CO($J=2-1$)/$^{12}$CO($J=1-0$) ratio in the line wings
\citep{Seta:1998}, and temperature, absorption and ionization
variations in X-ray emission
\citep[e.g.,][]{Chen:2001,Koo:2005,Nakamura:2014}. SNR-cloud
interactions are often radiative, and produce optical, IR and sub-mm
line emission \citep[e.g., as seen in
IC~433;][]{Fesen:1980,Snell:2005,Bykov:2008,Kokusho:2013}.
$\gamma$-ray emission, which to date has been detected from about 25
SNRs \citep{Slane:2015}, may also arise when SNRs interact with
molecular clouds. In two cases, W44 and IC~433, this emission is
established to be from energetic ions, and so is unambiguously from
the SNR shock \citep{Abdo:2010,Giuliani:2011,Ackermann:2013}.  A list
of galactic SNRs thought to be interacting with surrounding clouds is
given in the appendix of \citet{Jiang:2010}.


   
In other SNRs much finer features indicate interaction with much
smaller clouds, either in the ISM or in the ejecta. Significant
observational evidence now exists for clumpy ejecta, especially in
core-collapse SNe
\citep[e.g.,][]{Filippenko:1989,Spyromilio:1991,Spyromilio:1994,Fassia:1998,Matheson:2000,Elmhamdi:2004}. The
best observed examples of ejecta clumps are those seen in the Vela
remnant \citep{Aschenbach:1995,Strom:1995,Tsunemi:1999,Katsuda:2006}
and in Cassiopeia~A (Cas A)
\citep[e.g.,][]{Kamper:1976,Chevalier:1979,Reed:1995,Fesen:2001,Fesen:2011,Milisavljevic:2013,Patnaude:2014,Alarie:2014}.
N132D \citep{Lasker:1978,Lasker:1980,Morse:1996,Blair:2000}, Puppis A
\citep{Winkler:1985,Katsuda:2008}, SNR~G292.0+1.8
\citep{Park:2004,Ghavamian:2005} and the SMC SNR~1E~0102-7219
\citep{Finkelstein:2006} also display ejecta bullets.

In Cas~A ejecta knots are seen both ahead of the main forward shock,
where they interact with the circumstellar or interstellar medium, and
just after their passage through the remnant's reverse shock.
Some of the outer ejecta knots in Cas~A show emission
trails indicative of mass ablation, which \citet{Fesen:2011}
argue form best if the cloud density contrast $\chi=10^{3}$ (clouds with $\chi=10^{2}$ are destroyed
too rapidly, while too little material is ablated when $\chi=10^{4}$).
\citet{Patnaude:2014} instead present evidence for mass-ablation from the
inner ejecta knots in Cas~A. 
Enhanced X-ray emission which extends $1-2\arcsec$ downstream of the shocked clumps is interpreted
as stripped material which is heated to X-ray emitting
temperatures in the tail. For knot sizes of $a_{\rm 0} \sim
10^{15}-10^{16}\,$cm, this equates to $5-100\,a_{\rm 0}$, and is
compatible with the tail lengths found in
\citet{Pittard:2010}
A number of other studies have considered the ejecta clumps in Vela,
with the particular goal of reproducing the protrusions ahead of the
blast shock. \citet{Wang:2002} found that the survival of the ejecta
clumps through the reverse shock and out past the forward shock
required an initial $\chi \sim 10^{3}$.  \citet{Miceli:2013} find that
lower values of $\chi$ are acceptable if the effects of radiative
cooling and thermal conduction are included.

The middle-aged \citep[$\sim 3700\,$yr;][]{Winkler:1988} SNR Puppis~A
is interacting with several interstellar clouds, of which the most
prominent is known as the bright eastern knot. \cite{Hwang:2005}
present a {\em Chandra} observation of this region, and identify two
main morphological components. The first is a bright compact knot that
lies directly behind an indentation in the main shock front. The
second component lies about $1\arcsec$ downstream of the shock and
consists of a curved vertical structure (the ``bar'') separated from a smaller
bright cloud (the ``cap'') by faint diffuse emission. Based on hardness images
and spectra, and comparing to the ``voided sphere'' structures seen by
\citet{Klein:2003}, the bar and cap structure is identified as a
single shocked interstellar cloud. The interaction is inferred to have
$\chi=10$, and to be at a relatively late stage of evolution
($t\sim3\,t_{\rm cc}$, where $t_{\rm cc}$ is the characteristic cloud
crushing timescale - see Sec.~\ref{sec:numerics}). The compact knot
directly behind the shock front is identified as a more recent
interaction with another cloud\footnote{In a more general study of the
  X-ray emission resulting from numerical models of shock-cloud
  interactions, \citet{Orlando:2006} determined that the emission is
  brightest at $t\sim t_{\rm cc}$, and is dominated by the cloud core
  where the shocks transmitted into the cloud collide. They also find
  that the X-ray morphology is strongly affected by the strength of
  thermal conduction and evaporation.  This work was extended by
  \citet{Orlando:2010}, where diagnostic tools for interpreting X-ray
  observations of shock-cloud interactions were presented.}. Another
well-studied interaction between a SNR and a small ($< 1\,$pc)
interstellar cloud is FilD in the Vela SNR. \citet{Miceli:2006}
estimate that a Mach 57 shock is in the early stages of interacting
with an ellipsoidal cloud with $\chi=30$.

Numerical studies of shock-cloud interactions have now been performed
for many decades. However, the motivation for the current work comes
from our realization that, barring the work in the code development
paper of \citet{Schneider:2015}, {\em all} 3D ``pure-hydrodynamic''
shock-cloud calculations (i.e. those without additional physical
processes such as cooling, magnetic fields, thermal conduction and
gravity) in the astrophysics literature are for one set of parameters
only: $M=10$ and $\chi=10$ (see Sec.~\ref{sec:numerical_studies} and
Table~\ref{tab:previous}). We therefore extend the 2D work in
\citet{Pittard:2009,Pittard:2010} to 3D. The extension to 3D is
necessary for two reasons: i) non-axisymmetric perturbations can only
be obtained in 3D; ii) the late-time flow in shock-cloud interactions
can acquire characteristics similar to turbulence, which has a
fundamentally different behaviour in 2D due to the absence of
vortex-stretching (in 2D, vortices are well-defined and long-lasting).

We investigate 3D shock-cloud interactions for Mach numbers
$M=1.5,\,3$ and~10, and for density contrasts $\chi = 10,\, 10^{2}$,
and $10^{3}$. This extends the $\chi$ parameter space to $10\times$
higher values than any previously published 3D simulation that we are
aware, and a factor of 25 higher than any previously published 3D
adiabatic simulation. As in our previous 2D work we present
``inviscid'' simulations and simulations with a $k$-$\epsilon$ subgrid
turbulence model. In Sec.~\ref{sec:shockcloud_interaction} we review
the numerical and experimental work which currently exists. In
Sec.~\ref{sec:numerics} we describe the simulation setup and in
Sec.~\ref{sec:results} we present our results. As well as describing
the 3D nature of the interaction, we compare our 3D results to those
from 2D simulations. In Sec.~\ref{sec:conclusions} we summarize our
results and draw conclusions. A detailed resolution test is presented
in an appendix. In a follow-up paper \citep{Pittard:2016}, an
investigation of a shock striking a filament (as opposed to a
spherical cloud) will be presented.

\section{The interaction of a shock with a cloud}
\label{sec:shockcloud_interaction}
\subsection{Numerical studies}
\label{sec:numerical_studies}
The idealized problem of the hydrodynamical interaction of a planar
adiabatic shock with a single isolated cloud was first studied
numerically in the 1970s.  The evolution of the cloud can be described
in terms of a characteristic cloud-crushing timescale, and is scale-free
for strong shocks. The cloud is first compressed, becomes
overpressured, and then re-expands, and is subject to a variety of
dynamical instabilities, including Kelvin-Helmholtz (KH),
Rayleigh-Taylor (RT) and Richtmyer-Meshkov (RM).  Strong vorticity is
deposited at the surface of the cloud and this vorticity aids in the
subsequent mixing of cloud and ambient material.  Detailed
two-dimensional (2D) axisymmetric calculations by \citet*{Klein:1994}
showed that a numerical resolution of about 120 cells per cloud radius
(hereafter referred to as $R_{120}$) was required in order to properly
capture the main features of the interaction.  The effects of smooth
cloud boundaries, radiative cooling, thermal conduction and magnetic
fields have now been considered \citep[see][for a summary of work up
until 2010]{Pittard:2010}. The interaction is milder at lower shock
Mach numbers \citep[see, e.g.,][]{Nakamura:2006}, and when the
post-shock gas is subsonic with respect to the cloud a bow-wave
instead of a bow-shock forms.

A dedicated study of how the adiabatic interaction of a shock with a
cloud depends on $M$, $\chi$ and the numerical resolution was recently
presented by \citet{Pittard:2009,Pittard:2010}. Using 2D axisymmetric
simulations, the results from ``inviscid'' models with no explicit
artificial viscosity were compared against results when a
$k$-$\epsilon$ subgrid turbulence model is included.  The 2D inviscid
models confirmed that a resolution of approximately $R_{100}$ is
necessary for convergence in simple adiabatic simulations. However,
this requirement was found to reduce to $\sim R_{32}$ when a subgrid
turbulence model is included. The cloud lifetime, defined as the point
when material from the core of the cloud is well mixed with the
ambient material, is about $t_{\rm mix} \sim 6\,t_{\rm KHD}$, where
$t_{\rm KHD}$ is the growth-timescale for the most disruptive,
long-wavelength, KH instabilities.  Cloud density contrasts $\chi
\gtsimm 10^{3}$ are required for the cloud to form a long tail-like
feature.

The first three-dimensional (3D) shock-cloud calculation was presented
by \citet{Stone:1992}. The simulation was adiabatic, had $M=10$ and
$\chi=10$, and used a numerical resolution of $R_{60}$. More rapid
mixing was observed since 3D hydrodynamical instabilities are able to
fragment the cloud in all directions, although this was not
quantified\footnote{In contrast, \citet{Nakamura:2006} claim that
  global quantities from 2D and 3D simulations are within 10\% for $t
  < 10\,t_{\rm cc}$ when the cloud has a smooth boundary (their
  $n=8$).}. Subsequently, \cite{Klein:2003} noted that 2D
hydrodynamical simulations did not compare well against experimental
results obtained with the Nova laser, which showed a ``voiding'' of
the shocked cloud, and break up of the vortex ring by the azimuthal
bending-mode instability \citep{Widnall:1974}. Crucially, a 3D
simulation reproduced both of these features. Other 3D work in the
astrophysics literature (summarized in Table~\ref{tab:previous}) has
investigated the effects of additional physics, including the cloud
shape and edges, radiative cooling, thermal conduction and magnetic
fields
\citep*{Xu:1995,Orlando:2005,Nakamura:2006,Shin:2008,VanLoo:2010,Johansson:2013,Vaidya:2013,Li:2013,Schneider:2015}.

Additional 3D simulations have been used to study the behaviour of
clouds accelerated by winds
\citep[e.g.,][]{Gregori:2000,Agertz:2007,Raga:2007,Kwak:2011,McCourt:2015,Scannapieco:2015},
by finite-thickness supernova blast waves
\citep[e.g.,][]{Leao:2009,Obergaulinger:2014}, or by dense shells
\citep{Pittard:2011}. The ram-pressure stripping of the interstellar
medium from galaxies
\citep[e.g.,][]{Close:2013,Shin:2013,Shin:2014,Tonnesen:2014,Roediger:2015a,Roediger:2015b,Vijayaraghavan:2015}
has also been considered. Though each of these scenarios are similar
in some ways to a shock-cloud interaction, the details differ in each
case, and therefore we do not discuss these works further. 

Outside of the astrophysics literature, the shock-cloud interaction is
commonly referred to as a shock-bubble interaction (the bubble can be
lighter or denser than the surrounding medium). Simulations carried
out by the fluid dynamics community have focused
on a similar region of parameter space as their experiments, which for
practical reasons tend to have lower $\chi$ and $M$ than the work noted in
Table~\ref{tab:previous}. In the most comprehensive 3D study to date (performed at
a resolution of $R_{128}$), \citet{Niederhaus:2008} examined shock
Mach numbers up to 5, and cloud density contrasts up to 4.2. They
  also considered different (fixed) values of the ratio of specific
  heats, $\gamma$, for the ambient and cloud gas.
The work by \citet{Niederhaus:2008} is also noteable for its detailed 
study into the development and behaviour of vorticity.  In a 2D
axisymmetric simulation, the vorticity can only have a
$\theta$-component, and the late-time flow is dominated by large and
distinct vortex rings. However, in 3D simulations, this restriction
no-longer applies, and the vorticity develops components in the
axial and radial directions. 
\citet{Niederhaus:2008} find that when
$\chi\gtsimm 1.5$, the axial and radial components of the vorticity
grow to a similar magnitude as the azimuthal component - it is this
growth which accounts for the differences in the late-time flow-field
in 2D and 3D simulations.
\citet{Niederhaus:2008} also find that the degree of mixing of cloud
and ambient material increases as $\chi$ increases, due to the greatly
increased complexity and intensity of scattered shocks and rarefaction
waves, which ultimately cause the formation of the turbulent wake.

Finally, we note that \citet{Ranjan:2008} present 3D hydrodynamical
simulations of a Mach 5 interaction of an R12 bubble in air
($\chi=4.17$) with a resolution of $R_{134}$. They find that the
vorticity field becomes so complex that the primary vortex core
becomes almost indistinguishable.



\subsection{Shock-cloud laboratory experiments}
\label{sec:lab_expts}
There are two broad types of laboratory experiments: those which use a
conventional shock-tube, and those which are laser-driven. The
literature has recently been reviewed by \citet{Ranjan:2011}. Of most
relevance to this work are the shock-bubble experiments of
\citet{Layes:2009}, who reported shock waves ($M=1.05$, 1.16, 1.4 and
1.61) through air striking a krypton gas bubble ($\chi=2.93$), the
experimental results of \citet{Ranjan:2008} for $M=2.05$ and 3.38
shocks striking an argon bubble in nitrogen ($\chi=1.43$), an $M=2.03$
shock striking a bubble of R22 refrigerant gas ($\chi=3.13$), and
$M=2.07$ and 3.0 shocks striking a sulfur-hexafluoride bubble
($\chi=5.27$), and the experiments of \citet{Zhai:2011} of a
sulfur-hexafluoride bubble in air, with $M=1.23$ and $\chi=5.04$.
Also noteable are the experiments by \citet{Ranjan:2005} of an argon
bubble in nitrogen struck by an $M=2.88$ shock, which provided the
first detection of distinct secondary vortex rings.

Laser-driven experiments of strong shocks ($M\approx10$) interacting
with a copper, aluminium or sapphire sphere embedded in a low-density
plastic or foam ($\chi\approx 8-10$) have been reported by
\citet{Robey:2002}, \citet{Hansen:2007} and \citet{Rosen:2009} for the
Omega laser at the Laboratory of Laser Energetics, and by
\citet{Klein:2000,Klein:2003} for the Nova laser at the Lawrence
Livermore National Laboratory. In particular, \citet{Klein:2000} found
the first evidence of 3D bending mode instabilities in high Mach
number shock-cloud interactions.  The experimental results also
indicate much faster destruction and mixing of the cloud at late times
than occurs in 2D simulations \citep[see Figs.~12 and~13
in][]{Klein:2003}.  \citet{Robey:2002} detect a double vortex ring
structure with a dominant azimuthal mode number of $\approx5$ for the
inner ring, and $\approx15$ for the outer ring. 3D numerical
simulations are found to be in very good agreement.
\citet{Hansen:2007} are able to estimate the mass of the cloud as a
function of time. They find that a laminar model overestimates the
stripping time by an order of magnitude, and conclude that the
mass-stripping must be turbulent in nature.  \citet{Rosen:2009} find
that their 3D simulations broadly agree with the gross features in
their experimental data, but differ in the finer-scale structure.


\begin{table*}
  \centering
  \caption[]{A summary of previous 3D numerical investigations of
    shock-cloud interactions in the {\em astrophysics} literature. $\chi$ is
    the density contrast of the cloud with respect to the ambient
    medium and $M$ is the shock Mach number. The references are as
    follows: $^a$\citet{Stone:1992}; $^b$\citet{Xu:1995};
    $^c$\citet{Klein:2000,Klein:2003}; $^d$\citet{Orlando:2005};
    $^e$\citet{Nakamura:2006}; $^f$\citet{Shin:2008};
    $^g$\citet{VanLoo:2010}; $^h$\citet{Johansson:2013};
    $^i$\citet{Vaidya:2013}; $^j$\citet{Li:2013};
    $^k$\citet{Schneider:2015}.}
\label{tab:previous}
\begin{tabular}{llllccc}
\hline
Authors & Typical (max) & $\chi$ & $M$ & Cooling? & Conduction? & Magnetic \\
 & resolution & & & & & fields?\\
\hline
SN92$^a$ & 60 (60) & 10 & 10 & & & \\
XS95$^b$ & 25 (53) & 10 & 10 & & & \\
K00/03$^c$ & 90 & 10 & 10 & & & \\
O05$^d$ & 132 (132) & 10 & 30,50 & \tick & \tick & \\
N06$^e$ & 60 & 10 & 10 & & & \\
SSS08$^f$ & 68 (68) & 10 & 10 & & & \tick \\
VL10$^g$ & 120 & 45 & 2.5 & \tick & & \tick \\
JZ13$^h$ & 100 & 100 & 30 & \tick & anisotropic & \tick \\
V13$^i$ & 225+ & 17.8 & 1.5,2 & isothermal & & \tick \\
L13$^j$ & 54 & 100 & 10 & \tick & & \tick \\
SR15$^k$ & 54 & 10,20,40 & 50 & & & \\
\hline
\end{tabular}
\flushleft{Notes: \citet{Xu:1995} also consider prolate clouds. \citet{Nakamura:2006} present a wide
  range of 2D simulations but also one 3D simulation (see their
  Sec.~9.2.2). The simulations of \citet{Vaidya:2013} are
  quite different to the others: they include
  self-gravity, the cloud does not have a uniform density and is not in equilibrium. Initially, $\chi=17.8$,
  but this increases as the cloud collapses (aspherically) to give a maximum density
  contrast at the core of 37. The Alfv\'{e}nic Mach number is reported
  in this case. \citet{Schneider:2015} also consider clouds with substructure.\\}
\end{table*}

\section{The Numerical Setup}
\label{sec:numerics}
Our calculations were performed on a 3D {\it XYZ} cartesian grid using the
{\sc MG} adaptive mesh refinement (AMR) hydrodynamic code.  {\sc MG}
uses piece-wise linear cell interpolation to solve the Eulerian
equations of hydrodynamics. The Riemann problem is solved at cell
interfaces to obtain the conserved fluxes for the time update. A
linear Riemann solver is used for most cases, with the code switching
to an exact solver when there is a large difference between the two
states \citep{Falle:1991}. Refinement is performed on a cell-by-cell
basis and is controlled by the difference in the solutions on the
coarser grids. The flux update occurs for all directions
simultaneously. The time integration proceeds first with a half
time-step to obtain fluxes at this point. The conserved variables are
then updated over the full time-step. The code is 2$^{nd}$-order
accurate in space and time.  The full set of equations solved
(including when the subgrid turbulence model is employed) is given in
\citet{Pittard:2009}. We limit ourselves to a purely hydrodynamic
study in this work, and ignore the effects of magnetic fields, thermal
conduction, cooling and self-gravity.  All calculations were perfomed
for an ideal gas with $\gamma=5/3$ and are adiabatic. Our calculations
are thus scale-free and can be easily converted to any desired physical
scales. 

The cloud is initially in pressure equilibrium with its surroundings
and is assumed to have soft edges (typically over about 10 per cent of
its radius). The equation for the cloud profile is noted in
\citet{Pittard:2009} - in keeping with the results from this earlier
work we again adopt $p_{1}=10$ (i.e. a reasonably hard-edged
cloud)\footnote{A purely numerical reason for adopting a smooth-edge to
  the cloud is that it minimizes the effects of ill-posed phenomena
  \citep[e.g.,][]{Samtaney:1996,Niederhaus:2008}. Of course, actual
  astrophysical clouds are unlikely to have hard edges.}. An advected
scalar is used to distinguish between cloud and ambient material, and
can be used to track the ablation and mixing of the cloud, and the
cloud's acceleration by the passage of the shock and subsequent
exposure to the post-shock flow.

The cloud is initially centered at the grid origin $(x,y,z)=(0,0,0)$.
The grid has zero gradient conditions on each boundary and is set
large enough so that the cloud is well-dispersed and mixed into the
post-shock flow before the shock reaches the downstream boundary. The
grid extent is dependent on $\chi$ (clouds with larger density
contrasts take longer to be destroyed) and is noted in
Table~\ref{tab:gridextent}. Our grid extent is often significantly
greater than previously adopted in the literature. Note that we also
do not impose any symmetry constraints on the interaction
\citep[unlike some of the 3D work in the literature,
e.g.,][]{Stone:1992,Xu:1995,Orlando:2005,Niederhaus:2008}, and thus
all quadrants are calculated. The simulations are generally evolved
until $t \sim 20\,t_{\rm cc}$, though at lower Mach numbers they are
run to $t \sim80\,t_{\rm cc}$.

We also perform new 2D axisymmetric calculations at $R_{128}$
resolution and with a similar grid extent.  To compare our 3D
simulations against these, we define motion in the direction of shock
propagation as ``axial'' (the shock propagates along the {\it
  X}-axis), and refer to it with a subscript ``{\it z}'',
while we collapse the {\it Y} and {\it Z} directions to obtain a
``radial'', or ``{\it r}'' coordinate.

\begin{table}
\centering
\caption[]{The grid extent for the 3D simulations, which depends
  on the shock Mach number, $M$, and the cloud density contrast, $\chi$. The unit of
  length is the cloud radius, $r_{\rm c}$.}
\label{tab:gridextent}
\begin{tabular}{lllc}
\hline
$M$ & $\chi$& \centering X & Y,Z \\
\hline
10 & 10 & $-5 < X < 65$ & $-10 < Y,Z < 10$ \\ 
   & $10^{2}$ & $-5 < X < 95$ & $-16 < Y,Z < 16$ \\ 
   & $10^{3}$ & $-5 < X < 475$ & $-24 < Y,Z < 24$ \\ 
3  & 10 & $-6 < X < 154$ & $-16 < Y,Z < 16$ \\ 
   & $10^{2}$ & $-6 < X < 474$ & $-16 < Y,Z < 16$ \\ 
   & $10^{3}$ & $-6 < X < 474$ & $-16 < Y,Z < 16$ \\ 
1.5 & 10 & $-150 < X < 300$ & $-20 < Y,Z < 20$ \\ 
    & $10^{2}$ & $-200 < X < 600$ & $-20 < Y,Z < 20$ \\ 
    & $10^{3}$ & $-290 < X < 910$ & $-20 < Y,Z < 20$ \\ 
\hline
\end{tabular}
\end{table}

\begin{table}
\centering
\caption[]{The maximum resolution $N$ (defined as the number of cells
  per could radius) used as a function of $M$ and $\chi$.}
\label{tab:maxres}
\begin{tabular}{lccc}
\hline
$\chi$/M& 1.5 & 3 & 10 \\
\hline
10 & 64 & 64 & 128\\ 
$10^{2}$ & 32 & 64 & 128\\ 
$10^{3}$ & 32 & 64 & 64 \\ 
\hline
\end{tabular}
\end{table}
 
Various integrated quantities are monitored to study the evolution of
the interaction \citep[see][]{Klein:1994,Nakamura:2006,Pittard:2009}.
Averaged quantities 
$\langle f \rangle$, are constructed by
\begin{equation}
\langle f\rangle = \frac{1}{m_{\beta}}\int_{\kappa \geq \beta} \kappa \rho f \;dV,
\end{equation}
where the mass identified as being part of the cloud is
\begin{equation}
m_{\beta} = \int_{\kappa \geq \beta} \kappa \rho \;dV.
\end{equation}
$\kappa$ is an advected scalar, which has an initial value of
$\rho/(\chi \rho_{\rm amb})$ for cells within a distance of $2 r_{\rm
  c}$ from the centre of the cloud, and a value of zero at greater
distances. Hence, $\kappa=1$ in the centre of the cloud, and declines
outwards. The above integrations are performed only over cells in
which $\kappa$ is at least as great as the threshold value,
$\beta$. Setting $\beta = 0.5$ probes only the densest parts of the
cloud and its fragments (identified with the subscript ``core''),
while setting $\beta = 2/\chi$ probes the whole cloud including its
low density envelope, and regions where only a small percentage of
cloud material is mixed into the ambient medium (identified with the
subscript ``cloud'').

The integrated quantities which are monitored in the calculations
include the effective radii of the cloud in the radial ($a$) and axial ($c$)
directions\footnote{Note that
  \citet{Shin:2008} instead adopt $a$ as the axial direction, with $b$
  and $c$ transverse to this. Thus their ratio $b/a$ plotted in their
  Fig.~1 is equivalent to the inverse of the ratio $c/a$ plotted in
  other works and in Fig.~\ref{fig:3Dresults2} in our present
  work.}. These are defined as 
\begin{equation}
a = \left(\frac{5}{2}\langle r^{2}\rangle \right)^{1/2}, \;\;\;\;\;\; 
c = \left[5\left(\langle z^{2}\rangle - \langle z\rangle^{2}\right)\right]^{1/2},
\end{equation}
where we convert our 3D {\it XYZ} coordinate system into a 2D {\it rz} coordinate
system through $r = \sqrt(Y^2 + Z^2)$ and $z = X$.

We also monitor the velocity dispersions in the radial and axial
directions, defined respectively as
\begin{equation}
\delta v_{\rm r} = \left<v_{\rm r}^{2}\right>^{1/2},\;\;\;\;\;
\delta v_{\rm z} = \left(\left<v_{\rm z}^{2}\right> - \langle v_{\rm z} \rangle^{2}\right)^{1/2},
\end{equation}
the cloud mass ($m$), and its mean velocity in the axial direction
($\langle v_{\rm z}\rangle$, measured in the frame of the unshocked
cloud).  The whole of the cloud and the densest part of its core are
distinguished by the value of the scalar variable $\kappa$ associated
with the cloud \citep[see][]{Pittard:2009}. In this way, each global
statistic can be computed for the region associated only with the core
(e.g., $a_{\rm core}$) or with the entire cloud (e.g., $a_{\rm
  cloud}$).

The characteristic time for the cloud to be crushed by the shocks
driven into it is the ``cloud crushing'' time, $t_{\rm cc} =
\chi^{1/2} r_{\rm c}/v_{\rm b}$, where $v_{\rm b}$ is the velocity of
the shock in the intercloud (ambient) medium \citep{Klein:1994}.
Several other timescales are obtained from the simulations. The time
for the average velocity of the cloud relative to that of the
postshock ambient flow to decrease by a factor of 1/$e$ is defined as
the ``drag time'', $t_{\rm drag}$\footnote{This is obtained when
  $\langle v_{\rm z}\rangle_{\rm cloud} = v_{\rm ps}/e$, where $v_{\rm
    ps}$ is the postshock speed in the frame of the unshocked
  cloud. Note that this definition differs from that in
  \citet{Klein:1994}, where $t_{\rm drag}$ corresponds to $\langle
  v_{\rm z}\rangle_{\rm cloud} = (1 - 1/e)v_{\rm ps}$. We refer to
  this latter definition as $t_{\rm drag,KMC}$, and quote values for
  both definitions in Table~\ref{tab:timescales}.}. The ``mixing
time'', $t_{\rm mix}$, is defined as the time when the mass of the
core of the cloud, $m_{\rm core}$, reaches half of its initial
value. The ``life time'', $t_{\rm life}$, is defined as the time when
$m_{\rm core} = 0$. The zero-point of all time measurements occurs
when the intercloud shock is level with the centre of the cloud.

An effective or grid-scale Reynolds number can be derived for our
inviscid simulations. The largest eddies have a length scale, $l$,
which is comparable to the size of the cloud ($l \sim 2r_{\rm c}$),
while the minimum eddy size, $\eta \approx 2 \Delta x$, where $\Delta
x$ is the cell size. Since the Reynolds number, $Re = (l/\eta)^{4/3}$,
we find that $Re \sim 650$ in our $R_{128}$ 3D simulations. The
effective Reynolds number is likely to be $\gtsimm 10^{3}$ in
the tail region of some simulations, where the tail is several times
broader than the original cloud.

\section{Results}
\label{sec:results}
We begin by examining the level of convergence in our simulations:
i.e. that the calculations are performed at spatial resolutions that
are high enough to resolve the key features of the
interaction. Increasing the resolution in inviscid calculations leads
to smaller scales of instabilities. Quantities which are sensitive to
these small scales (such as the mixing rate between cloud and ambient
gas) may not be converged, while quantities which are insensitive to
gas motions at small scales (e.g., the shape of the cloud) are more
likely to show convergence. Previous 2D studies
\citep[e.g.,][]{Klein:1994, Nakamura:2006} have indicated that about
100 cells per cloud radius are needed for convergence of the
simulations. \citet{Pittard:2009} demonstrated that $k$-$\epsilon$
simulations converged at lower resolution. 

In Appendix~\ref{sec:restest} we carry out a similar study for 3D
calculations with and without the use of a subgrid turbulence model.
Appendix~\ref{sec:restest_morphology} shows that resolutions of at
least $R_{64}$ are necessary to properly capture the nature of the
interaction in terms of the appearance and morphology of the
cloud. However, Appendix~\ref{sec:restest_evolution}
and~\ref{sec:restest_convergence} show that the broad evolution of the
cloud can often be adequately captured at lower resolution, for
instance at $R_{32}$. This is helpful considering the much greater
computational demands of 3D simulations. Another key finding is that
3D inviscid and $k$-$\epsilon$ models are often in very good
agreement. Comparing the morphology, we typically find that the core
structure is almost identical (being dominated by shocks and
rarefaction waves) until later times. Instead, the $k$-$\epsilon$
model tends to reveal its presence in the cloud tail and wake, where
it tends to smooth out the flow (this region is dominated by eddies
and vortices). This is a surprising result given that 2D calculations
can show significant differences
\citep[see][]{Pittard:2009,Pittard:2010}, but must be related to the
different way that vortices behave and evolve in 2D and 3D flows.
Since we find no compelling benefit from using the $k$-$\epsilon$
model in 3D calculations, in the rest of this work our focus will
therefore be on the inviscid simulations.

\begin{figure*}
\resizebox{170mm}{!}{\includegraphics{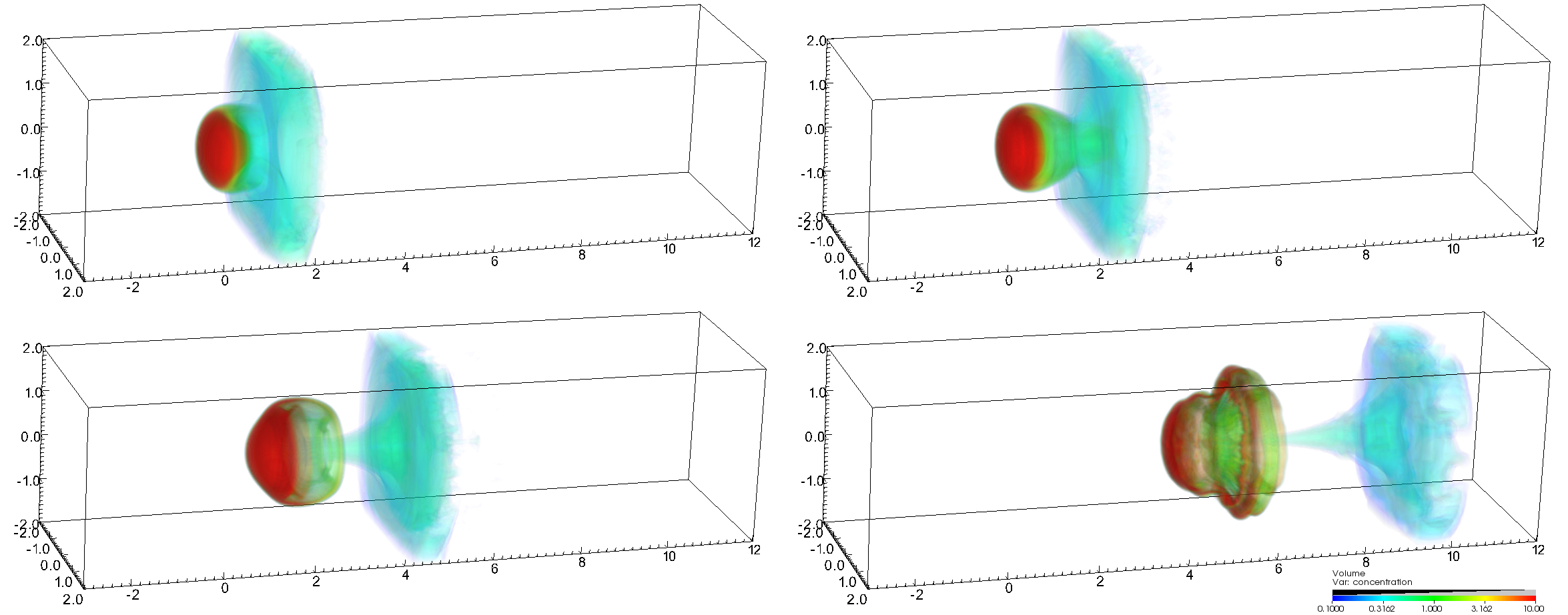}}
\caption{A 3D volumetric rendering of the time evolution of the
  $M=10$, $\chi=10$ simulation. From left to right and top to bottom
  the times are $t=0.65, 1.08, 1.94$ and $4.09\,t_{\rm cc}$ ($t=0$ is
  defined as the time when the intercloud shock is level with the
  centre of the cloud). The colour indicates the density of the cloud
  material, normalized by the ambient density (i.e. the initial cloud
  density is 10). Since the ambient material is not shown the bowshock
  upstream of the cloud is not visible. The actual grid extends much
  further than the bounding box shown.}
    \label{fig:chi1e1_3Devolplot}
\end{figure*}

\begin{figure*}
\resizebox{175mm}{!}{\includegraphics{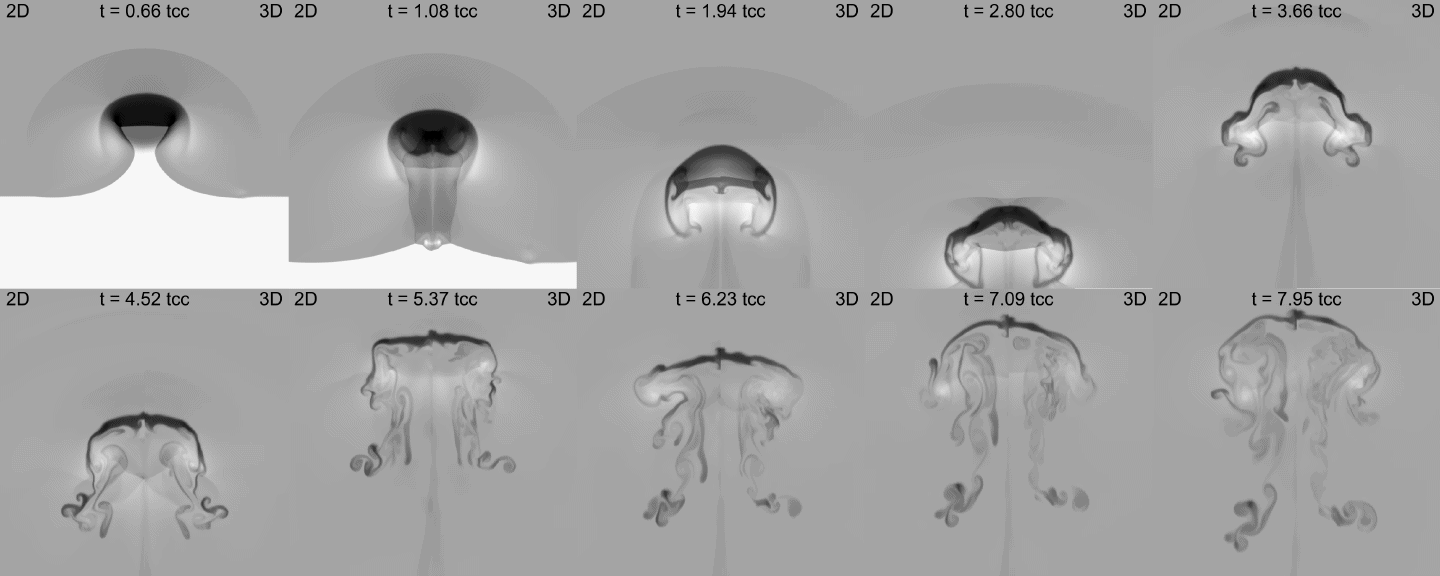}}
\caption{2D versus 3D comparison of the time evolution of the $M=10$,
  $\chi=10$ simulation. In each frame, 2D axisymmetric results are
  shown on the left, and part of the $+Y$, $Z=0$ plane from the 3D
  simulation is shown on the right. The grayscale shows the logarithm
  of the mass density, from $\rho_{\rm amb}$ (white) to $5 \rho_{\rm
    c}$ (black). Each frame is labelled with the time, and extends
  $3\,r_{\rm c}$ off-axis. The first 4 frames show the same region
  ($-2 < X < 4$, in units of $r_{\rm c}$) so that the motion of the
  cloud is clear. The displayed region is shifted in the other frames
  in order to show the cloud. The frames at $t=3.66$ and $4.52\,t_{\rm
    cc}$ show $2 < X < 8$,
  while the remaining frames show $5 < X < 11$, $6 < X < 12$, $8 < X <
  14$ and $9.5 < X < 15.5$ at $t=5.37$, 6.23, 7.09 and $7.95\,t_{\rm
    cc}$, respectively. Note that in this and similar figures the
  $X$-axis is plotted vertically, with positive down.}
    \label{fig:chi1e1_2dvs3d}
\end{figure*}

\begin{figure*}
\resizebox{170mm}{!}{\includegraphics{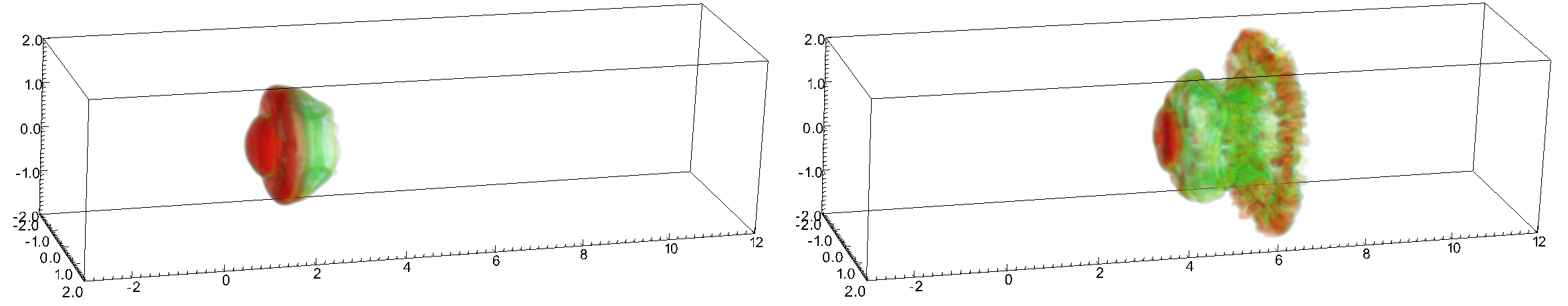}}
\caption{A 3D volumetric rendering of the time evolution of the
  $M=10$, $\chi=10$ hard-edged-cloud simulation. Left:
  $t=1.94\,t_{\rm cc}$. Right: $t = 4.09\,t_{\rm cc}$. Other details
  are as in Fig.~\ref{fig:chi1e1_3Devolplot}.}
    \label{fig:chi1e1he_3Devolplot}
\end{figure*}

\begin{figure*}
\resizebox{175mm}{!}{\includegraphics{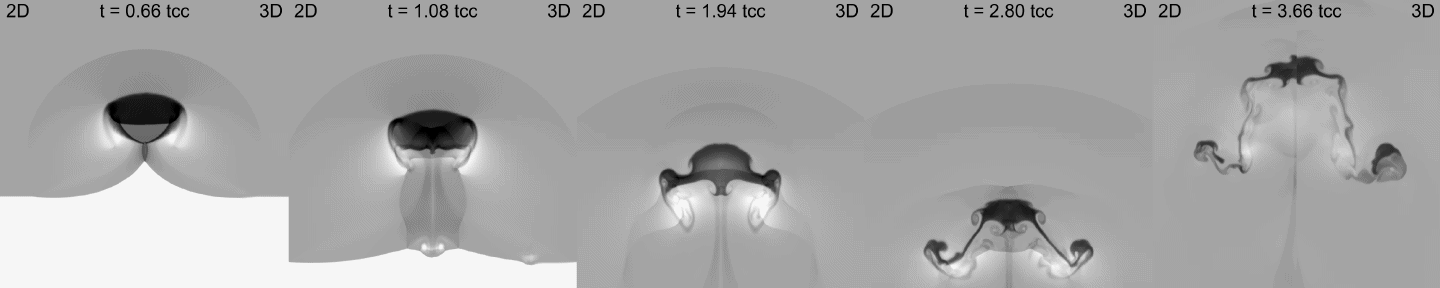}}
\caption{2D versus 3D comparison of the time evolution of the $M=10$,
  $\chi=10$, hard-edged-cloud simulation. All details are as
  Fig.~\ref{fig:chi1e1_2dvs3d}.}
    \label{fig:chi1e1_2dvs3dhe}
\end{figure*}

\begin{figure*}
\resizebox{175mm}{!}{\includegraphics{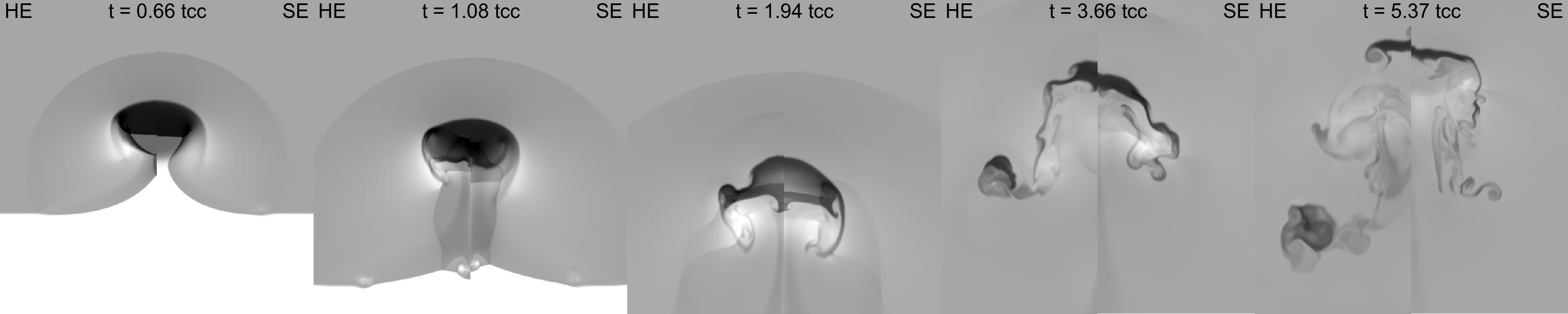}}
\caption{Comparison of the time evolution of the $M=10$,
  $\chi=10$, 3D simulations for hard-edged clouds (left ``HE'' plot in
  each panel) and soft-edged clouds (right ``SE'' plot in each
  panel). All other details are as in 
  Fig.~\ref{fig:chi1e1_2dvs3d}.}
    \label{fig:chi1e1_3dHEvsSE}
\end{figure*}

\begin{figure*}
\resizebox{170mm}{!}{\includegraphics{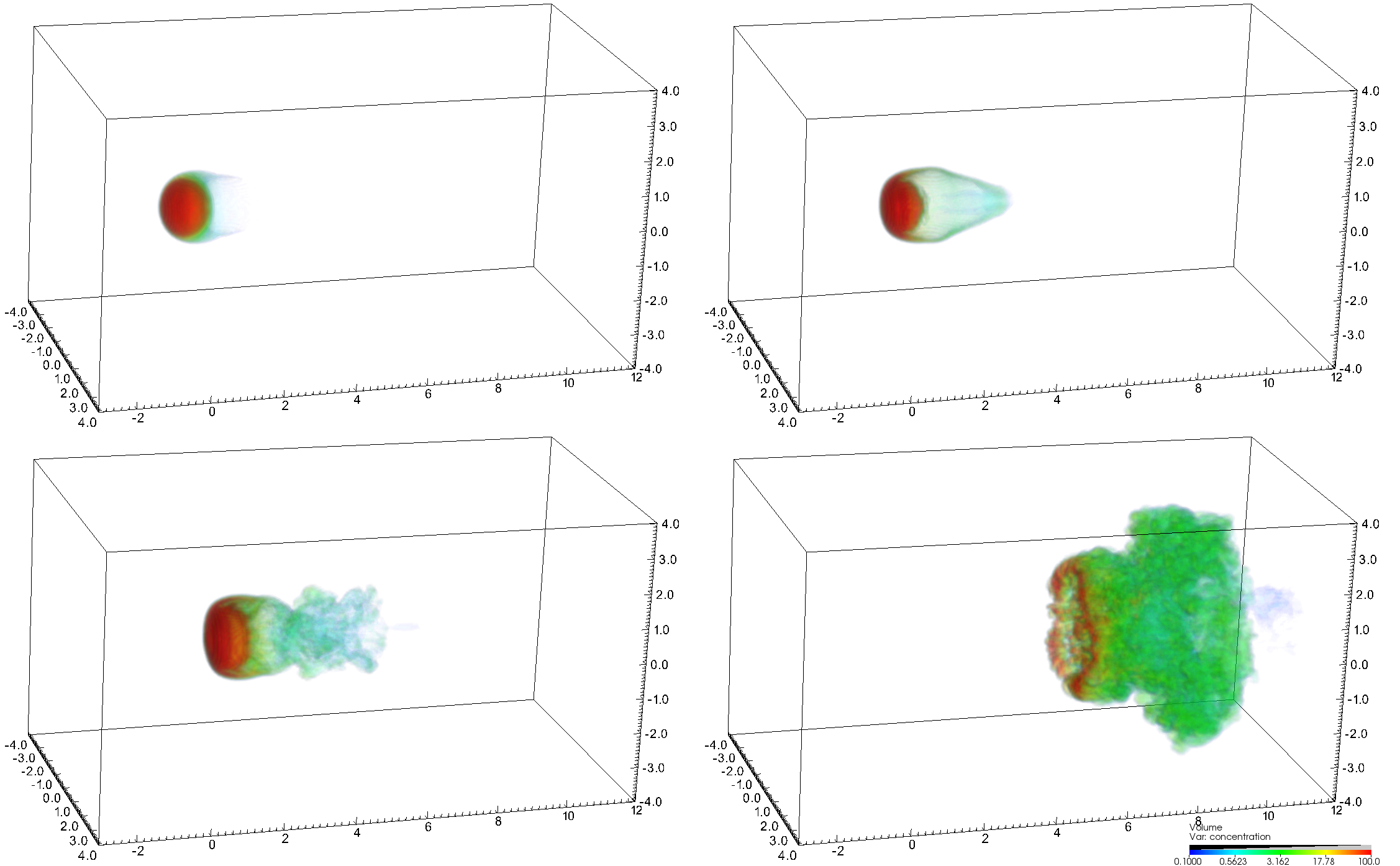}}
\caption{As Fig.~\ref{fig:chi1e1_3Devolplot} but for $M=10$ and
  $\chi=10^{2}$. The panels are at $t=0.48, 1.16, 1.84$ and
  $3.87\,t_{\rm cc}$. The initial cloud density is 100.}
    \label{fig:chi1e2_3Devolplot}
\end{figure*}

\begin{figure*}
\resizebox{175mm}{!}{\includegraphics{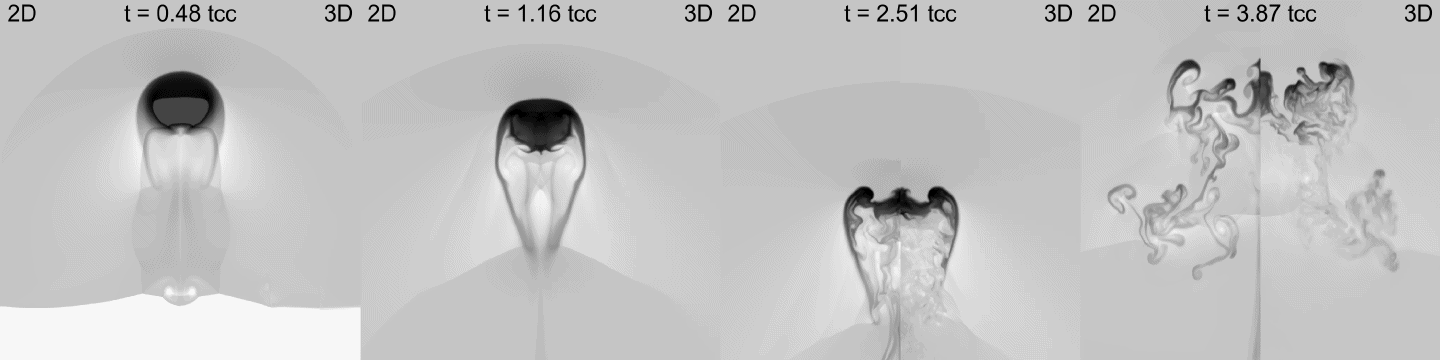}}
\caption{As Fig.~\ref{fig:chi1e1_2dvs3d} but for $M=10$ and
  $\chi=10^{2}$. Each frame extends $4\,r_{\rm c}$ off-axis. The first
  3 frames show the same region ($-2 < X < 6$, in units of $r_{\rm
    c}$) so that the motion of the cloud is clear. The final frame
  shows $4 < X < 12$.}
    \label{fig:chi1e2_2dvs3d}
\end{figure*}

\begin{figure*}
\resizebox{170mm}{!}{\includegraphics{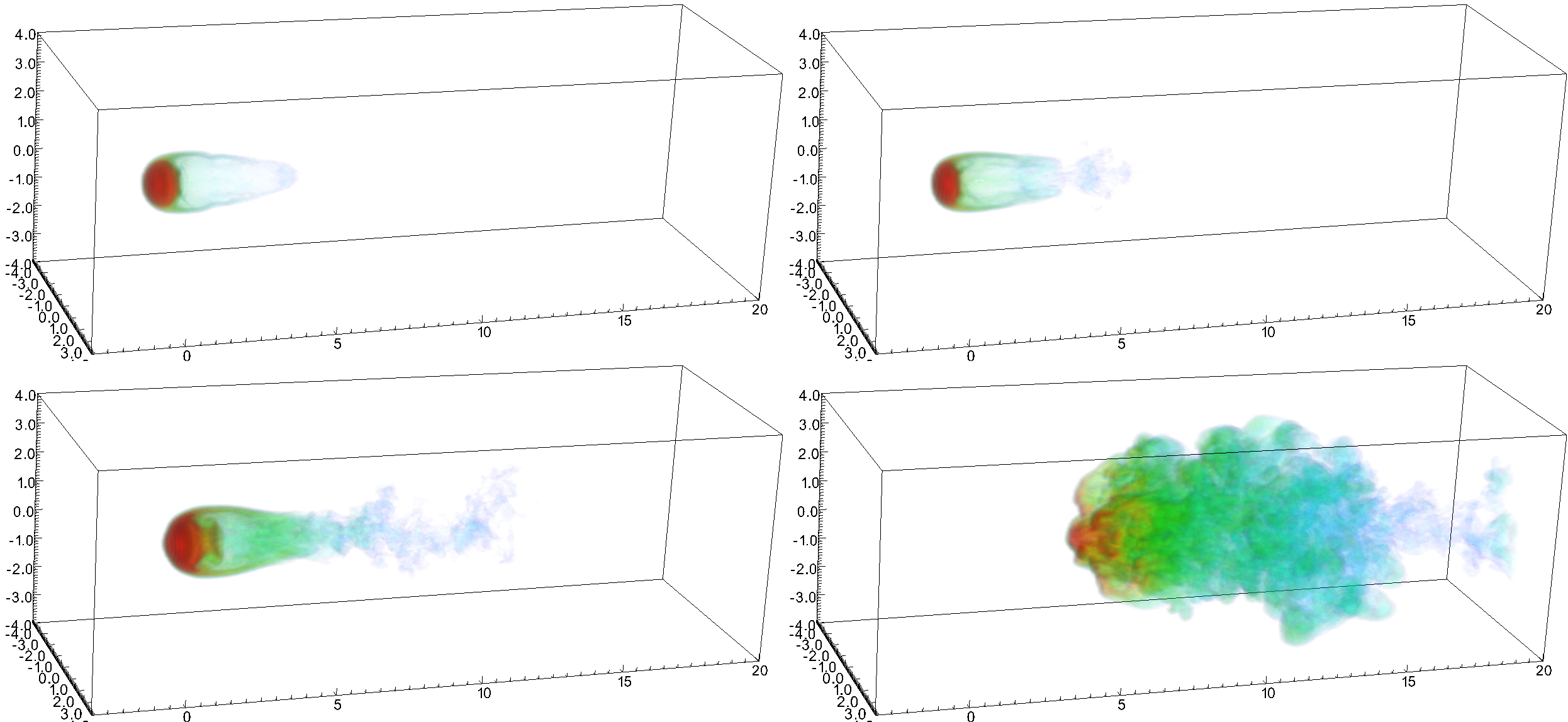}}
\caption{As Fig.~\ref{fig:chi1e1_3Devolplot} but for $M=10$ and
  $\chi=10^{3}$. The panels are at $t=0.58, 0.80, 1.65$ and
  $3.80\,t_{\rm cc}$. The initial cloud density is $10^{3}$.}
    \label{fig:chi1e3_3Devolplot}
\end{figure*}

\begin{figure*}
\resizebox{175mm}{!}{\includegraphics{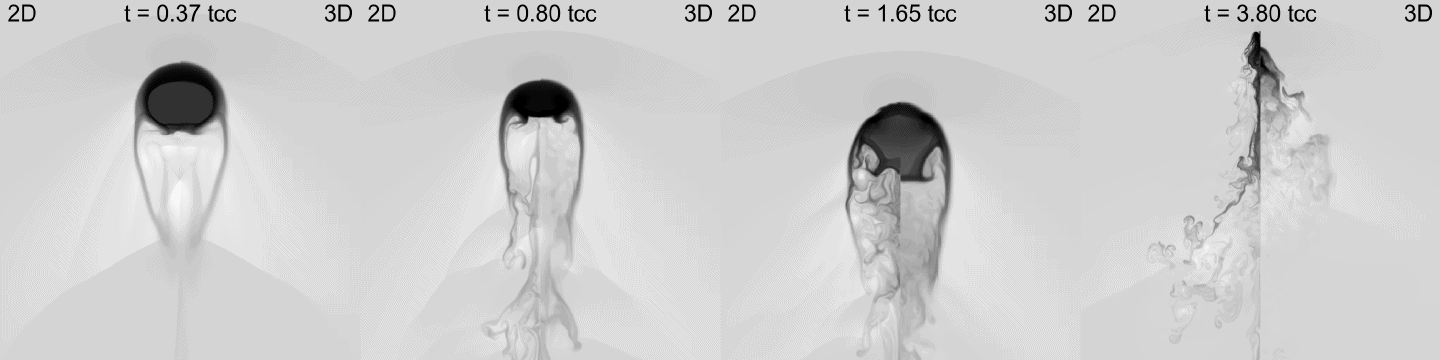}}
\caption{As Fig.~\ref{fig:chi1e1_2dvs3d} but for $M=10$ and $\chi=10^{3}$. The
  first 3 frames show the same region ($-2 < X < 6$, $0 < Y < 4$ in
  units of $r_{\rm c}$) so that the motion of the cloud is clear. The
  final frame shows $2 < X < 20$, $0 < Y < 9$. Note that the 2D
  simulation has a resolution of $R_{128}$, while the 3D simulation is
  at the lower resolution of $R_{64}$.}
    \label{fig:chi1e3_2dvs3d}
\end{figure*}

\subsection{Cloud Morphology}
\subsubsection{The interaction of an $M=10$ shock with a $\chi=10$ cloud}
\begin{figure}
\resizebox{80mm}{!}{\includegraphics{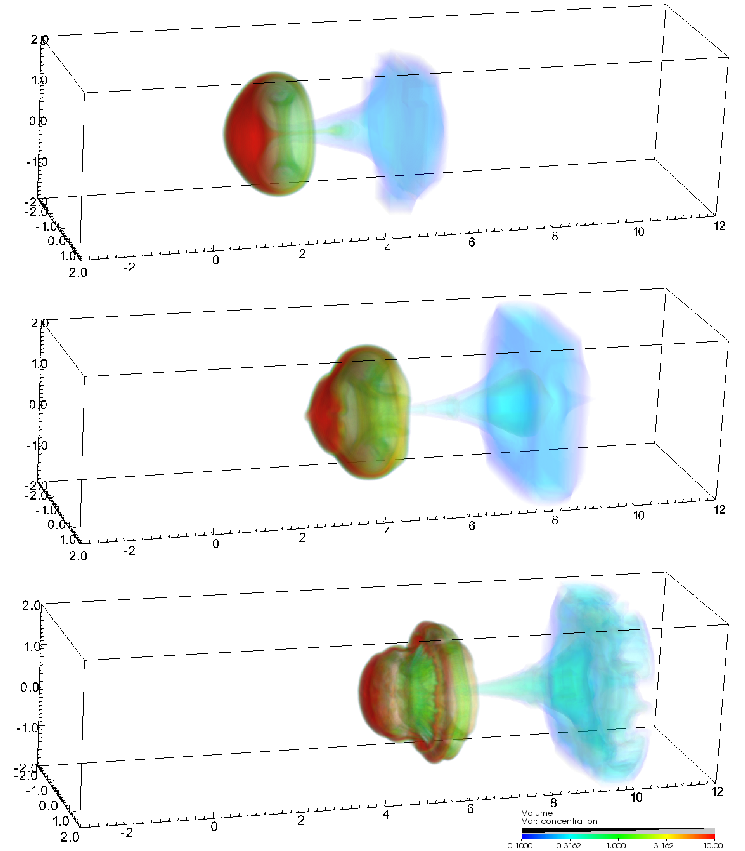}}
   \caption{3D volumetric rendering of $\chi=10$ simulations.
     Top: $M=1.5$ ($t=3.87\,t_{\rm cc}$). Middle: $M=3$ ($t=3.87\,t_{\rm
       cc}$). Bottom: $M=10$ ($t=4.09\,t_{\rm cc}$). The initial cloud
     density is 10.}
    \label{fig:chi1e1_3Dplot}
\end{figure}

We begin by examining the time evolution of the {\em cloud} material
in the 3D $M=10$, $\chi=10$ simulation
(Fig.~\ref{fig:chi1e1_3Devolplot}). The bowshock and other features in
the ambient medium are not visible in this plot due to this focus.
The blue ``block'' of material advected rapidly downstream represents
some ``trace'' material which highlights roughly where the main shock
is. It is added only for visualisation purposes to
Figs.~\ref{fig:chi1e1_3Devolplot},~\ref{fig:chi1e1_3Dplot}
and~\ref{fig:M10chi1e2_3Dresplot_3.87tcc}.

In the initial stages of the interaction, the transmitted shock front
becomes strongly concave and undergoes shock focusing, with the cloud
acting like a strongly convergent lens, refracting the transmitted
shock towards the axis (see also Fig.~\ref{fig:chi1e1_2dvs3d}).
Meanwhile the external shock diffracts around the cloud, remaining
nearly normal to the cloud surface as it sweeps from the equator to
the downstream pole. A dramatic pressure jump occurs as it is focused
onto the axis, and secondary shocks are driven into the back of the
cloud. Shortly after, the transmitted shock moving through the cloud
reaches the back of the cloud. It then accelerates downstream into the
lower density ambient gas, and a rarefaction wave is formed which
moves back towards the front of the cloud.  The secondary shocks
deposit further baryoclinic vorticity as they pass through the cloud,
and together with the reflected rarefaction waves cause the cloud to
reverberate. Shocks leaving the cloud introduce reflected rarefaction
waves into the cloud, while diffraction and focusing processes
introduce additional shocks into the cloud. Rarefaction waves within
the cloud which reach the cloud boundary introduce a transmitted
rarefaction wave into the external medium and a reflected shock which
moves back into the cloud.

At this point, the centre of the cloud becomes hollow for a moment
(see the panel at $t=1.94\,t_{\rm cc}$ in
Fig.~\ref{fig:chi1e1_2dvs3d}), before the upstream and downstream
surfaces collide together and the cloud attains an ``arc-like''
morphology (see the panel at $t=2.80\,t_{\rm cc}$ in
Fig.~\ref{fig:chi1e1_2dvs3d}). Some lateral expansion of the cloud has
occurred as a result of the lower pressure which exists at the sides
of the cloud. The strong shear across the surface of the arc-like
cloud also causes instabilities to grow and results in material being
stripped off. The cloud then deforms, resulting in a ring of material
being ripped off the rest of the cloud. The final panel in
Fig.~\ref{fig:chi1e1_3Devolplot} shows this occuring.

The misalignment of the local pressure and density gradients results
in the generation of vorticity in the flow field. The maximum
misalignment occurs at the sides of the cloud, and this is where the
maximum vorticity is deposited. The vortex sheet deposited on the
surface of the cloud rolls up into a torus to form a vortex ring. This
torus later disintegrates into many vortex filaments under the action
of hydrodynamic instabilities.
 
The features seen in Fig.~\ref{fig:chi1e1_3Devolplot} bear some
resemblance to those in Fig.~1 of \citet{Stone:1992} and in Fig.~2 of
\citet{Xu:1995}, but it is clear that there are some differences. In
both works the initial shock position is at $x=-1.2$ (this is true in
Xu \& Stone's work for their high resolution simulation), so the
elapsed time before the shock is level with the centre of the cloud is
$\approx 0.09\,t_{\rm cc}$. To compare with our results we deduct this
interval from the times noted in their works.  The middle panel of
Fig.~1 of \citet{Stone:1992} is hence at $t=1.91\,t_{\rm cc}$, while
Fig.~2b of \citet{Xu:1995} is at $t=1.81\,t_{\rm cc}$. Both are close
enough in time to be compared to the third panel in our
Figs.~\ref{fig:chi1e1_3Devolplot} and~\ref{fig:chi1e1_2dvs3d} at
$t=1.94\,t_{\rm cc}$. We can also compare against Fig.~19 of
\citet{Klein:2003}, which shows that by $t\sim3\,t_{\rm cc}$, strong
instabilities are shaping the vortex ring into a ``multimode fluted
structure''.

In comparison to these works, we find that our results do not show
such rapid development of KH instabilities, and of nonaxisymmetric
filaments/fluting in the vortex ring. We attribute these differences
to the softer edge of our cloud, which delays the onset of KH
instabilities.  \citet{Nakamura:2006} note that the development of KH
instabilities on the surface of the cloud takes longer than $t \sim
2\,t_{\rm cc}$ when the cloud has a smooth envelope, which is
confirmed in our 3D simulations also.  \citet{Stone:1992} also see RM
fingers on the upstream surface of the cloud in the right panel of
their Fig.~1 (at $t=4.41\,t_{\rm cc}$ in our time
frame). \citet{Xu:1995} do not mention such features and they are not
visible in their Fig.~2. In comparison, we see a (small) central
RM finger by $t=3.66\,t_{\rm cc}$. We conclude, therefore, that a soft
edge to the cloud does more to hinder KH instabilities than the growth
of RT and RM instabilities.

Fig.~\ref{fig:chi1e1_2dvs3d} compares cross-sections through the 2D
and 3D simulations. We are interested in such a comparison given that
\citet{Klein:2003} claim that their 2D results do not show ``voiding''
(i.e. separation between the front part and the back part of the
shocked cloud). The voiding is believed to arise in their 3D
experimental results due to the breaking up of the vortex ring by
azimuthal bending mode instabilities, and is visible by
$t=3.35\,t_{\rm cc}$ (see the panel at 49.2\,ns in their
Fig.~15). However, Fig.~\ref{fig:chi1e1_2dvs3d} reveals very good
agreement between our 2D and 3D simulation results. The large-scale
structure of the cloud is very similar (on fine-scales there are some
differences, though these are barely perceptible until $t>3.6\,t_{\rm
  cc}$, and at late times there is more vigorous mixing in the 3D
simulation). We also see that the cloud is indeed ``void'' or
``hollow'', though there appears to be less separation between the
front and back of the cloud than the 3D simulation of
\citet{Klein:2003} at comparable times.

That we do not see the large differences between 2D and 3D simulations
that \citet{Klein:2003} note is extremely interesting. Clearly, the
smoother edge of the cloud delays the onset of instabilities in our
simulations, but it is not clear whether this also causes the 2D and
3D simulations to evolve more closely. Therefore we have also
performed 2D and 3D simulations of an $M=10$, $\chi=10$ interaction
where the cloud has hard edges. Fig.~\ref{fig:chi1e1he_3Devolplot}
shows 3D volumetric renderings of the cloud material from an
interaction with a hard-edged cloud. Fig.~\ref{fig:chi1e1_2dvs3dhe}
also compares cross-sections through the 2D and 3D hard-edged
simulations. We see that the 2D and 3D simulation results are still in
good agreement with each other, with almost identical behaviour up to
$t=2.8\,t_{\rm cc}$ and very little difference at $t=3.66\,t_{\rm
  cc}$. At later times the level of agreement decreases as
non-azimuthal instabilities grow in the 3D simulations.

In Fig.~\ref{fig:chi1e1_3dHEvsSE} the results of the 3D soft-edged and
hard-edged simulations are directly compared. This figure, and
Figs.~\ref{fig:chi1e1_3Devolplot}-\ref{fig:chi1e1_2dvs3dhe}, indicate
the dramatic differences which can occur in the evolution of
soft-edged and hard-edged clouds. As noted by \citet{Nakamura:2006},
we see that the interaction can be significantly milder for soft-edged
clouds. The most dramatic difference in the hard-edge case is the
stronger and more rapid development of the vortex ring, which pulls
material off the sides of the cloud more quickly (compare the
morphology at $t=1.94\,t_{\rm cc}$). This leads to greater separation
between the head of the cloud and the vortex ring at later times.
Differences can, however, be seen as early as $t=0.66\,t_{\rm cc}$. In
the hard-edged case the external shock has already converged behind
the cloud at this time, whereas in the soft-edged case it has yet to
do so. A key factor behind the different evolution of the hard- and
soft-edged clouds is the stronger focussing of the transmitted shock
through the hard-edged cloud. This causes doubly shocked material,
formed behind the focussed shock moving in from the side of the cloud
as it overruns material behind the roughly planar transmitted shock,
to occur at a greater off-axis distance. This high-density region
kinks and becomes separated from the main cloud, particularly as the
shock transmitted into the back of the cloud, which becomes very
curved, first encounters the upstream surface of the cloud when
on-axis.  At $t=1.94\,t_{\rm cc}$ Fig.~\ref{fig:chi1e1_2dvs3dhe}
clearly shows two shocks in the ambient upstream environment. The
inner shock (created from the shock transmitted into the back of the
cloud) is not seen in the soft-edged case.

We conclude that 2D axisymmetric and fully 3D simulations of
shock-cloud interactions are in good agreement until non-axisymmetric
instabilities become important. We note that there are a number of
differences in the 2D and 3D simulations performed by
\citet{Klein:2003}: i) the 2D calculations were computed with CALE, an
arbitrary Lagrangian-Eulerian code with interface tracking, which was
used in pure-Eulerian mode, while the 3D calculations were computed
with a patch-based AMR code\footnote{In other work, \citet{Kane:2000}
  note that ``fine structure [is] somewhat suppressed by the interface
  tracking in CALE'' \citep[relative to that produced by the
  PROMETHEUS code which uses the piecewise-parabolic-method - see
  also][]{Kane:1997}.}; ii) the 2D simulations were run at a lower
resolution ($R_{50}$, versus $R_{90}$ for the 3D simulations); iii)
the 2D simulations were for $\chi=8$ (versus $\chi=10$ for the 3D
simulation). We emphasize that we do not see substantial differences
between 2D and 3D simulations (until non-axisymmetric instabilities
develop) when the same code and initial conditions are used.

\subsubsection{$\chi$ dependence when $M=10$}
The nature of the interaction changes with $\chi$ \citep[see, e.g.,
Sec.~4.1.2 of][]{Pittard:2010}. Fig.~\ref{fig:chi1e2_3Devolplot} shows
the time evolution of the $M=10$, $\chi=10^{2}$ simulation.  The
higher density contrast reduces the speed of the transmitted shock,
such that it does not pass the centre of the cloud before
the diffracted external shock converges on the axis behind the
cloud. The cloud is therefore compressed from all sides for a
significant period of time before the transmitted shock reaches the back of
the cloud, and launches a reflected rarefaction wave back towards the
front of the cloud. At this point further shocks are driven into the back of the
cloud, causing the cloud to have a distinctly hollow centre. The front
surface of the cloud kinks due to the RT instability as the cloud is
accelerated downstream and the resulting collapse of the cloud as its
front and back regions pancake together cause a large ring of material
to break off and accelerate downstream. This ring is readily apparent
in the last panel of Fig.~\ref{fig:chi1e2_3Devolplot}. It is
significantly larger by this time as its vorticity drives it away from
the axis.

\begin{figure}
\resizebox{80mm}{!}{\includegraphics{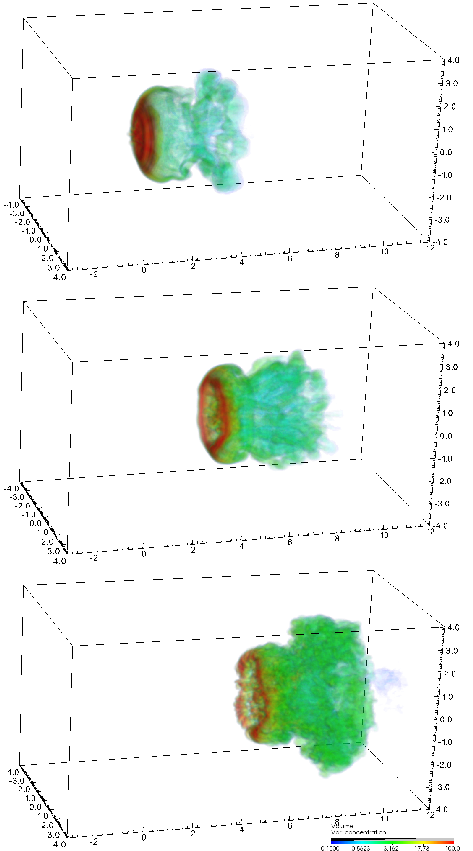}}
\caption{3D volumetric rendering of $\chi=10^{2}$ simulations at
  $t=3.87\,t_{\rm cc}$.  Top: $M=1.5$. Middle: $M=3$.  Bottom:
  $M=10$. The initial cloud density is $10^{2}$.}
    \label{fig:chi1e2_3Dplot}
\end{figure}

\begin{figure}
\resizebox{80mm}{!}{\includegraphics{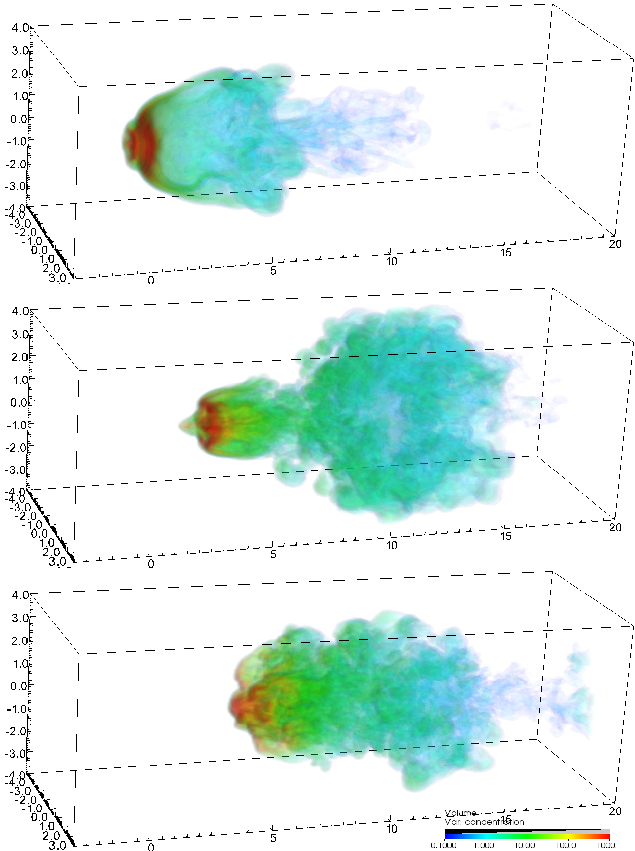}}
\caption{3D volumetric rendering of $\chi=10^{3}$ simulations at
  $t=3.80\,t_{\rm cc}$.  Top: $M=1.5$. Middle: $M=3$.  Bottom:
  $M=10$. The initial cloud density is $10^{3}$.}
    \label{fig:chi1e3_3Dplot}
\end{figure}

The behaviour of the 3D simulation is again similar to a 2D
axisymmetric simulation. Fig.~\ref{fig:chi1e2_2dvs3d} shows that the
large-scale morphology of the cloud is similar at the selected time
frames, but that the interior of the cloud has undergone substantially
more mixing by $t=2.51\,t_{\rm cc}$ in the 3D simulation, as witnessed
by the ``blurring'' of structure within the centre of the tail. In
fact, the tail in the 3D simulations bears characteristics of
``turbulence'', as is apparent also from the third panel in
Fig.~\ref{fig:chi1e2_3Devolplot}. By $t=3.87\,t_{\rm cc}$, mixing is
more advanced throughout the whole cloud structure in the 3D
simulation, and particularly in the vortex ring (note the ``blurring''
of structure in the downstream off-axis part of the cloud in the 3D
panel compared to the 2D panel). We attribute this speed-up to the
azimuthal instabilities which develop in the 3D simulation. This
faster mixing is visible as a slightly earlier decline in $m_{\rm
  core}$ in the 3D simulations compared to the 2D simulations (see
Fig.~\ref{fig:mcore_2Dvs3D}).

Fig.~\ref{fig:chi1e3_3Devolplot} shows the time evolution of the
$M=10$, $\chi=10^{3}$ simulation, in which the cloud is even more
resistant to the flow. Parts of the tail show characteristics of
turbulence (i.e. rapid spatial and temporal variation in the fluid
properties) by $t=0.8\,t_{\rm cc}$, though the main part of the cloud
only becomes ``turbulent'' between $t\sim1.65$ and $3.8\,t_{\rm
  cc}$. It is again interesting to see the dramatic lateral broadening
of the cloud between $t=0.8$, 1.65 and $3.8\,t_{\rm cc}$.

A comparison between 2D and 3D simulations reveals somewhat greater
differences this time, especially at the later stages of the
interaction (see Fig.~\ref{fig:chi1e3_2dvs3d}, and also Fig.~4 in
\citet{Pittard:2009}). For instance, the part of the tail nearest to
the cloud core is narrower in the 2D simulation at $t=0.80\,t_{\rm
  cc}$, while it is wider at $t=1.65\,t_{\rm cc}$. A KH instability is
visible on the front surface of the cloud in the 2D simulation at
$t=1.65\,t_{\rm cc}$, which is not seen in the 3D simulation. The
shape of the back of the cloud is also clearly different. However,
these differences may be due to the difference in resolution this
time, rather than changes due to the dimensionality.  At
$t=3.80\,t_{\rm cc}$, the 3D simulation shows a greater initial
flaring of the tail and the more rapid mixing of material within
it. In the 2D simulation the tail is noticeably longer, and stays
narrower as it leaves the cloud, before rapidly growing in width in
its bottom half.

\subsubsection{$M$ and $\chi$ dependence}
\begin{figure*}
\resizebox{175mm}{!}{\includegraphics{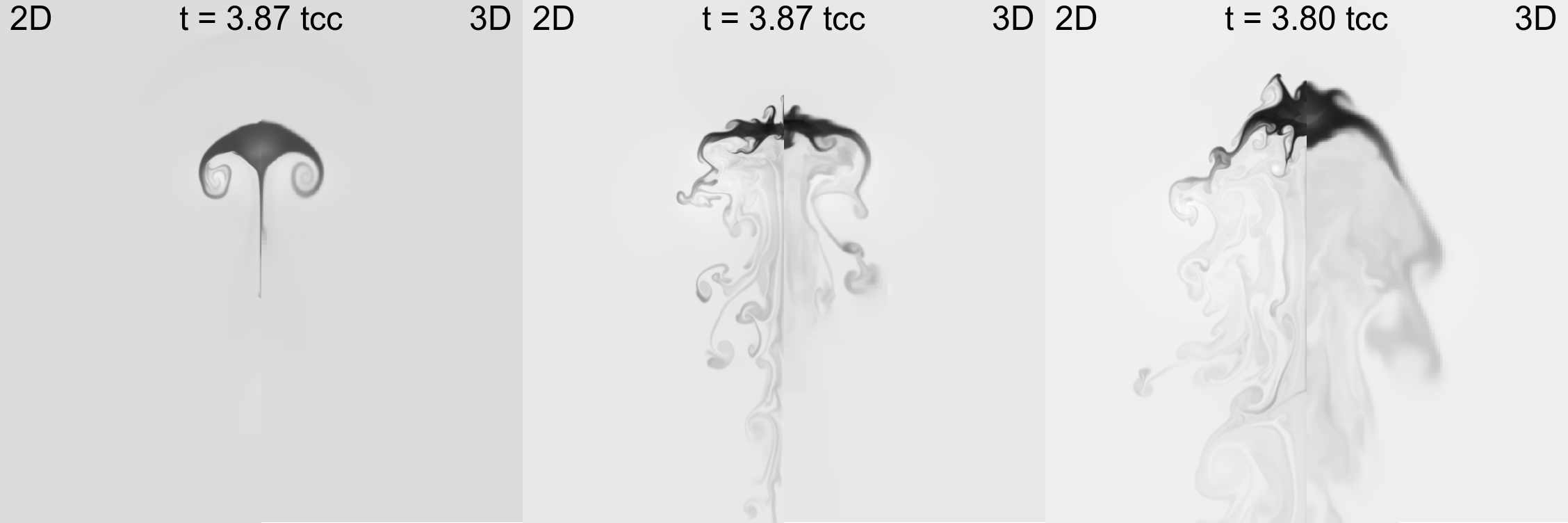}}
\caption{Comparison of 2D and 3D simulations for $M=1.5$. Left:
  $\chi=10$; Middle: $\chi=10^{2}$; Right: $\chi=10^{3}$. Each frame
  shows the region $-2 < X < 10$, $0 < Y < 6$. Note that each 2D
  simulation is at resolution $R_{128}$, while the 3D simulations are
  at the lower resolutions of $R_{64}$ ($\chi=10$) and $R_{32}$
  ($\chi=10^{2}$ and $10^{3}$).}
    \label{fig:m1.5_2dvs3d}
\end{figure*}

Fig.~\ref{fig:chi1e1_3Dplot} shows the Mach number dependance of the
interaction of a shock with a cloud of $\chi=10$. The interaction is
clearly much milder when $M=1.5$, with the cloud being accelerated
more slowly and instabilities taking longer to develop. The flow past
the cloud appears to be reasonably laminar at the times shown since
there is a lack of obvious instabilities in the cloud material, except
perhaps when $M=10$. While the $M=3$ and $M=10$ simulations evolve in
a near identical fashion, the $M=1.5$ simulation is markedly
different. Firstly, an axial jet forms behind the cloud in the
downstream direction. Such jets are often seen in shock-cloud
interactions (e.g., \citet{Niederhaus:2008} note that a particularly
strong downstream jet forms in the air-R12 $M=1.14$ case). Secondly,
there are fewer and weaker shocks and rarefaction waves in the cloud
and its environment. The rarefaction wave reflected into the cloud
when the transmitted shock reaches its back is quickly followed by a
shock so the cloud does not become as hollow, or for as long, as in
the higher $M$ cases. Finally, the reduced compression that the cloud
experiences means that it does not collapse into such a thin pancake,
and it is instead more readily shaped by the primary vortex which
pulls material off the sides of the cloud (see also
Fig.~\ref{fig:m1.5_2dvs3d}). This stream of gas is then subject to KH
instabilities, and is mixed into ambient material in the cloud wake.

The Mach number dependence for density contrasts of $\chi=10^{2}$ is
shown in Fig.~\ref{fig:chi1e2_3Dplot}. The increase in $\chi$ means
that the cloud better resists the shock and immersion in the
post-shock flow. This increases the velocity shear over the surface of
the cloud relative to the $\chi=10$ case, which in turn increases the
growth rate of KH instabilities. The result is that the interaction
becomes more turbulent. In the $M=1.5$ simulation, the transmitted
shock into the cloud moves slowly compared to the external shock, with
the result that the cloud is compressed from all sides. The shocks
driven into the cloud converge just downstream of its
centre. Secondary shocks which pass through the cloud and encounter
its upsteam surface cause the development of RM instabilites on the
leading edge of the cloud, which are just visible in the top panel of
Fig.~\ref{fig:chi1e2_3Dplot}, and can also be seen in the middle panel of
Fig.~\ref{fig:m1.5_2dvs3d}. The cloud pancakes and material is pulled
off it by vortical motions and KH instabilites.

The $M=3$, $\chi=10^{2}$ interaction is more violent. The rarefaction
waves which pass through the cloud in the early stages of the
interaction cause the cloud to hollow out, just as in the $M=10$
case. The cloud subsequently pancakes, and plumes of material soar off
the upstream surface, which in turn rapidly kink and fragment under
the action of KH and RT instabilities and the surrounding flow field.
A large number of smaller vortices form in the downstream wake. Some
non-axisymmetric structure can be seen in all of the panels in
Fig.~\ref{fig:chi1e2_3Dplot}.

Fig.~\ref{fig:chi1e3_3Dplot} shows the Mach number dependance for
cloud density contrasts of $\chi=10^{3}$. These clouds are very
resistant to the shock. In the $M=1.5$ case, the transmitted shock
initially converges just downstream of the cloud centre. When the
shock driven from the back of the cloud reaches the upstream surface a
prominent RM finger forms off of which secondary vortices occur. RM
fingers also grow off the back of the cloud and are ablated by the
swirling gas in the cloud wake (see the right panel of
Fig.~\ref{fig:m1.5_2dvs3d}). In the $M=3$ case, examination of a movie
of the interaction reveals that the initial transmitted shock moving
down through the cloud pushes out the back, creating a plume of
material. Shortly afterwards, a secondary vortex ring grows on the
front surface of the cloud (visible also in a plot of the magnitude of
the vorticity) as the cloud starts to pancake. The growth of this
secondary vortex ring stretches and shreds the outer part of the
cloud, causing it to detach from the main part of the cloud, whereupon
it is rapidly accelerated and mixed into the downstream turbulent
wake. It acquires considerable transverse velocity as it does so, such
that the wake extends significantly further off-axis. A large number
of secondary shocks and waves fills the wake, and the head of the
cloud suffers significant ablation via KH instabilities.

Fig.~\ref{fig:m1.5_2dvs3d} compares 2D and 3D simulations at $M=1.5$
for $\chi=10, 10^{2}$ and $10^{3}$. Note that the 3D simulations are
at lower resolution. Despite this, they clearly capture the main
features of the interaction, and again display faster mixing of
stripped material which clearly benefits from the development of
non-axisymmetric modes.

We conclude with two general observations. First, the $M=3$ simulation
tends to behave more closely to the $M=10$ simulation than to the
$M=1.5$ simulation. This is due to the fact that the post-shock flow
for $M=1.5$ is subsonic with respect to the cloud, whereas for $M=3$
and $M=10$ it is supersonic. We also find broad agreement between our
3D results and previously published 3D simulations, and between 3D and
2D calculations. However, it is clear that the 3D simulations better
capture the true nature of the interaction, which involves
non-axisymmetric instabilites.

\subsection{Statistics}
\begin{figure*}
\resizebox{130mm}{!}{\includegraphics{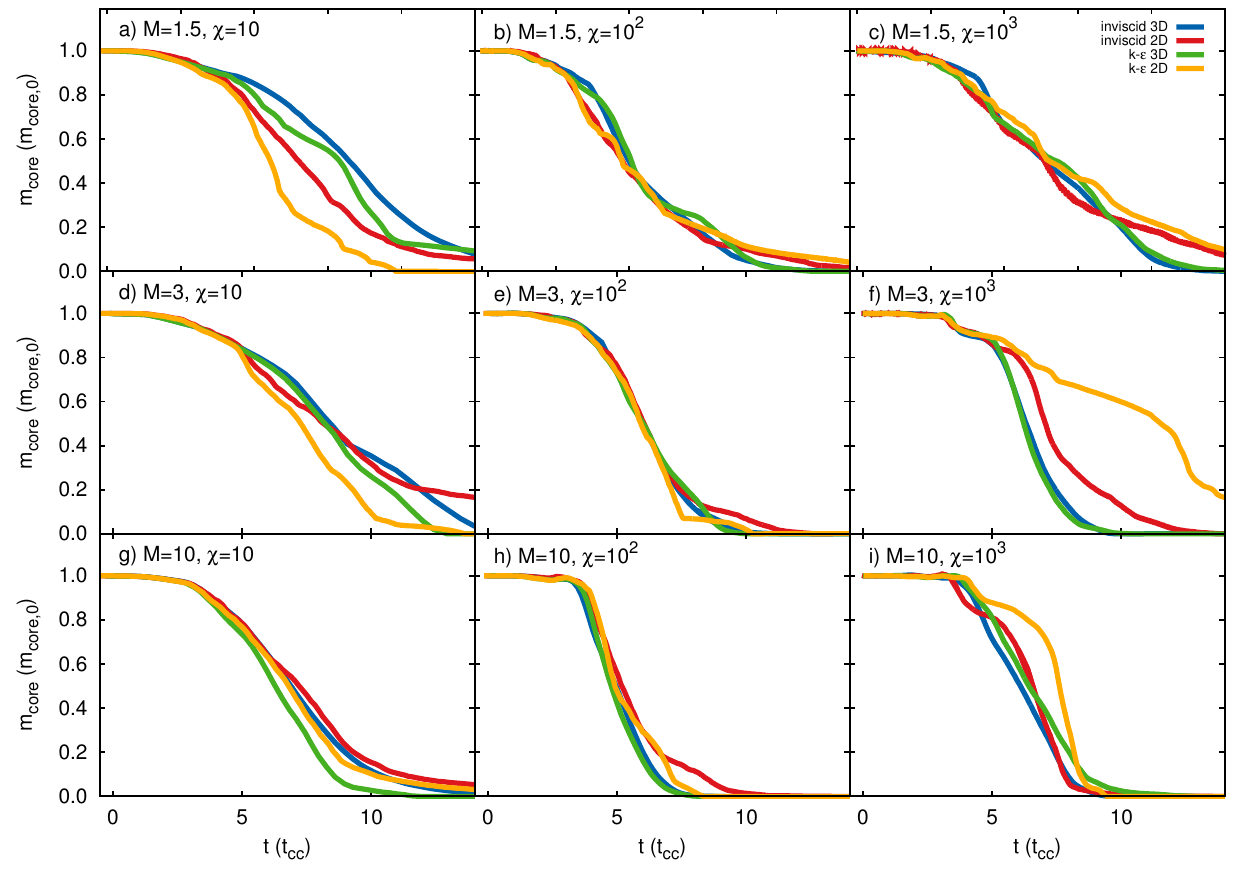}}
  \vspace{-4mm}
  \caption{Comparison of the time evolution of the normalized core
    mass, $m_{\rm core}/m_{\rm core,0}$, for 2D (red and yellow) and
    3D (blue and green) simulations with $M=1.5, 3$ and 10 (top,
    middle and bottom panels, respectively), and $\chi=10, 10^{2}$ and
    $10^{3}$ (left, centre and right panels, respectively).  Results
    from inviscid and $k$-$\epsilon$ simulations are shown. Note the
    difference in the time-scale for the top panels (tick marks are at
    intervals of $5\,t_{\rm cc}$). The resolution of
    each of the 3D calculations is noted in Table~\ref{tab:maxres}.
    All the 2D calculations were performed at resolution $R_{128}$.}
    \label{fig:mcore_2Dvs3D}
\end{figure*}

Fig.~\ref{fig:mcore_2Dvs3D} shows the evolution of $m_{\rm core}$ as a
function of $M$ and $\chi$ for 2D and 3D simulations, with and without
the subgrid turbulence model. This figure reveals that the 2D and 3D
calculations are generally in very good agreement with each other. The
most obvious differences occur between the $M=1.5$, $\chi=10$
simulations. The $M=3$, $\chi=10^{3}$, 2D $k$-$\epsilon$ simulation
shown in panel f) is also surprisingly different from the
others. Examination of this simulation shows that it proceeds
similarly to the others, but that at later times the cloud and its
core remains more compact than in the 2D inviscid or the 3D
calculations. This ultimately leads to slower ablation and
acceleration. It is not obvious why the cloud behaves so differently
in this case, but we note similar behaviour in a 3D simulation at
resolution $R_{32}$, which is examined in more detail in the appendix.
The 4 models are most closely aligned when $\chi=10^{2}$ (for all
$M$), and agreement is also good for the $M=10$, $\chi=10$
simulations. It is also interesting that the core is destroyed
noticeably quicker in 3D simulations when $M=3$ and $\chi=10^{3}$.

\begin{figure*}
\resizebox{130mm}{!}{\includegraphics{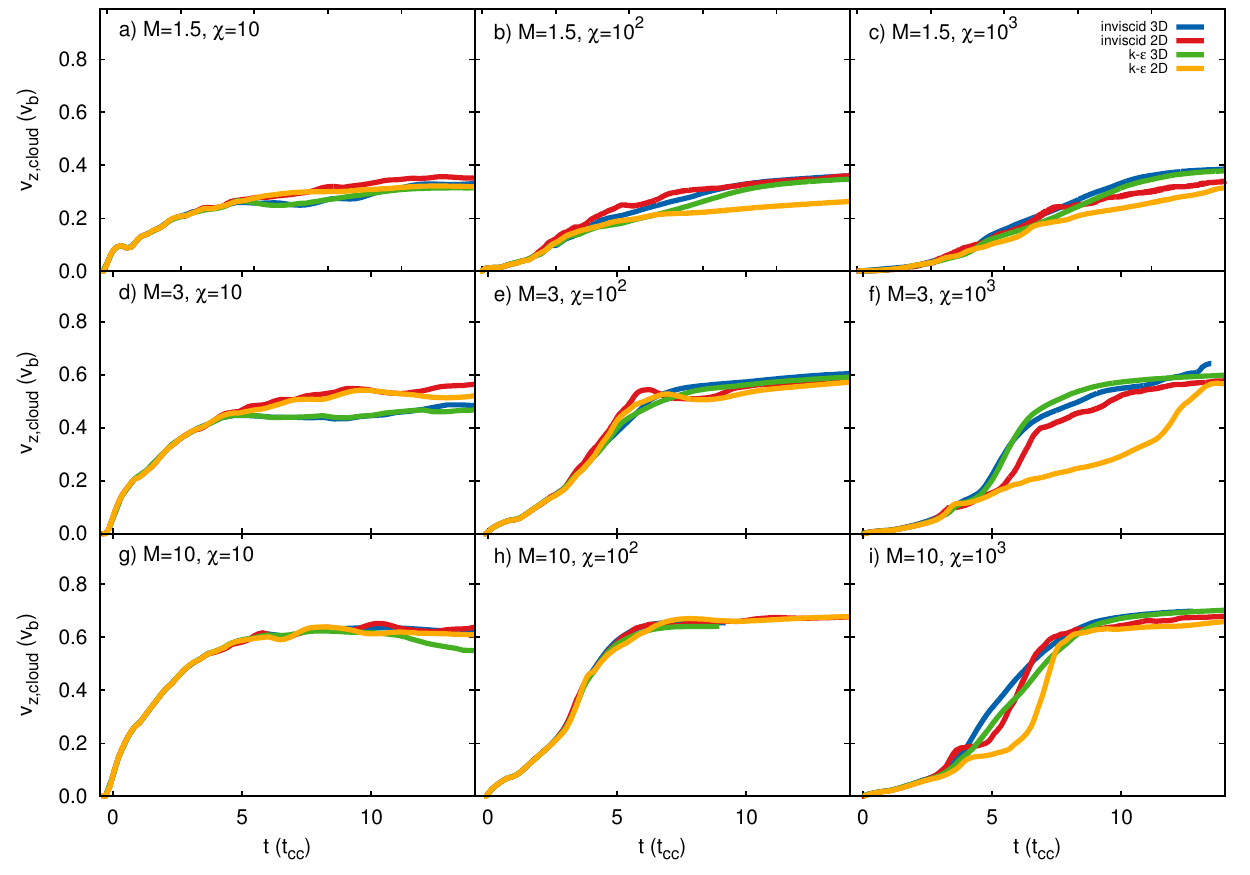}}
  \vspace{-4mm}
   \caption{As Fig.~\ref{fig:mcore_2Dvs3D} but showing the time
     evolution of the mean cloud velocity, $<v_{\rm
       z,cloud}>$.}
    \label{fig:vzcloud_2Dvs3D}
\end{figure*}

\begin{figure*}
\resizebox{130mm}{!}{\includegraphics{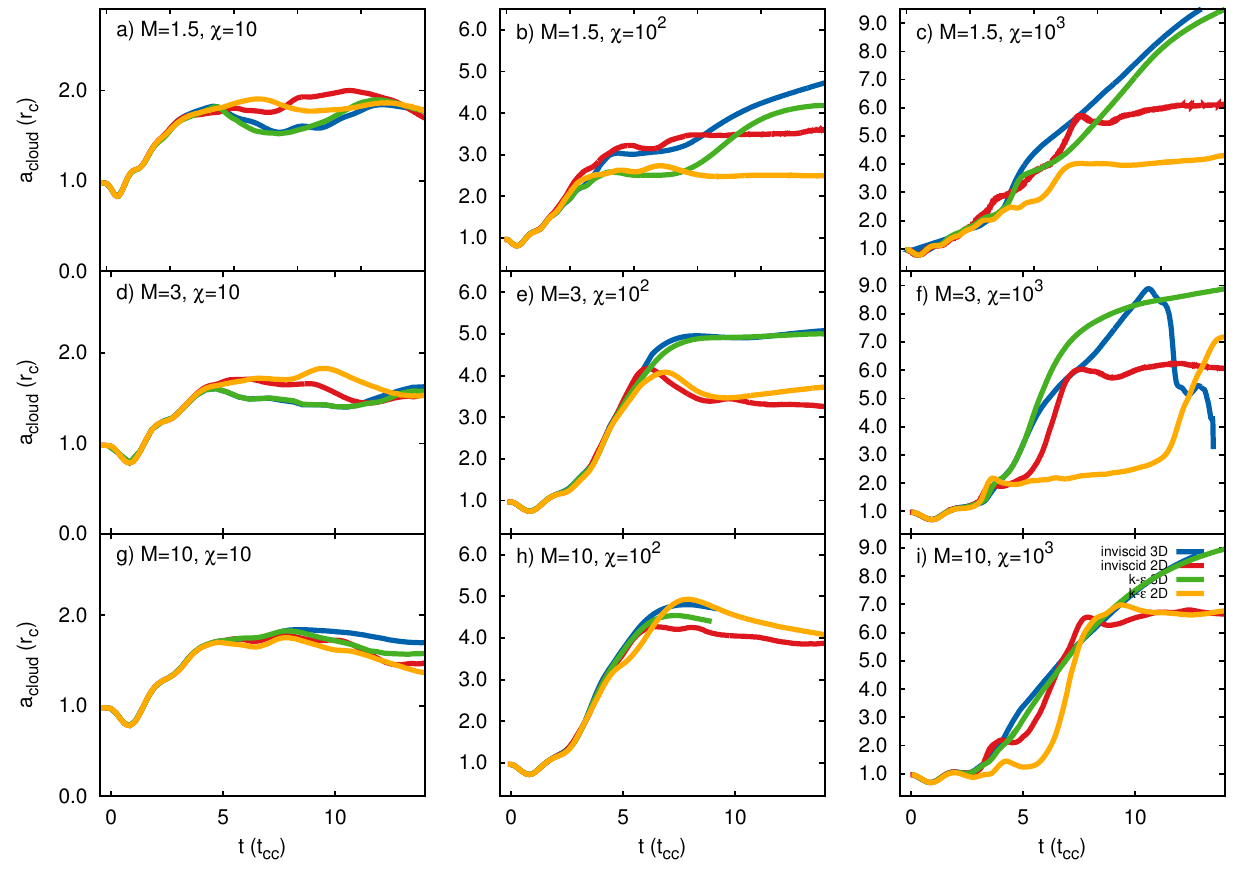}}
  \vspace{-4mm}
  \caption{As Fig.~\ref{fig:mcore_2Dvs3D} but showing the time
    evolution of the effective transverse radius of the cloud, $a_{\rm
      cloud}$.}
    \label{fig:acloud_2Dvs3D}
\end{figure*}

\begin{figure*}
\resizebox{130mm}{!}{\includegraphics{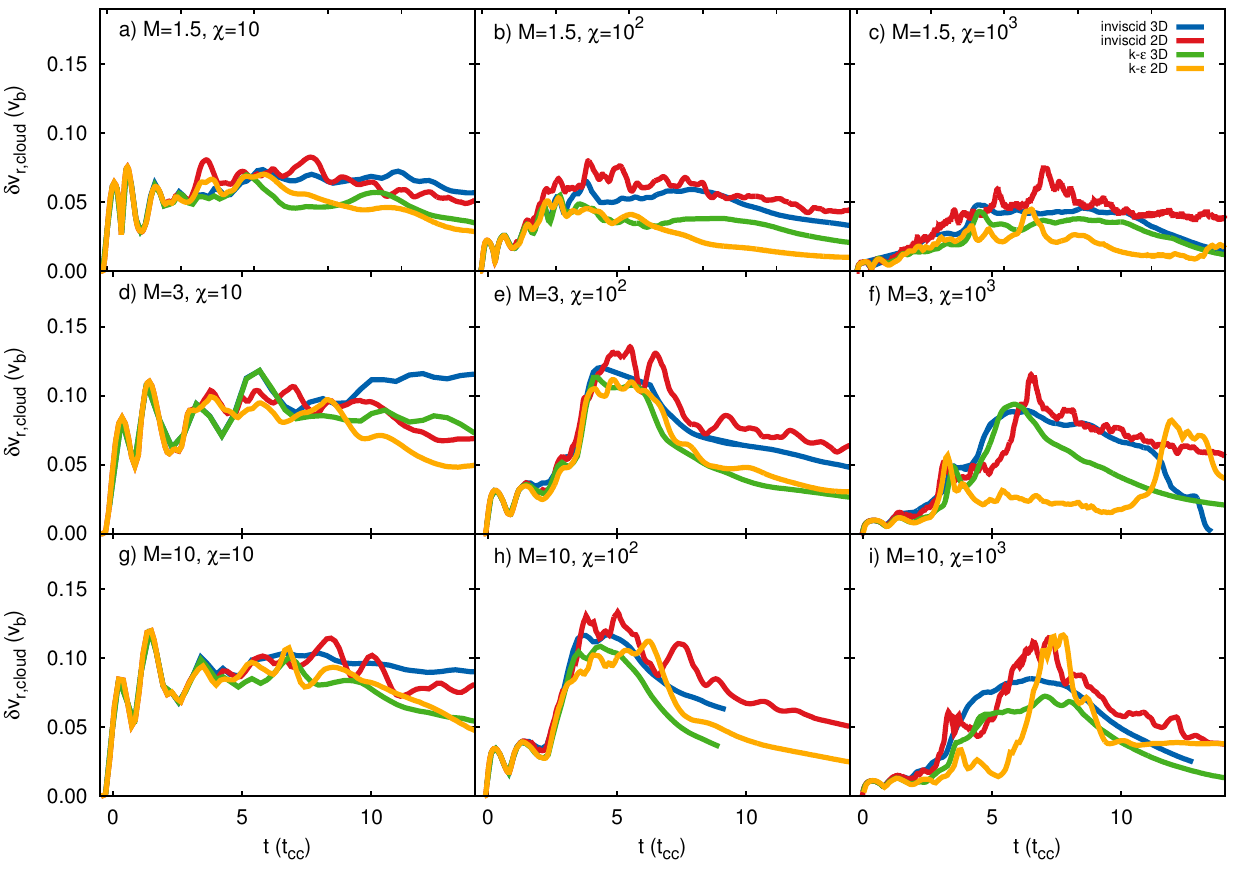}}
  \vspace{-4mm}
  \caption{As Fig.~\ref{fig:mcore_2Dvs3D} but showing the time
    evolution of the cloud velocity dispersion in the radial
    direction, $\delta v_{\rm r,cloud}$.}
    \label{fig:dvrcloud_2Dvs3D}
\end{figure*}

\begin{figure*}
\resizebox{130mm}{!}{\includegraphics{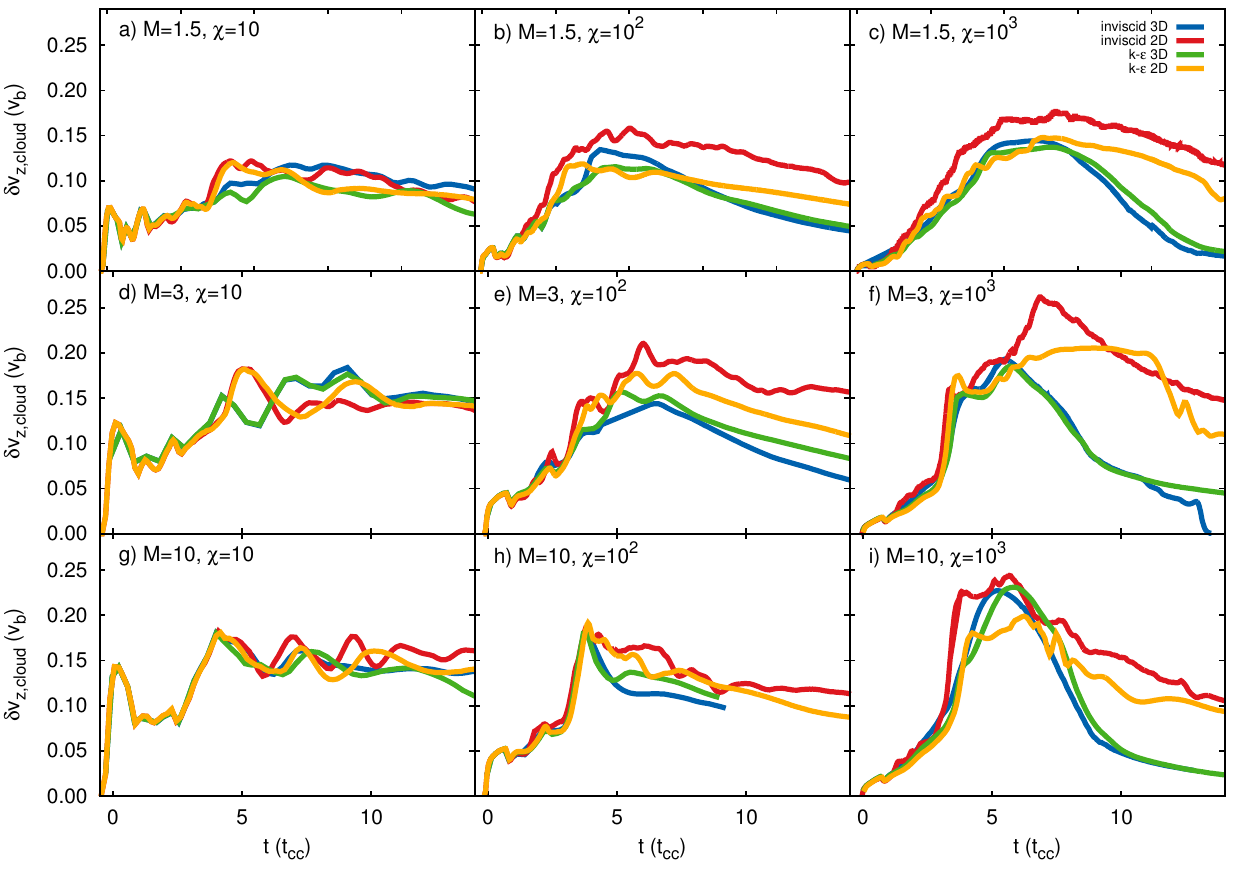}}
  \vspace{-4mm}
  \caption{As Fig.~\ref{fig:mcore_2Dvs3D} but showing the time
    evolution of the cloud velocity dispersion in the axial direction,
    $\delta v_{\rm z,cloud}$.}
    \label{fig:dvzcloud_2Dvs3D}
\end{figure*}

Previously, \citet{Nakamura:2006} reported that global quantities from
a single 3D simulation of a shock striking a relatively hard-edged
cloud ($n=8$) with $M=10$, $\chi=10$, at resolution $R_{60}$, are
within 10\% of an equivalent 2D calculation for $t < 10\,t_{\rm cc}$
(see their Sec.~9.2.2). In
Figs.~\ref{fig:mcore_2Dvs3D}-\ref{fig:dvzcloud_2Dvs3D} we compare our
2D and 3D results against each other. Examination of panel g) in
Figs.~\ref{fig:mcore_2Dvs3D}-\ref{fig:dvzcloud_2Dvs3D} reveals that
our 2D and 3D simulations are comparably similar for such parameters.

Fig.~\ref{fig:vzcloud_2Dvs3D} shows the acceleration of the cloud.
Good agreement between the simulations is again seen, with the 2D
$M=3$, $\chi=10^3$ $k$-$\epsilon$ simulation again significantly
discrepant. Clouds appear to generally be accelerated marginally
faster in 3D calculations compared to 2D calculations when $\chi$ is
high. This is caused by a faster and/or greater increase in the
transverse radius of the cloud in 3D simulations (see
Fig.~\ref{fig:acloud_2Dvs3D}). In contrast, the acceleration of clouds
in the 3D simulations appears to be slightly slower when $\chi$ is low
(particularly for $M=3$). Again, this appears to be related to
differences in the transverse radius of the cloud.  \citet{Xu:1995}
note that their average cloud velocity reaches 0.85 of the postshock
velocity (so $0.64 v_{\rm b}$) by $t\approx 4\,t_{\rm cc}$ for M=10,
$\chi=10$, so our results are in good agreement with theirs.

Fig.~\ref{fig:dvrcloud_2Dvs3D} shows the evolution of the transverse
cloud velocity dispersion, $\delta v_{\rm r,cloud}$. Of note is that
$\delta v_{\rm r,cloud}$ is almost always greater in the inviscid
simulations than in simulations that use the subgrid turbulence
model. This is irrespective of the dimensionality, and likely
indicates the damping of velocity motions by the turbulent viscosity
in the subgrid model. Again the $M=3$, $\chi=10^3$ $k$-$\epsilon$
simulation is noticeably discrepant.

The longitudinal velocity dispersion of the cloud is shown in
Fig.~\ref{fig:dvzcloud_2Dvs3D}. The simulation results are broadly
comparable, but for low to moderate $M$ and moderate to high $\chi$,
$\delta v_{\rm z,cloud}$ appears to peak higher and decay more slowly in the 2D
simulations. This behaviour may be related to the lower resolution
used in the 3D simulations in this region of parameter space.

\begin{figure*}
\resizebox{130mm}{!}{\includegraphics{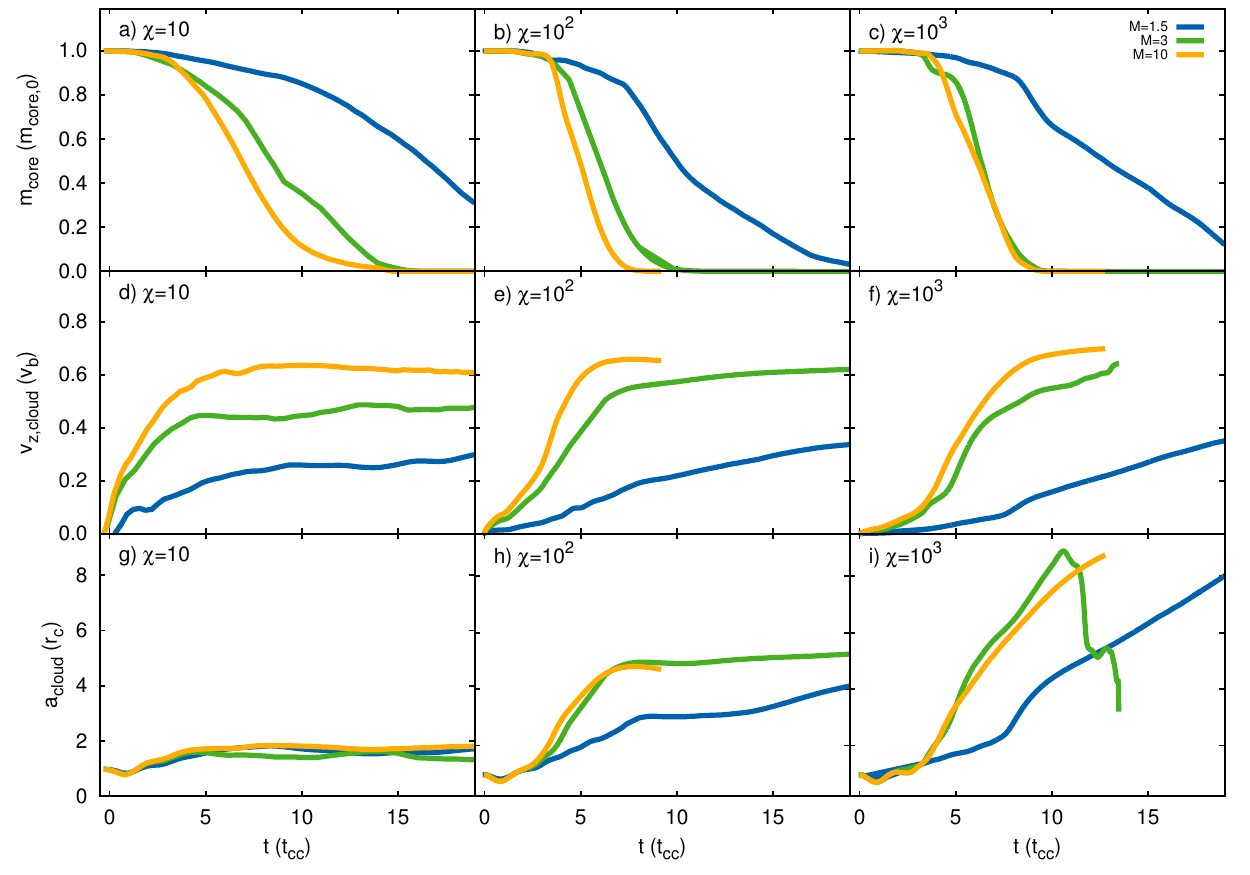}}
  \vspace{-4mm}
  \caption{The time
    evolution of the normalized core mass, cloud mean velocity, and cloud
    transverse radius as a function of Mach number and cloud density
    contrast, for 3D inviscid simulations.}
    \label{fig:3Dresults}
\end{figure*}

Fig.~\ref{fig:3Dresults} summarizes the Mach and density contrast
dependence of the 3D inviscid results. These results can be compared
against the 2D $k$-$\epsilon$ results in Figs.~5, 8 and 9 in
\citet{Pittard:2010}. The same behaviour is seen but there are some
qualitative differences. Compared to the 2D results, the 3D behaviour
of $a_{\rm cloud}$ at $\chi=10^{3}$ shows much more variation with
$M$. The major difference concerning the behaviour of $m_{\rm core}$
is the much less rapid ablation of the cloud when $M=1.5$ and
$\chi=10$ in the 3D simulation compared to that in the 2D simulation.

\begin{figure*}
\resizebox{130mm}{!}{\includegraphics{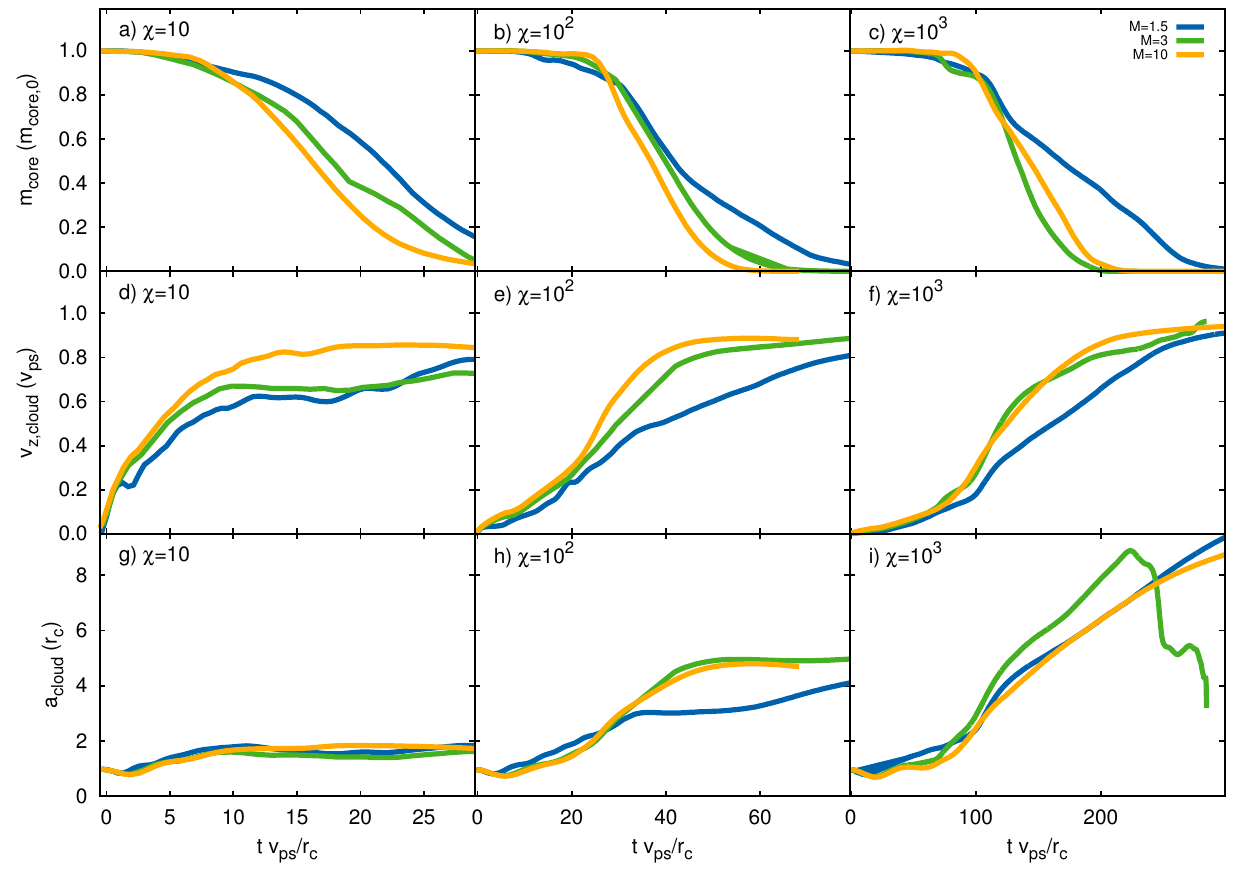}}
  \vspace{-4mm}
  \caption{As Fig.~\ref{fig:3Dresults}, but plotting on the
    dimensionless timescale $tv_{\rm ps}/r_{\rm c}$. $v_{\rm z,cloud}$
  is also scaled to the post-shock velocity, $v_{\rm ps}$.}
    \label{fig:3Dresults3}
\end{figure*}

Fig.~\ref{fig:3Dresults3} takes the results in
Fig.~\ref{fig:3Dresults} and plots them on a dimensionless timescale
based on the post-shock velocity. Since the mixing and acceleration of
the cloud is driven by the velocity gradients in the post-shock flow,
we see that the data collapses to a tighter trend. This extends the
behaviour previously noted by \citet{Niederhaus:2008} to higher $\chi$
and $M$.

\begin{figure*}
\resizebox{130mm}{!}{\includegraphics{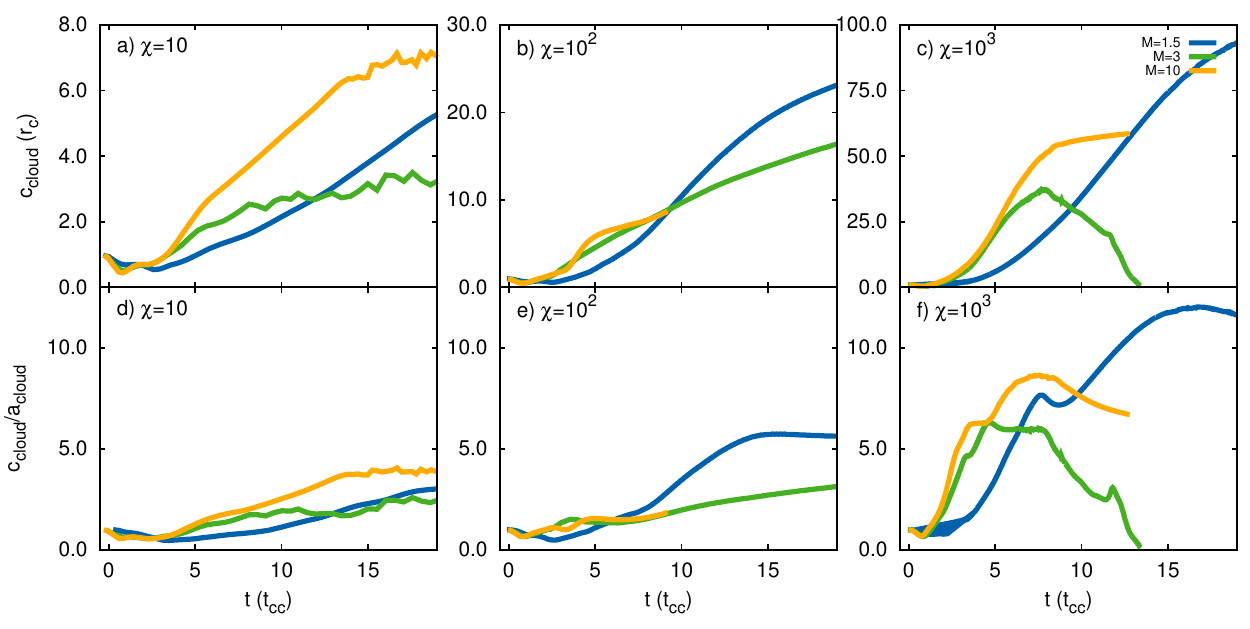}}
  \vspace{-4mm}
  \caption{The time evolution of the cloud axial radius and shape as a
    function of Mach number and cloud density contrast, for 3D
    inviscid simulations.}
    \label{fig:3Dresults2}
\end{figure*}

Fig.~\ref{fig:3Dresults2} also shows the variation of $c_{\rm cloud}$
and $c_{\rm cloud}/a_{\rm cloud}$ for the 3D inviscid
calculations. As previously noted by \citet{Pittard:2010}, a long
``tail-like'' feature is formed only when $\chi \gtsimm 10^{3}$.
Comparison of $a_{\rm cloud}$, $c_{\rm cloud}$ and $c_{\rm
  cloud}/a_{\rm cloud}$ with Fig.~4 in \citet{Xu:1995} reveals good
agreement for $M=10$ and $\chi=10$.

\begin{table}
\centering
  \caption[]{Various timescales (in units of $t_{\rm cc}$) calculated
    from the 3D inviscid simulations.}
\label{tab:timescales}
\begin{tabular}{llcccc}
\hline
$\chi$ & $M$ & $t_{\rm drag}$ & $t_{\rm drag,KMC}$ & $t_{\rm mix}$ & $t_{\rm life}$ \\
\hline
10 & 1.5 & 3.14 & 8.65 & 16.4 & $>30$ \\
   & 3   & 1.36 & 3.86 & 8.42 & 16.6\\
   & 10  & 0.98 & 2.69 & 6.89 & 15.4 \\
$10^{2}$ & 1.5 & 6.85 & 13.3 & 9.97 & 26.1\\
        & 3   & 3.65 & 5.38 & 6.00 & 12.8\\
        & 10  & 3.06 & 4.03 & 4.95 & 9.50\\
$10^{3}$ & 1.5 & 9.79 & 14.8 & 12.8 & 28.4\\
        & 3   & 5.13 & 6.57 & 6.31 & 10.7\\
        & 10  & 4.55 & 6.18 & 6.10 & 10.4\\ 
\hline
\end{tabular}
\end{table}

\subsection{Timescales}
\begin{figure}
\includegraphics{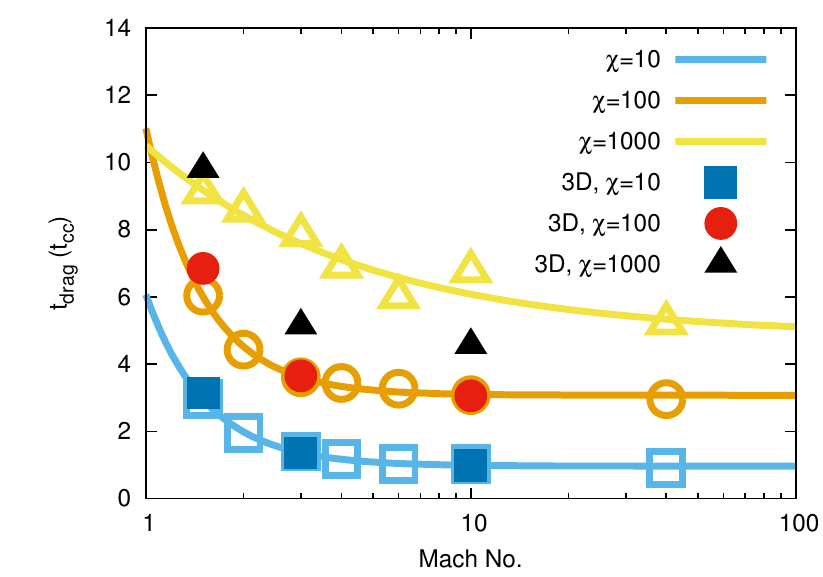}
\includegraphics{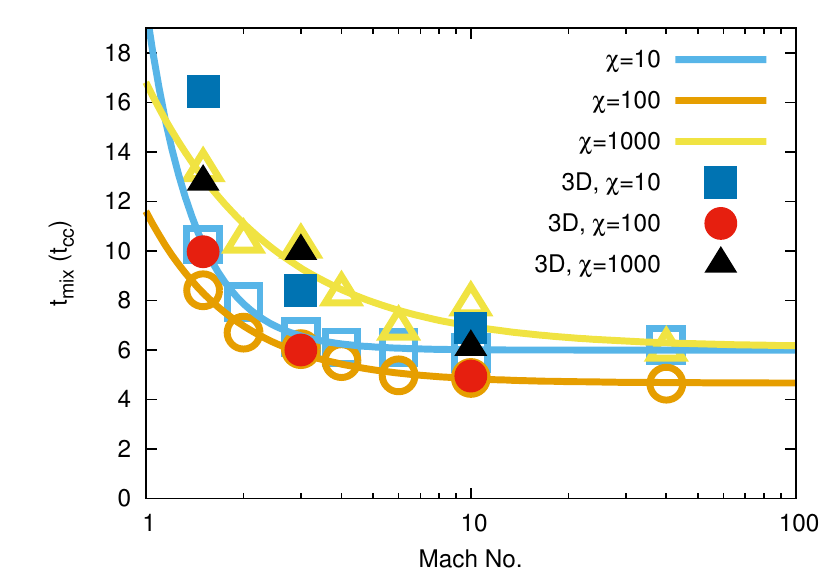}
\includegraphics{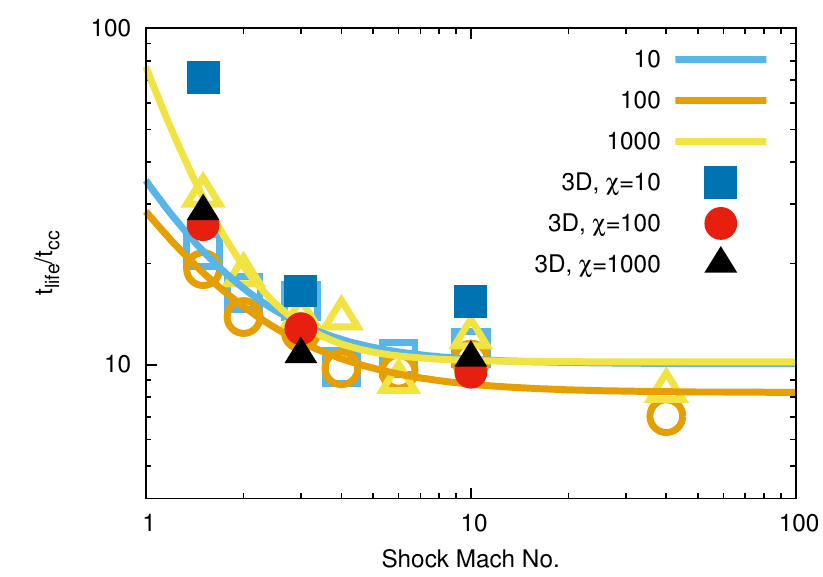}
    \caption{Top: $t_{\rm drag}$ (for the cloud); middle: $t_{\rm
        mix}$; and bottom: $t_{\rm
        life}$, as functions of the Mach number $M$ and
      cloud density contrast $\chi$. The 2D results are plotted using
      the open symbols, and fits to the 2D data are also shown
      \citep[cf.][]{Pittard:2010}.}
    \label{fig:tdragtmix}
\end{figure}

Values of $t_{\rm drag}$, $t_{\rm mix}$ and $t_{\rm life}$ are noted
in Table~\ref{tab:timescales}. In all cases $t_{\rm drag} < t_{\rm
  mix} < t_{\rm life}$ (though sometimes $t_{\rm drag,KMC} > t_{\rm
  mix}$).  Fig.~\ref{fig:tdragtmix} shows the values of
$t_{\rm drag}$, $t_{\rm mix}$ and $t_{\rm life}$ as a function of $M$
and $\chi$ for the 3D inviscid simulations. Also shown are the
corresponding values from the 2D $k$-$\epsilon$ simulations in
\citet{Pittard:2010} and the fits made to this latter data.  There is
more scatter in $t_{\rm drag}$ and $t_{\rm mix}$ when $\chi=10^{3}$
due to spontaneous and random fragmentation.

Excellent agreement is found between the 2D and 3D results for $t_{\rm
  drag}$ when $\chi=10$ and 100, but clouds with $\chi=10^{3}$
accelerate more rapidly in the 3D calculations when $M \gtsimm
3$. Since in the strong shock limit $t_{\rm drag}/t_{\rm cc} \propto
\chi^{1/2}$ \citep[see Eq.~9 in][]{Pittard:2010}, one wonders whether
the lower than expected drag time for the $M=10$, $\chi=10^{3}$ 3D
simulation is a result of the lower resolution used. Alternatively,
this may instead just be a result of the larger scatter when
$\chi=10^{3}$ and the small number of simulations performed in 3D.

\citet{Xu:1995} postulated that clouds may be mixed more rapidly in 3D
simulations due to the non-axisymmetric instabilities which develop,
but this has not been tested prior to this work. We have already shown
generally good agreement between our 2D and 3D calculations, both in
terms of the morphology, and in terms of various global
quantities. Fig.~\ref{fig:mcore_2Dvs3D} shows that this is the case
for $m_{\rm core}$, and Fig.~\ref{fig:tdragtmix} now lends further
support by revealing that the 2D and 3D results have similar values of
$t_{\rm mix}$ for $\chi=10^{2}$ and $10^{3}$. However, there is one
set of simulations which stand out: for $M=1.5$ and $\chi=10$ it seems
that the cloud takes {\em longer} to mix in the 3D
simulations. Similar behaviour is found for $t_{\rm life}$.
Fig.~\ref{fig:mcore_2Dvs3D} shows that $m_{\rm core}$ declines
increasingly slowly at late times in the 3D inviscid simulation,
whereas in the 2D $k$-$\epsilon$ simulation $m_{\rm core}$ declines
much more rapidly, reaching zero by $t\approx 20\,t_{\rm
  cc}$. Although not quite as rapid, the 2D inviscid simulation also
has $m_{\rm core}$ declining faster than the 3D simulations.

\begin{figure*}
\resizebox{175mm}{!}{\includegraphics{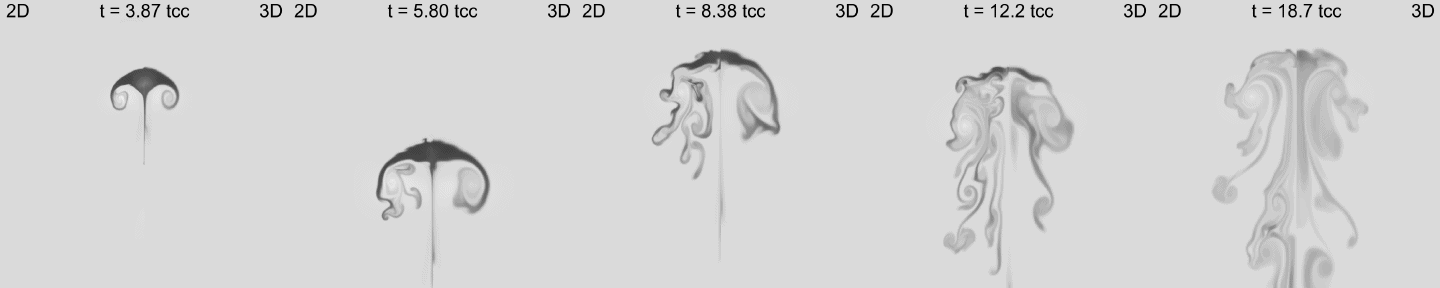}}
\caption{As Fig.~\ref{fig:chi1e1_2dvs3d} but for the $M=1.5$,
  $\chi=10$ simulations. The
  first 2 frames show the same region ($-2 < X < 6$, $0 < Y < 4$ in
  units of $r_{\rm c}$) so that the motion of the cloud is clear. The
  other frames shift the $X$-axis to show $2 < X < 10$, $4 < X < 12$,
  and $9 < X < 17$ at $t=8.38$, 12.2, and $18.7\,t_{\rm cc}$,
  respectively. Note that the 2D
  simulation has a resolution of $R_{128}$, while the 3D simulation is
  at the lower resolution of $R_{64}$.}
    \label{fig:m1.5chi10_2dvs3d}
\end{figure*}

Fig.~\ref{fig:m1.5chi10_2dvs3d} examines the 2D and 3D {\em inviscid}
simulations side-by-side. It is clear that secondary vortices form
earlier and are more prevelant in the higher resolution 2D simulation,
and this may be the cause of the faster decline in $m_{\rm core}$. We
also raise the possibility that the subgrid turbulence model is
perhaps overly efficient at mixing the core material into low Mach
number flows, given that the 2D inviscid simulation shows a slightly
less rapid decline in $m_{\rm core}$ (see
Fig.~\ref{fig:mcore_2Dvs3D}).

\section{Conclusions}
\label{sec:conclusions}

This is the third of a series of papers investigating the turbulent
destruction of clouds. Our first paper \citep{Pittard:2009} noted the
benefits of using a sub-grid turbulence model in simulations of
shock-cloud interactions and found that clouds could be destroyed more
rapidly when overrun by a highly turbulent flow. The inviscid and
$k$-$\epsilon$ simulations were found to be in good agreement when the
cloud density contrast $\chi \ltsimm 100$, but they became
increasingly divergent as $\chi$ increased. The $k$-$\epsilon$
simulations also displayed significantly better convergence
properties, such that $\sim 30$ grid cells per cloud radius is needed
for reasonable convergence (compared to the $\sim 120$ needed in
inviscid simulations).

Our second paper \citep{Pittard:2010} investigated how the nature of
the interaction changed with the Mach number $M$ and density contrast
$\chi$. For $M\gtsimm7$, the lifetime of the cloud, $t_{\rm life}$,
showed little variation with $M$ or $\chi$ and we found that $t_{\rm
  life} \sim 10\,t_{\rm cc}$. Due to the gentler nature of the
interaction, $t_{\rm life}$ increases significantly at lower Mach
numbers.  A popular analytical formula for the mass-loss rate due to
hydrodynamic ablation \citep{Hartquist:1986} was shown to predict
cloud lifetimes which were inconsistent with Mach scaling and which
had a $\chi$ dependence which was not supported by the simulation
results.

In this third paper we have examined whether the conclusions in
\citet{Pittard:2010} remain valid for three dimensional simulations,
and whether the nature of the interaction is different in 2D
axisymmetric and fully 3D simulations.  This was motivated by previous
reports that clouds are destroyed more rapidly in 3D due to the
additional development of non-axisymmetric instabilites. However, our
detailed investigation, covering Mach numbers from $1.5-10$ and cloud
density contrasts from $10-10^{3}$, has instead revealed that the
interaction proceeds very similarly in 2D and 3D. Although
non-azimuthal modes lead to different behaviour in the later stages of
the interaction, they have very little effect on key global quantities
such as the lifetime of the cloud and its acceleration.

In particular, we are not able to confirm differences in the hollowing
or ``voiding'' of the cloud between 2D and 3D simulations with $M=10$
and $\chi=10$. This contrasts with the findings in \citet{Klein:2003},
where 3D experimental data and 3D simulations display such voiding but
synthetic shadowgrams based on 2D simulations do not.  We note that
the 2D and 3D simulations in \citet{Klein:2003} are computed with
different numerical codes and different initial conditions. Our work
shows that when the same code and initial conditions are used the
interaction evolves almost identically.

The biggest differences between our 2D and 3D simulations occur for
$M=1.5$ and $\chi=10$ - the destruction is noticeably slower in 3D. It
is not clear why this is so, though secondary vortices form earlier
and are more prevelant in the higher resolution 2D simulations. Having
said this, our resolution tests indicate that increasing the
resolution of the 3D simulation is likely to slow the destruction of
the cloud yet further (see Fig.~\ref{fig:mcore_restest_M10}).
Additional 3D simulations at higher resolution are necessary to
resolve this issue.

We have also shown how the cloud acceleration (through $t_{\rm drag}$)
and mixing (through $t_{\rm mix}$) are affected by low resolution. We
find that these timescales are up to $5\times$ shorter for clouds at
resolution $R_{1}$ (i.e. {\em very} poorly resolved clouds).  This is
relevant to simulations of the mixing and entrainment of cold clouds
in multiphase-flows: simulations which do not adequately resolve the
cold clouds in the flow will underestimate $t_{\rm drag}$ and $t_{\rm
  mix}$, often to a significant degree.

Our work has also highlighted that 3D inviscid and $k$-$\epsilon$
simulations give typically very similar results.  This is somewhat
surprising given that 2D calculations can show significant differences
\citep[see][]{Pittard:2009,Pittard:2010}, but must be related to the
different way that vortices behave and evolve in 2D and 3D flows.
Unlike in 2D, we find no evidence for convergence at lower resolution
when employing the $k$-$\epsilon$ model. Hence, there seems to be no
compelling reason to use the $k$-$\epsilon$ model in 3D calculations,
but clearly it remains very useful in 2D calculations.

In future work we will examine the dependence of the interaction on
the shape and orientation of the cloud, and in particular whether the
nature of the interaction changes when the cloud is
elongated/filamentary. By examining the destruction of spherical
clouds in 3D, the present work has laid the necessary groundwork for
this forthcoming study.

\section*{Acknowledgements}
We would like to thank the referee for a timely and useful report. 
JMP and ERP thank STFC for funding, and Kathryn Goldsmith for comments on an earlier
draft. We would also like to thank S.\,Falle for the use of the {\sc
  MG} hydrodynamics code used to calculate the simulations in this
work and S.\,van~Loo for adding SILO output to it. The calculations
for this paper were performed on the DiRAC Facility jointly funded by
STFC, the Large Facilities Capital Fund of BIS and the University of
Leeds. This paper made use of VisIt \citep{Childs:2012}. 






\begin{thebibliography}{99}
\bibitem[\protect\citeauthoryear{Abdo et al.}{2010}]{Abdo:2010}
Abdo A.~A., et al., 2010, Science, 327, 1103
\bibitem[\protect\citeauthoryear{Ackermann et
    al.}{2013}]{Ackermann:2013}
Ackermann M., et al., 2103, Science, 339, 807
\bibitem[\protect\citeauthoryear{Agertz et al.}{2007}]{Agertz:2007}
Agertz O., et al., 2007, MNRAS, 380, 963
\bibitem[\protect\citeauthoryear{Alarie, Bilodeau \& Drissen}{Alarie
    et al.}{2014}]{Alarie:2014}
Alarie A., Bilodeau A., Drissen L., 2014, MNRAS, 441, 2996
\bibitem[\protect\citeauthoryear{Al\={u}zas et al.}{2012}]{Aluzas:2012}
Al\={u}zas R., Pittard J.~M., Hartquist T.~W., Falle S.~A.~E.~G., Langton
R., 2012, MNRAS, 425, 2212
\bibitem[\protect\citeauthoryear{Arthur \& Henney}{1996}]{Arthur:1996} 
Arthur, S.~J., \& Henney, W.~J.\ 1996, ApJ, 457, 752
\bibitem[\protect\citeauthoryear{Aschenbach, Egger \&
    Trumper}{Aschenbach et al.}{1995}]{Aschenbach:1995}
Aschenbach B., Egger R., Trumper J., 1995, Nature, 373, 587
\bibitem[\protect\citeauthoryear{Blair et al.}{2000}]{Blair:2000}
Blair W.~P., et al., 2000, ApJ, 537, 667
\bibitem[\protect\citeauthoryear{Brogan et al.}{2013}]{Brogan:2013}
Brogan C.~L., et al., 2013, ApJ, 771, 91
\bibitem[\protect\citeauthoryear{Bykov et al.}{2008}]{Bykov:2008}
Bykov A.~M., 2008, ApJ, 676, 1050
\bibitem[\protect\citeauthoryear{Cecil et al.}{2001}]{Cecil:2001}
Cecil G., Bland-Hawthorn J., Veilleux S., Filippenko A.~V., 2001, ApJ,
555, 338
\bibitem[\protect\citeauthoryear{Ceverino \&
    Klypin}{2009}]{Ceverino:2009}
Ceverino D., Klypin A., 2009, ApJ, 695, 292
\bibitem[\protect\citeauthoryear{Chen \& Slane}{2001}]{Chen:2001}
Chen Y., Slane P.~O., 2001, ApJ, 563, 202
\bibitem[\protect\citeauthoryear{Chevalier \&
    Kirshner}{1979}]{Chevalier:1979}
Chevalier R.~A., Kirshner R.~P., 1979, ApJ, 233, 154
\bibitem[\protect\citeauthoryear{Chi\`{e}ze \& Lazareff}{1981}]{Chieze:1981}
Chi\`{e}ze, J.~P., \& Lazareff, B.\ 1981, A\&A, 95, 194
\bibitem[\protect\citeauthoryear{Childs et al.}{2012}]{Childs:2012}
Childs H., et al., 2012, ``VisIt: An End-User Tool For Visualizing and
Analyzing Very Large Data'', in High Performance
Visualization--Enabling Extreme-Scale Scientific Insight (eds. E. Wes
Bethel, Hank Childs, Charles Hansen), p.357-372, CRC Press
\bibitem[\protect\citeauthoryear{Close et al.}{2013}]{Close:2013}
Close J.~L., Pittard J.~M., Hartquist T.~W., Falle S.~.A.~E.~G., 2013,
MNRAS, 436, 3021 
\bibitem[\protect\citeauthoryear{Cooper et al.}{2008}]{Cooper:2008}
Cooper J.~L., Bicknell G.~V., Sutherland R.~S., Bland-Hawthorn J., 2008, ApJ, 674, 157
\bibitem[\protect\citeauthoryear{Cowie, McKee \& Ostriker}{Cowie et al.}{1981}]{Cowie:1981}
Cowie, L.~L., McKee, C.~F., \& Ostriker, J.~P.\ 1981, ApJ, 247, 908
\bibitem[\protect\citeauthoryear{Creasey, Theuns \& Bower}{Creasey et
    al.}{2013}]{Creasey:2013}
Creasey P., Theuns T., Bower R.~G., 2013, MNRAS, 429, 1922
\bibitem[\protect\citeauthoryear{Dale et al.}{2014}]{Dale:2014}
Dale J.~E., Ngoumou J., Ercolano B., Bonnell I.~A., 2014, MNRAS, 442,
694
\bibitem[\protect\citeauthoryear{de Avillez \&
    Breitschwerdt}{2005}]{deAvillez:2005}
de Avillez M.~A., Breitschwerdt D., 2005, A\&A, 436, 585
\bibitem[\protect\citeauthoryear{Dubois \&
    Teyssier}{2008}]{Dubois:2008}
Dubois Y., Teyssier R., 2008, A\&A, 477, 79
\bibitem[\protect\citeauthoryear{Dursi \& Pfrommer}{2008}]{Dursi:2008}
Dursi L.~J., Pfrommer C., 2008, ApJ, 677, 993
\bibitem[\protect\citeauthoryear{Dyson, Arthur \& Hartquist}{Dyson et al.}{2002}]{Dyson:2002}
Dyson, J.~E., Arthur, S.~J., \& Hartquist, T.~W.\ 2002, A\&A, 390, 1063
\bibitem[\protect\citeauthoryear{Elmegreen \& Scalo}{2004}]{Elmegreen:2004}
Elmegreen B.~G., Scalo J., 2004, ARA\&A, 42, 211
\bibitem[\protect\citeauthoryear{Elmhamdi et
    al.}{2004}]{Elmhamdi:2004}
Elmhamdi A., Danziger I.~J., Cappellaro E., Della Valle M., Gouiffes
C., Phillips M.~M., Turatto M., 2004, A\&A, 426, 963
\bibitem[\protect\citeauthoryear{Falle}{1991}]{Falle:1991}
Falle S.~A.~E.~G., 1991, MNRAS, 250, 581
\bibitem[\protect\citeauthoryear{Farris \& Russell}{1994}]{Farris:1994}
Farris M.~H., Russell C.~T., 1994, J. Geophys. Research, 99, 17681
\bibitem[\protect\citeauthoryear{Fassia et al.}{1998}]{Fassia:1998}
Fassia A., Meikle W.~P.~S., Geballe T.~R., Walton N.~A., Pollacco
D.~L., Rutten R.~G.~M., Tinney C., 1998, MNRAS, 299, 150
\bibitem[\protect\citeauthoryear{Fesen \& Kirshner}{1980}]{Fesen:1980}
Fesen R.~A., Kirshner R.~P., 1980, ApJ, 242, 1023
\bibitem[\protect\citeauthoryear{Fesen}{2001}]{Fesen:2001}
Fesen R.~A., Morse J.~A., Chevalier R.~A., Borkowski K.~J., Gerardy
C.~L., Lawrence S.~S., van den Bergh S., 2001, AJ, 122, 2644
\bibitem[\protect\citeauthoryear{Fesen et al.}{2011}]{Fesen:2011}
Fesen R.~A., Zastrow J.~A., Hammell M.~C., Shull J.~M., Silvia D.~W.,
2011, ApJ, 736, 109
\bibitem[\protect\citeauthoryear{Filippenko \&
    Sargent}{1989}]{Filippenko:1989}
Filippenko A.~V., Sargent W.~L.~W., 1989, ApJ, 345, L43
\bibitem[\protect\citeauthoryear{Finkelstein et
    al.}{2006}]{Finkelstein:2006}
Finkelstein S.~L., et al., 2006, ApJ, 641, 919
\bibitem[\protect\citeauthoryear{Fujita et al.}{2009}]{Fujita:2009}
Fujita A., Martin C.~L., Mac Low M.-M., New K.~C.~B., Weaver R., 2009,
ApJ, 698, 693
\bibitem[\protect\citeauthoryear{Ghavamian, Hughes \&
    Williams}{Ghavamian et al.}{2005}]{Ghavamian:2005}
Ghavamian P., Hughes J.~P., Williams T.~B., 2005, ApJ, 635, 365
\bibitem[\protect\citeauthoryear{Girichidis et
    al.}{2015}]{Girichidis:2015}
Girichidis P., et al., 2015, MNRAS (arXiv:1508.06646)
\bibitem[\protect\citeauthoryear{Graham et al.}{1995}]{Graham:1995}
Graham J.~R., Levenson N.~A., Hester J.~J., Raymond J.~C., Petre R.,
1995, ApJ, 444, 787
\bibitem[\protect\citeauthoryear{Gregori et al.}{2000}]{Gregori:2000}
Gregori G., Miniati F., Ryu D., Jones T.~W., 2000, ApJ, 543, 775
\bibitem[\protect\citeauthoryear{Giuliani et
    al.}{2011}]{Giuliani:2011}
Giuliani A., et al., 2011, ApJL, 742, 30
\bibitem[\protect\citeauthoryear{Hansen et al.}{2007}]{Hansen:2007}
Hansen J.~F., Robey H.~F., Klein R.~I., Miles A.~R., 2007,
Phys. Plasmas, 14, 056505
\bibitem[\protect\citeauthoryear{Hartquist et al.}{1986}]{Hartquist:1986}
Hartquist T.~W., Dyson J.~E., Pettini M., Smith L.J., MNRAS, 1986,
221, 715
\bibitem[\protect\citeauthoryear{Hennebelle \&
    Iffrig}{2014}]{Hennebelle:2014}
Hennebelle P., Iffrig O., 2014, A\&A (arXiv:1405.7819)
\bibitem[\protect\citeauthoryear{Hill et al.}{2012}]{Hill:2012}
Hill A.~S., et al., 2012, ApJ, 750, 104
\bibitem[\protect\citeauthoryear{Hopkins, Quataert \& Murray}{Hopkins
    et al.}{2012}]{Hopkins:2012}
Hopkins P.~F., Quataert E., Murray N., 2012, MNRAS, 421, 3522
\bibitem[\protect\citeauthoryear{Hwang, Flanagan \& Petre}{Hwang et
    al.}{2005}]{Hwang:2005}
Hwang U., Flanagan K.~A., Petre R., 2005, ApJ, 635, 355
\bibitem[\protect\citeauthoryear{Jiang et al.}{2010}]{Jiang:2010}
Jiang B., Chen Y., Wang J., Su Y., Zhou X., Safi-Harb S., DeLaney T.,
2010, ApJ, 712, 1147
\bibitem[\protect\citeauthoryear{Johansson \& Ziegler}{2013}]{Johansson:2013}
Johansson E.~P.~G., Ziegler U., 2013, ApJ, 766, 45
\bibitem[\protect\citeauthoryear{Joung \& Mac Low}{2006}]{Joung:2006}
Joung M.~K.~R., Mac Low M.-M., 2006, ApJ, 653, 1266
\bibitem[\protect\citeauthoryear{Joung, Mac Low \& Bryan}{Joung et al.}{2009}]{Joung:2009}
Joung M.~K.~R., Mac Low M.-M., Bryan G.~L., 2009, ApJ, 704, 137
\bibitem[\protect\citeauthoryear{Jun, Jones \& Norman}{1996}]{Jun:1996}
Jun B.-I., Jones T.~W., Norman M.~L., 1996, ApJ, 468, L59
\bibitem[\protect\citeauthoryear{Kamper \&
    van~den~Bergh}{1976}]{Kamper:1976}
Kamper K., van den Bergh S., 1976, ApJS, 32, 351
\bibitem[\protect\citeauthoryear{Kane et al.}{1997}]{Kane:1997}
Kane J., et al., 1997, ApJL, 478, L75
\bibitem[\protect\citeauthoryear{Kane et al.}{2000}]{Kane:2000}
Kane J., Arnett D., Remington B.~A., Glendinning S.~G., Bazan G.,
Drake R.~P., Fryxell B.~A., 2000, ApJSS, 127, 365
\bibitem[\protect\citeauthoryear{Katsuda \& Tsunemi}{2006}]{Katsuda:2006}
Katsuda S., Tsunemi H., 2006, ApJ, 642, 917
\bibitem[\protect\citeauthoryear{Katsuda et al.}{2008}]{Katsuda:2008}
Katsuda S., et al., 2008, ApJ, 678, 297
\bibitem[\protect\citeauthoryear{Katsuda et al.}{2010}]{Katsuda:2010}
Katsuda S., et al., 2010, ApJ, 709, 1387
\bibitem[\protect\citeauthoryear{Kim, Ostriker \& Kim}{Kim et al.}{2013}]{Kim:2013}
Kim C.-G., Ostriker E.~C., Kim W.-T., 2013, ApJ (arXiv:1308.3231) 
\bibitem[\protect\citeauthoryear{Kimm et al.}{2015}]{Kimm:2015}
Kimm T., Cen R., Devriendt J., Dubois Y., Slyz A., 2015, MNRAS (arXiv:1501.05655)
\bibitem[\protect\citeauthoryear{Klein, McKee \& Colella}{Klein et al.}{1994}]{Klein:1994}
Klein R.~I., McKee C.~F., Colella P., 1994, ApJ, 420, 213
\bibitem[\protect\citeauthoryear{Klein et al.}{2000}]{Klein:2000}
Klein R.~I., Budil K.~S., Perry T.~S., Bach D.~R., 2000, ApJS, 127, 379
\bibitem[\protect\citeauthoryear{Klein et al.}{2003}]{Klein:2003}
Klein R.~I., Budil K.~S., Perry T.~S., Bach D.~R., 2003, ApJ, 583, 245
\bibitem[\protect\citeauthoryear{Kokusho et al.}{2013}]{Kokusho:2013}
Kokusho T., Nagayama T., Kaneda H., Ishihara D., Lee H.-G., Onaka T.,
2013, ApJL, 768, 8
\bibitem[\protect\citeauthoryear{Koo et al.}{2005}]{Koo:2005}
Koo B.-C., Lee J.-J., Seward F.~D., Moon D.-S., 2005, ApJ, 633, 946
\bibitem[\protect\citeauthoryear{Kwak, Henley \& Shelton}{Kwak et al.}{2011}]{Kwak:2011}
Kwak K., Henley D.~B., Shelton R., 2011, ApJ, 739, 30
\bibitem[\protect\citeauthoryear{Lasker}{1978}]{Lasker:1978}
Lasker B.~M., 1978, ApJ, 223, 109
\bibitem[\protect\citeauthoryear{Lasker}{1980}]{Lasker:1980}
Lasker B.~M., 1980, ApJ, 237, 765
\bibitem[\protect\citeauthoryear{Layes, Jourdan \& Houas}{Layes et
    al.}{2009}]{Layes:2009}
Layes G., Jourdan G., Houas L., 2009, Phys. Fluids, 21:074102
\bibitem[\protect\citeauthoryear{Lea\~{o} et al.}{2009}]{Leao:2009}
Lea\~{o} M.~R.~M., de Gouveia Dal Pino E.~M., Falceta-Gon\c{c}alves D., Melioli
C., Geraissate F.~G., 2009, MNRAS, 394, 157
\bibitem[\protect\citeauthoryear{Levenson, Graham \& Snowden}{Levenson
    et al.}{1999}]{Levenson:1999}
Levenson N.~A., Graham J.~R., Snowden S.~L., 1999, ApJ, 526, 874
\bibitem[\protect\citeauthoryear{Li, Frank \& Blackman}{Li et al.}{2013}]{Li:2013}
Li S., Frank A., Blackman E.~G., 2013, ApJ, 774, 133
\bibitem[\protect\citeauthoryear{Marinacci, Pakmor \&
    Springel}{Marinacci et al.}{2014}]{Marinacci:2014}
Marinacci F., Pakmor R., Springel V., 2014, MNRAS, 437, 1750
\bibitem[\protect\citeauthoryear{Matheson et
    al.}{2000}]{Matheson:2000}
Matheson T., Filippenko A.~V., Ho L.~C., Barth A.~J., Leonard D.~C.,
2000, AJ, 120, 1499
\bibitem[\protect\citeauthoryear{McCourt et al.}{2015}]{McCourt:2015} 
McCourt M., O'Leary R.M., Madigan A.-M., Quataert E., 2015, MNRAS, 449, 2
\bibitem[\protect\citeauthoryear{McKee \& Ostriker}{1977}]{McKee:1977} 
McKee, C.~F., \& Ostriker, J.~P.\ 1977, ApJ, 218, 148
\bibitem[\protect\citeauthoryear{Melioli, de Gouveia Dal Pino \& Raga}{Melioli et al.}{2005}]{Melioli:2005}
Melioli C., de Gouveia Dal Pino E.~M., Raga A., 2005, A\&A, 443, 495
\bibitem[\protect\citeauthoryear{Miceli et al.}{2006}]{Miceli:2006}
Miceli M., Reale F., Orlando S., Bocchino F., 2006, A\&A, 458, 213
\bibitem[\protect\citeauthoryear{Miceli et al.}{2013}]{Miceli:2013}
Miceli M., Orlando S., Reale F., Bocchino F., Peres G., 2013, MNRAS,
430, 2864
\bibitem[\protect\citeauthoryear{Miceli et al.}{2014}]{Miceli:2014}
Miceli M., Acero F., Dubner G., Decourchelle A., Orlando S., Bocchino
F., 2014, ApJL, 782, 33
\bibitem[\protect\citeauthoryear{Milisavljevic \&
    Fesen}{2013}]{Milisavljevic:2013}
Milisavljevic D., Fesen R.~A., 2013, ApJ, 772, 134
\bibitem[\protect\citeauthoryear{Morse et al.}{1996}]{Morse:1996}
Morse J.~A., et al. 1996, AJ, 112, 509
\bibitem[\protect\citeauthoryear{Nakamura et al.}{2006}]{Nakamura:2006}
Nakamura F., McKee C.~F., Klein R.~I., Fisher R.~T., 2006, ApJSS, 164,
477
\bibitem[\protect\citeauthoryear{Nakamura et al.}{2014}]{Nakamura:2014}
Nakamura R., et al., 2014, PASJ, 66, 62
\bibitem[\protect\citeauthoryear{Niederhaus et
    al.}{2007}]{Niederhaus:2007} 
Niederhaus J.~H.~J., 2007, PhD thesis, University of Wisconsin - Madison
\bibitem[\protect\citeauthoryear{Niederhaus et
    al.}{2008}]{Niederhaus:2008} 
Niederhaus J.~H.~J., Greenough J.~A., Oakley J.~G., Ranjan D.,
Anderson M.~H., Bonazza R., 2008, J. Fluid Mech., 594, 85
\bibitem[\protect\citeauthoryear{Obergaulinger et al.}{2014}]{Obergaulinger:2014}
Obergaulinger, M., Iyudin A.~F., M\"{u}ller E., Smoot G.~F., 2014,
MNRAS, 437, 976
\bibitem[\protect\citeauthoryear{Ohyama et al.}{2002}]{Ohyama:2002}
Ohyama Y., et al., 2002, PASJ, 54, 891
\bibitem[\protect\citeauthoryear{Orlando et al.}{2005}]{Orlando:2005}
Orlando S., Peres G., Reale F., Bocchino F., Rosner R., Plewa T., Siegel A.,
2005, A\&A, 444, 505
\bibitem[\protect\citeauthoryear{Orlando et al.}{2006}]{Orlando:2006}
Orlando S., Bocchino F., Peres G., Reale F., Plewa T., Rosner R.,
2006, A\&A, 457, 545
\bibitem[\protect\citeauthoryear{Orlando et al.}{2010}]{Orlando:2010}
Orlando S., Bocchino F., Miceli M., Zhou X., Reale F., Peres G., 2010,
A\&A, 514, A29
\bibitem[\protect\citeauthoryear{Park et al.}{2004}]{Park:2004}
Park S., Hughes J.~P., Slane P.~O., Burrows D.~N., Roming P.~W.~A.,
Nousek J.~A., Garmire G.~P., 2004, ApJL, 602, 33
\bibitem[\protect\citeauthoryear{Parkin et al.}{2011}]{Parkin:2011}
Parkin E.~R., Pittard J.~M., Corcoran M.~F., Hamaguchi K., 2011, ApJ,
726, 105
\bibitem[\protect\citeauthoryear{Patnaude \&
    Fesen}{2014}]{Patnaude:2014}
Patnaude D.~J., Fesen R.~A., 2014, ApJ, 789, 138
\bibitem[\protect\citeauthoryear{Pittard}{2007a}]{Pittard:2007a}
Pittard J.~M., 2007a, ApJ, 660, L141
\bibitem[\protect\citeauthoryear{Pittard}{2007b}]{Pittard:2007b}
Pittard J.~M., 2007b, in Hartquist T.~W., Pittard J.~M., Falle
S.~A.~E.~G., eds., Astrophys. \& Space Sci. Proc., Diffuse Matter From
Star Forming Regions to Active Galaxies - A Volume Honouring John
Dyson. Springer, Dordrecht, p.~245
\bibitem[\protect\citeauthoryear{Pittard}{2009}]{Pittard:2009}
Pittard J.~M., 2009, MNRAS, 396, 1743
\bibitem[\protect\citeauthoryear{Pittard et al.}{2003}]{Pittard:2003}
Pittard, J.~M., Arthur, S.~J., Dyson, J.~E., Falle, S.~A.~E.~G., Hartquist,
T.~W., Knight M.~I., \& Pexton M.\ 2003, A\&A, 401, 1027
\bibitem[\protect\citeauthoryear{Pittard et al.}{2009}]{Pittard:2009}
Pittard J.~M., Falle S.~A.~E.~G., Hartquist T.~W., Dyson J.~E., 2009, MNRAS, 394, 1351
\bibitem[\protect\citeauthoryear{Pittard et al.}{2010}]{Pittard:2010}
Pittard J.~M., Hartquist T.~W., Falle S.~A.~E.~G., 2010, MNRAS, 405,
821
\bibitem[\protect\citeauthoryear{Pittard}{2011}]{Pittard:2011}
Pittard J.~M., 2011, MNRAS, 411, L41
\bibitem[\protect\citeauthoryear{Pittard \& Goldsmith}{2016}]{Pittard:2016}
Pittard J.~M., Goldsmith K.~J.~A., 2016, MNRAS, submitted
\bibitem[\protect\citeauthoryear{Poludnenko, Frank \& Blackman}{Poludnenko et al.}{2002}]{Poludnenko:2002}
Poludnenko A.~Y., Frank A., Blackman E.~G., 2002, ApJ, 576, 832
\bibitem[\protect\citeauthoryear{Raga et al.}{2007}]{Raga:2007}
Raga A.~C., Esquivel A., Riera A., Vel\'{a}zquez P.~F., 2007, ApJ,
668, 310
\bibitem[\protect\citeauthoryear{Ranjan et al.}{2005}]{Ranjan:2005}
Ranjan D., Anderson M.~H., Oakley J.~G., Bonazza R., 2005,
Phys. Rev. Lett., 94, 184507
\bibitem[\protect\citeauthoryear{Ranjan et
    al.}{2008}]{Ranjan:2008}
Ranjan D., Niederhaus J.~H.~J., Oakley J.~G., Anderson M.~H.,
Greenough J.~A., Bonazza R., 2008, Phys. Scr., T132, 014020
\bibitem[\protect\citeauthoryear{Ranjan, Oakley \& Bonazza}{Ranjan et
    al.}{2011}]{Ranjan:2011}
Ranjan D., Oakley J., Bonazza R., 2011, Annu. Rev. Fluid Mech., 43, 117
\bibitem[\protect\citeauthoryear{Reed et al.}{1995}]{Reed:1995}
Reed J.~E., Hester J.~J., Fabian A.~C., Winkler P.~F., 1995, ApJ, 440,
706
\bibitem[\protect\citeauthoryear{Robey et al.}{2002}]{Robey:2002}
Robey H.~F., Perry T.~S., Klein R.~I., Kane J.~O., Greenough J.~A.,
Boehly T.~R., 2002, Phys. Rev. Lett., 89, 085001
\bibitem[\protect\citeauthoryear{Roediger et al.}{2014}]{Roediger:2014}
Roediger E., Br\"{u}ggen M., Owers M.~S., Ebeling H., Sun M., 2014,
MNRAS, 443, L114
\bibitem[\protect\citeauthoryear{Roediger et al.}{2015a}]{Roediger:2015a}
Roediger E., et~al., 2015a, ApJ, 806, 103
\bibitem[\protect\citeauthoryear{Roediger et al.}{2015b}]{Roediger:2015b}
Roediger E., et~al., 2015b, ApJ, 806, 104
\bibitem[\protect\citeauthoryear{Rogers \&
    Pittard}{2013}]{Rogers:2013}
Rogers H., Pittard J.~M., 2013, MNRAS, 431, 1337
\bibitem[\protect\citeauthoryear{Rosen et al.}{2009}]{Rosen:2009}
Rosen P.~A., et al., 2009, Astrophys. Space Sci., 322, 101
\bibitem[\protect\citeauthoryear{Sales et al.}{2010}]{Sales:2010}
Sales L.~V., Navarro J.~F., Schaye J., Dalla Vecchia C., Springel V.,
Booth C.~M., 2010, MNRAS, 409, 1541
\bibitem[\protect\citeauthoryear{Samtaney \&
    Pullin}{1996}]{Samtaney:1996}
Samtaney R., Pullin D.~I., 1996, Physics of Fluids, 8, 2650
\bibitem[\protect\citeauthoryear{Scalo \& Elmegreen}{2004}]{Scalo:2004}
Scalo J., Elmegreen B.~G., 2004, ARA\&A, 42, 275
\bibitem[\protect\citeauthoryear{Scannapieco \& Br\"{u}ggen}{2015}]{Scannapieco:2015}
Scannapieco E., Br\"{u}ggen M., 2015, ApJ, 805, 158
\bibitem[\protect\citeauthoryear{Schaye et al.}{2015}]{Schaye:2015}
Schaye J., et al., 2015, MNRAS, 446, 521
\bibitem[\protect\citeauthoryear{Schneider \& Robertson}{2015}]{Schneider:2015}
Schneider E.~E., Robertson B.~E., 2015, ApJSS, 217, 24
\bibitem[\protect\citeauthoryear{Seta et al.}{1998}]{Seta:1998}
Seta M., et al., 1998, ApJ, 505, 286
\bibitem[\protect\citeauthoryear{Shin \& Ruszkowski}{2013}]{Shin:2013}
Shin M.-S., Ruszkowski M., 2013, MNRAS, 428, 804
\bibitem[\protect\citeauthoryear{Shin \& Ruszkowski}{2014}]{Shin:2014}
Shin M.-S., Ruszkowski M., 2014, MNRAS, 445, 1997
\bibitem[\protect\citeauthoryear{Shin, Stone \& Snyder}{Shin et al.}{2008}]{Shin:2008}
Shin M.-S., Stone J.~M., Snyder G.~F., 2008, ApJ, 680, 336
\bibitem[\protect\citeauthoryear{Slane et al.}{2015}]{Slane:2015}
Slane P., Bykov A., Ellison D.~C., Dubner G., Castro D., 2015, Space
Sci. Rev., 188, 187
\bibitem[\protect\citeauthoryear{Snell et al.}{2005}]{Snell:2005}
Snell R.~L., 2005, ApJ, 620, 758
\bibitem[\protect\citeauthoryear{Spyromilio}{1991}]{Spyromilio:1991}
Spyromilio J., 1991, MNRAS, 253, 25
\bibitem[\protect\citeauthoryear{Spyromilio}{1994}]{Spyromilio:1994}
Spyromilio J., 1994, MNRAS, 266, L61
\bibitem[\protect\citeauthoryear{Steffen \& L\'{o}pez}{2004}]{Steffen:2004}
Steffen W., L\'{o}pez J.~A., 2004, ApJ, 612, 319
\bibitem[\protect\citeauthoryear{Stevens, Blondin \& Pollock}{Stevens
    et al.}{1992}]{Stevens:1992}
Stevens I.~R., Blondin J.~M., Pollock A.~M.~T., 1992, ApJ, 386, 265
\bibitem[\protect\citeauthoryear{Stone \& Norman}{1992}]{Stone:1992}
Stone J.~M., Norman M.~L., 1992, ApJ, 390, L17
\bibitem[\protect\citeauthoryear{Strickland \& Stevens}{2000}]{Strickland:2000}
Strickland D.~K., Stevens I.~R., 2000, MNRAS, 314, 511
\bibitem[\protect\citeauthoryear{Strom et al.}{1995}]{Strom:1995}
Strom R., Johnston H.~M., Verbunt F., Aschenbach B., 1995, Nature,
373, 590
\bibitem[\protect\citeauthoryear{Sutherland \&
    Bicknell}{2007}]{Sutherland:2007}
Sutherland R.~S., Bicknell G.~V., 2007, ApJSS, 173, 37
\bibitem[\protect\citeauthoryear{Tenorio-Tagle et al.}{2006}]{Tenorio-Tagle:2006}
Tenorio-Tagle G., Mu\~{n}oz-Tu\~{n}\'{o}n, C., P\'{e}rez E., Silich S., Telles E., 2006, ApJ, 643, 186
\bibitem[\protect\citeauthoryear{Tonnesen \& Bryan}{2009}]{Tonnesen:2009}
Tonnesen S., Bryan G.~L., 2009, ApJ, 694, 789
\bibitem[\protect\citeauthoryear{Tonnesen \& Stone}{2014}]{Tonnesen:2014}
Tonnesen S., Stone J., 2014, ApJ, 795, 148
\bibitem[\protect\citeauthoryear{Tsunemi, Miyata \&
    Aschenbach}{Tsunemi et al.}{1999}]{Tsunemi:1999}
Tsunemi H., Miyata E., Aschenbach B., 1999, PASJ, 51, 711
\bibitem[\protect\citeauthoryear{Vaidya, Hartquist \& Falle}{Vaidya et al.}{2013}]{Vaidya:2013}
Vaidya B., Hartquist T.~W., Falle S.~A.~E.~G., 2013, MNRAS, 433, 1258
\bibitem[\protect\citeauthoryear{Van Loo, Falle \& Hartquist}{Van Loo et al.}{2010}]{VanLoo:2010}
Van~Loo S., Falle S.~A.~E.~G., Hartquist T.~W., 2010, MNRAS, 406, 1260
\bibitem[\protect\citeauthoryear{Veilleux, Cecil \&
    Bland-Hawthorn}{Veilleux et al.}{2005}]{Veilleux:2005}
Veilleux S., Cecil G., Bland-Hawthorn J., 2005, ARA\&A, 43, 769
\bibitem[\protect\citeauthoryear{Vijayaraghavan \& Ricker}{2015}]{Vijayaraghavan:2015}
Vijayaraghavan R., Ricker P.~M., 2015, MNRAS, 449, 2312
\bibitem[\protect\citeauthoryear{Vorobyov, Recchi \& Hensler}{Vorobyov
    et al.}{2015}]{Vorobyov:2015}
Vorobyov E.~I., Recchi S., Hensler G., 2015, A\&A, 579, 9
\bibitem[\protect\citeauthoryear{Wagner, Bicknell \& Umemura}{Wagner
    et al.}{2012}]{Wagner:2012}
Wagner A.~Y., Bicknell G.~V., Umemura M., 2012, ApJ, 757, 136
\bibitem[\protect\citeauthoryear{Walch et al.}{2015}]{Walch:2015}
Walch S., et al., 2015, MNRAS, 454, 238
\bibitem[Walder \& Folini(2002)]{Walder:2002} 
Walder R., \& Folini D., 2002, in ASP Conf. Ser. 260, Interacting Winds from Massive Stars, ed. A. F. J. Moffat \& N. St.-Louis (San Francisco: ASP), 595
\bibitem[\protect\citeauthoryear{Wang \& Chevalier}{2002}]{Wang:2002}
Wang C.-Y., Chevalier R.~A., 2002, ApJ, 574, 155
\bibitem[\protect\citeauthoryear{Widnall, Bliss \& Tsai}{Widnall et
    al.}{1974}]{Widnall:1974}
Widnall S.~E., Bliss D.~B., Tsai C.~Y., 1974, J. Fluid Mech., 66, 35
\bibitem[\protect\citeauthoryear{Williams et
    al.}{2013}]{Williams:2013}
Williams B.~J., et al., 2013, ApJ, 770, 129
\bibitem[\protect\citeauthoryear{Winkler \&
    Kirshner}{1985}]{Winkler:1985}
Winkler P.~F., Kirshner R.~P., 1985, ApJ, 299, 981
\bibitem[\protect\citeauthoryear{Winkler et al.}{1988}]{Winkler:1988}
Winkler P.~F., Tuttle J.~H., Kirshner R.~P., Irwin M.~J., 1988, in IAU
Colloq. 101, Supernova Remnants and the Interstellar Medium,
ed. R.~S.~Roger \& T.~L.~Landecker (Cambridge: Cambridge Univ. Press), 65
\bibitem[\protect\citeauthoryear{Winkler et al.}{2014}]{Winkler:2014}
Winkler P.~F., Williams B.~J., Reynolds S.~P., Petre R., Long K.~S.,
Katsuda S., Hwang U., 2014, ApJ, 781, 65
\bibitem[\protect\citeauthoryear{White \& Long}{1991}]{White:1991}
White, R.~L., \& Long, K.~S.\ 1991, ApJ, 373, 543
\bibitem[\protect\citeauthoryear{Xu \& Stone}{1995}]{Xu:1995}
Xu J., Stone J.~M., 1995, ApJ, 454, 172
\bibitem[\protect\citeauthoryear{Yirak et al.}{2008}]{Yirak:2008}
Yirak K., Frank A., Cunningham A., Mitran S., 2008, ApJ, 672, 996
\bibitem[\protect\citeauthoryear{Yirak, Frank \& Cunningham}{Yirak et al.}{2010}]{Yirak:2010}
Yirak K., Frank A., Cunningham A., 2010, ApJ, 722, 412 
\bibitem[\protect\citeauthoryear{Zhai et al.}{2011}]{Zhai:2011}
Zhai Z., Si T., Luo X., Yang J., 2011, Phys. Fluids, 23, 084104
\end{thebibliography}




\appendix

\section{Resolution Test}
\label{sec:restest} 
In an actual shock-cloud interaction, the smallest instabilities have
a length scale, $\eta$, which is set by the damping of hydromagnetic
waves. This is typically through particle collisions, but can also be
through wave-particle interactions \citep[see Sec.~2.2
of][]{Pittard:2009}, and is dependent on the nature of the
problem. For instance, in astrophysical problems it depends on whether
the cloud is ionized, neutral or molecular, and the strength of the
magnetic field and thermal conductivity. The Reynolds number of a flow
past a cloud is ${\rm Re} = ur_{\rm c}/\nu$, where $u$ is the average
flow speed past the cloud, $r_{\rm c}$ is the radius of the cloud, and
$\nu$ is the kinematic viscosity. For astrophysical scenarios, Re can
easily exceed a value of $10^{5-6}$ \citep{Pittard:2009}. The size of
the smallest eddies, $\eta\sim{\rm Re}^{-3/4}l$, where the largest
eddies have a length scale, $l$, comparable in size to the
cloud. Resolving the smallest eddies in a numerical simulation can
thus be very challenging. Alternatively, a $k$-$\epsilon$ model can be
used to explicitly model the effects of sub-grid-scale turbulent
viscosity through the addition of turbulence-specific viscosity and
diffusion terms to the Euler equations
\citep{Pittard:2009,Pittard:2010}.

Without any prescription for the small-scale dissipative physics, new
unstable scales will be added as the resolution of the simulation is
increased. This is the case for simulations which simply solve the
Euler equations for inviscid fluid flow. For instance, simulations of
a shock striking a cloud will produce features which depend on the
resolution adopted\footnote{This is also true of simulations which
  specify small-scale dissipative physics but which do not have the
  resolution to resolve the smallest physical scales present.}. Higher
resolution simulations allow the development of smaller instabilities,
and surfaces and interfaces become sharper. These differences can
affect the rate at which material is stripped from the cloud and mixed
into the surrounding flow, and the acceleration that the cloud
experiences. Increasing the resolution simply creates finer and finer
structure as Re increases. In the shock-cloud scenario, the different
instabilities present at different resolution will break up the cloud
differently, thus eventually affecting the convergence of integral
quantities. Thus formal convergence may be impossible in ``inviscid''
simulations. \citet{Samtaney:1996} have shown that initial value
problems for the Euler equations involving shock-contact interactions
exhibit features indicating that such problems are ill-posed,
including non-convergence of the solution at a given time.

Simulations of problems for which there is no analytical solution
typically rely on a demonstration of self-convergence. Lower
resolution simulations are compared against the highest resolution
simulation performed, and a resolution is chosen which balances
accuracy against computational cost. \citet{Klein:1994} suggested that
$\sim100$ cells per cloud radius was required to adequately model the
adiabatic interaction of a Mach 10 shock with a $\chi=10$ cloud. Most
simulations in the astrophysics literature since then have adopted
resolutions matching or exceeding this requirement, though some 3D
studies have been performed at lower resolution. More recently,
\citet{Niederhaus:2007} examined the issue of convergence for 2D
calculations of the purely adiabatic interaction of a shock with a
spherical cloud. They find that although the solution is locally and
pointwise nonconvergent, some aspects of the computed flowfields,
particularly certain integrated and mean quantities, do reach a
converged grid-independent state. For instance, they show that the
maximum density in the flowfield continues to vary with the spatial
resolution (even for resolutions up to $R_{1024}$), while the mean
cloud density converges to a nearly grid-independent value for
resolutions $> R_{500}$.

At very low resolution, important features of the flow may not be
present, and ultimately the simulated interaction will compare poorly
to reality. Thus, rather than attempting to obtain a converged
solution, some previous work has instead focussed on resolving key
features of the flow. In purely hydrodynamic shock-cloud simulations
this includes the stand-off distance of the bowshock
\citep[e.g.,][]{Farris:1994} and the thickness of the turbulent
boundary layer on the cloud surface \citep[see][and references
therein]{Pittard:2009}; in radiative shock-cloud simulations it is the
cooling layer behind shocks \citep{Yirak:2010}, while in the MHD
simulations of \citet{Dursi:2008} it is the magnetic draping layer on
the upstream surface of the cloud.

To understand how the grid resolution affects our results we have run
a variety of simulations at different resolutions, with and without
inclusion of a $k$-$\epsilon$ sub-grid model.  In the following
subsections we examine the resolution dependence of the cloud
morphology, study some statistics of the interaction, determine how
certain integral quantities vary with resolution, and finally study
the impact of resolution on the cloud acceleration and mixing
timescales, $t_{\rm drag}$ and $t_{\rm mix}$.

\begin{figure}
\resizebox{80mm}{!}{\includegraphics{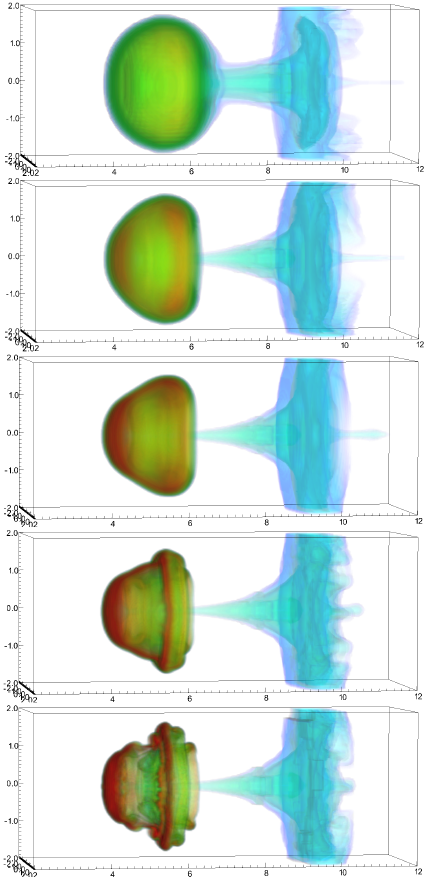}}
   \caption{Comparison with resolution of the ``inviscid'' $M=10$, $\chi=10$
     simulation at $t = 4.09\,t_{\rm cc}$. From
     top to bottom the resolutions are $R_{8},
     R_{16}, R_{32}, R_{64}$
     and $R_{128}$.}
    \label{fig:M10chi1e1_3Dresplot_4.09tcc}
\end{figure}

\subsection{Cloud Morphology}
\label{sec:restest_morphology}
We first study the resolution dependence of the cloud morphology for
``inviscid'' simulations with $M=10$ and $\chi=10$, which are the most
popular parameter choices in the astrophysical literature to date (see
Table~\ref{tab:previous}). We expect the bowshock to have a stand-off
distance of $\approx 0.28\,r_{\rm c}$ \citep{Farris:1994}. Hence the
bowshock will be resolved at resolutions $\gtsimm R_{16}$, while
resolving the turbulent boundary layer requires resolutions $\sim
R_{100}$.

Fig.~\ref{fig:M10chi1e1_3Dresplot_4.09tcc} shows volumetric plots of
the density of cloud material at $t=4.09\,t_{\rm cc}$ (this focus
means that features in the ambient medium - e.g., the bowshock - are
not visible). As the resolution increases we see that the shape of the
cloud changes, from rounded and relatively featureless at lower
resolutions, to displaying a torus of high vorticity at the highest
resolutions. The cloud and the vortex ring at the rear of the cloud
are merged together in the $R_{8}$ simulation, but become increasingly
separate and distinct as the resolution increases. At $R_{64}$
numerous density structures occur within the cloud interior (these are
not readily visible in Fig.~\ref{fig:M10chi1e1_3Dresplot_4.09tcc}, but
are clearly identifiable when this figure is rotated on the computer
screen), which break up into smaller structures in the $R_{128}$
simulation (these are clearly visible in the 2D slices shown in
Fig.~\ref{fig:chi1e1_2dvs3d}). At $R_{64}$ the vortex ring shows
azimuthal variations for the first time. In addition, the thickness of
the slip surface decreases and the maximum density of cloud material
increases as the resolution increases.

\begin{figure*}
\resizebox{170mm}{!}{\includegraphics{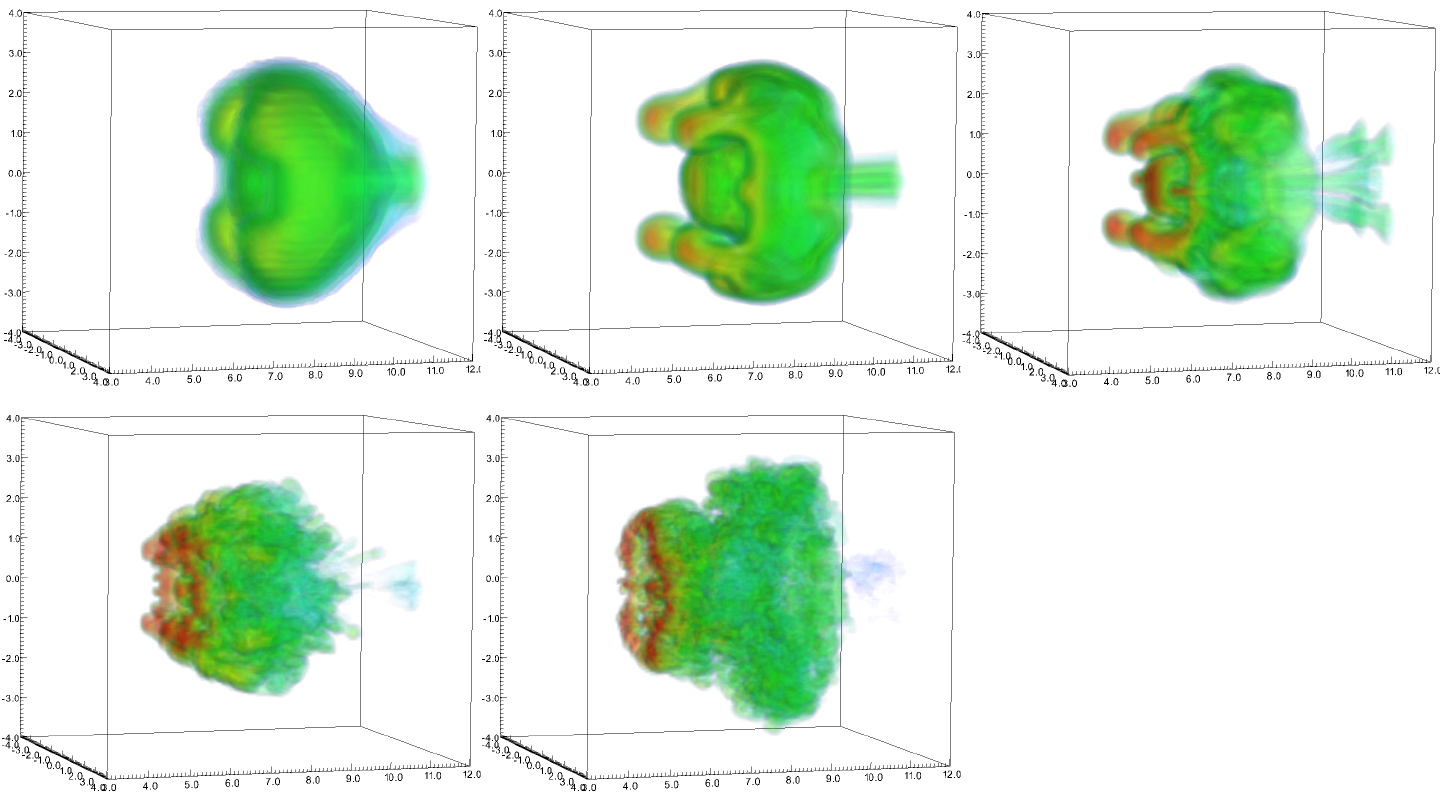}}
\caption{Comparison with resolution of the $M=10$, $\chi=10^{2}$
  simulation at $t = 3.87\,t_{\rm cc}$. From left to right and top to
  bottom the resolutions are $R_{8}, R_{16}, R_{32}, R_{64}$ and
  $R_{128}$.}
    \label{fig:M10chi1e2_3Dresplot_3.87tcc}
\end{figure*}

Fig.~\ref{fig:M10chi1e2_3Dresplot_3.87tcc} shows the
resolution-dependent behaviour of simulations with $M=10$,
$\chi=10^{2}$. It is interesting to see how the dominant scale of the
instabilities changes with resolution. For $R_{8}$, $R_{16}$ and
$R_{32}$, the cloud has 4 dominant fingers. At $R_{64}$ and $R_{128}$
smaller scale structures develop which change these fingers into a
single ring-like feature. The main effect of the resolution in
``inviscid'' simulations is to set the size of the instabilities which
develop: at low resolution only longer-wavelength instabilities can
develop. The tail appears to display some characteristics of
turbulence at $R_{64}$ and above.

\begin{figure*}
\resizebox{170mm}{!}{\includegraphics{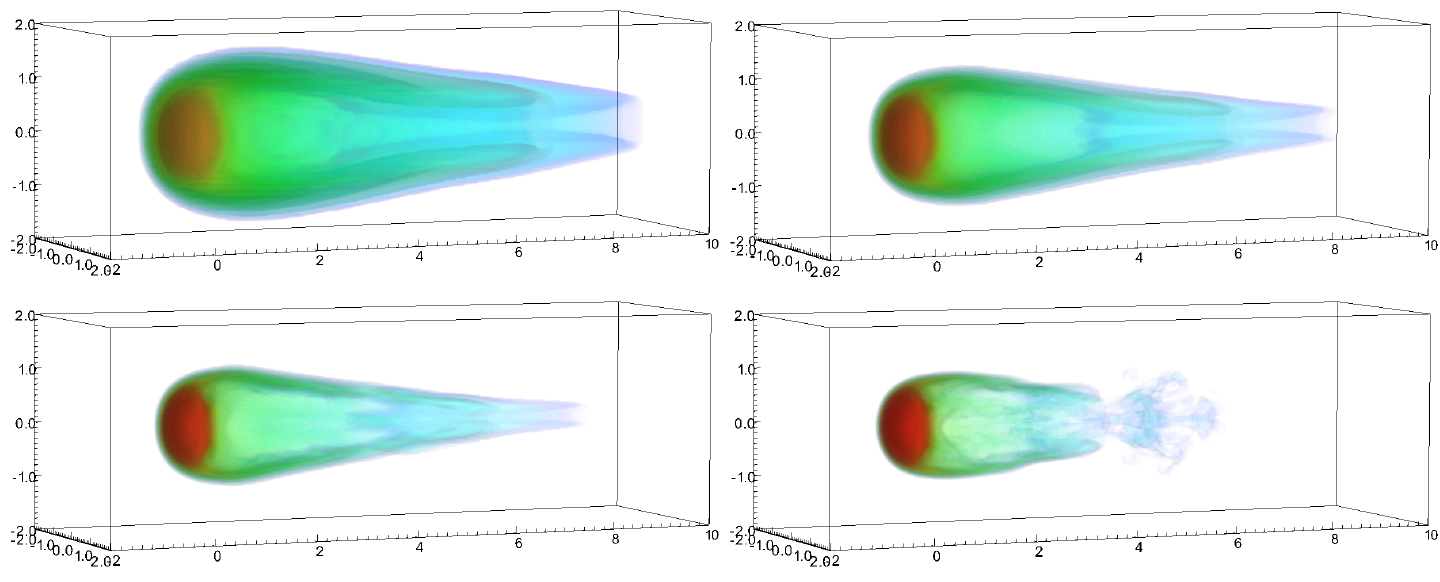}}
\caption{Comparison with resolution of the $M=10$, $\chi=10^{3}$
  simulation at $t = 0.79\,t_{\rm cc}$. From left to right and top to
  bottom the resolutions are $R_{8}, R_{16}, R_{32}$ and $R_{64}$. The
  initial cloud density is 1000.}
    \label{fig:M10chi1e3_3Dresplot_0.79tcc}
\end{figure*}

Fig.~\ref{fig:M10chi1e3_3Dresplot_0.79tcc} shows volumetric plots of
the cloud density at $t=0.79\,t_{\rm cc}$ for the $M=10$,
$\chi=10^{3}$ simulations as a function of the resolution. As the
resolution increases we again see that the thickness of
the slip surface decreases and the maximum density of the shocked
cloud increases. The tail of ablated material also becomes more hollow,
and its shape changes. At the highest resolution studied the tail is
disrupted by instabilities after about 3 cloud radii, and becomes
``turbulent''. At this time very little material has been stripped
from the cloud but there are already important qualitative and
quantitative differences in the flow.

\begin{figure*}
\resizebox{170mm}{!}{\includegraphics{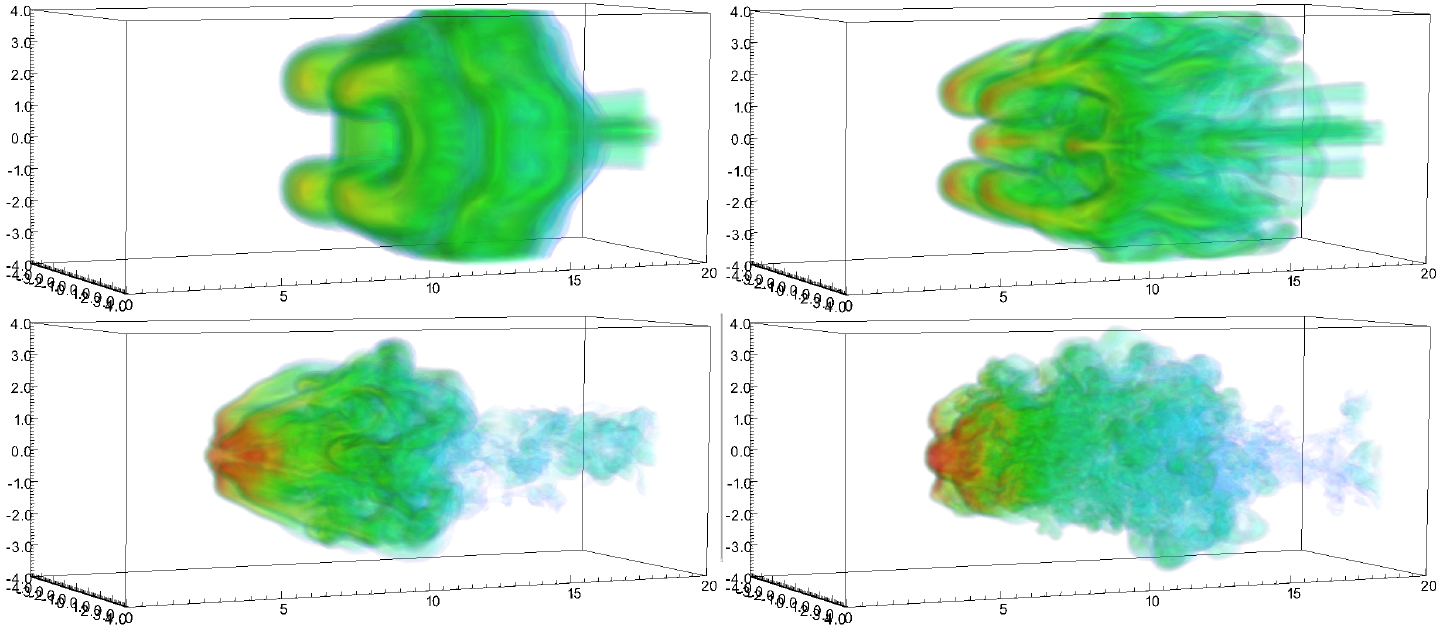}}
   \caption{As Fig.~\ref{fig:M10chi1e3_3Dresplot_0.79tcc} but at $t = 3.80\,t_{\rm cc}$.}
    \label{fig:M10chi1e3_3Dresplot_3.80tcc}
\end{figure*}

Fig.~\ref{fig:M10chi1e3_3Dresplot_3.80tcc} shows the resolution
dependent morphology at a later time ($t=3.80\,t_{\rm cc}$). Despite
the dramatic changes to the shape of the cloud, the core has yet to
suffer significant mass loss. As expected, the differences with
resolution are much more pronounced than in
Fig.~\ref{fig:M10chi1e3_3Dresplot_0.79tcc}. At $R_{8}$ the cloud has 4
dominant fingers while at $R_{16}$ a central finger is also seen. At
$R_{32}$ and $R_{64}$ we instead find that the bulk of the cloud
material forms a coherent structure located on the original cloud
axis. At these later times we see similar changes with resolution as
for the $M=10$, $\chi=10^{2}$ simulations discussed previously.

\begin{figure*}
\resizebox{170mm}{!}{\includegraphics{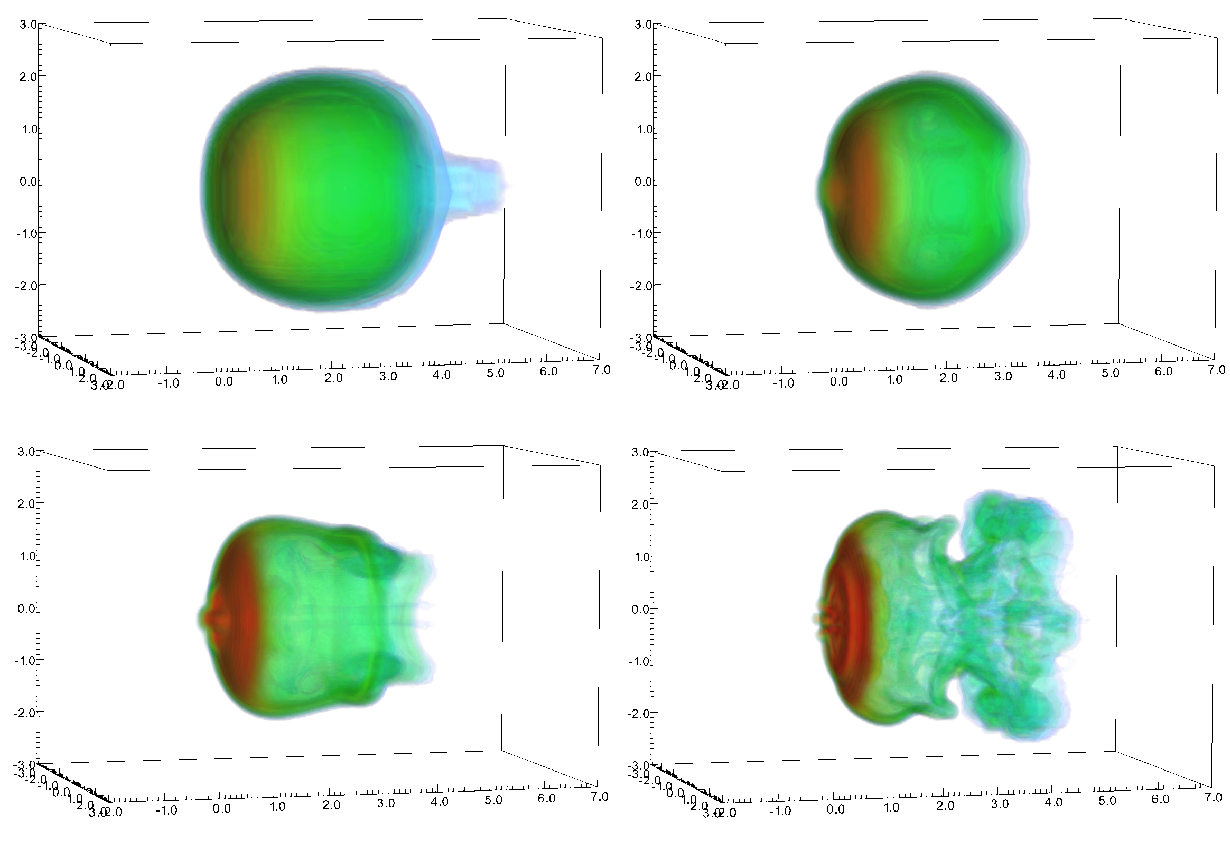}}
\caption{Comparison with resolution of the $M=1.5$, $\chi=10^{2}$
  simulation at $t = 3.87\,t_{\rm cc}$. From left to right and top to
  bottom the resolutions are $R_{8}, R_{16}, R_{32}$, and $R_{64}$.}
    \label{fig:M1.5chi1e2_3Dresplot_3.87tcc}
\end{figure*}

Fig.~\ref{fig:M1.5chi1e2_3Dresplot_3.87tcc} examines how the
resolution affects simulations with $\chi=10^{2}$ when the Mach number
of the shock is lowered to $M=1.5$. At $R_{8}$ the cloud is reasonably
featureless. At $R_{16}$ a small ``bump'' is visible on the leading
surface, and the cloud becomes both somewhat hollow and also less
extended in the axial direction. In the $R_{32}$ simulation the
leading ``bump'' is more extended, and in the $R_{64}$ simulation it
splits into 4 parts. We identify these features as RM instabilities
\citep[cf. Fig~1 in][]{Stone:1992}.  The remainder of the cloud has an
appearance which resembles a ``jelly-fish'' at the highest resolution
examined.

In terms of the morphology, the general impression that one gets from
Figs.~\ref{fig:M10chi1e1_3Dresplot_4.09tcc}-\ref{fig:M1.5chi1e2_3Dresplot_3.87tcc}
is that $R_{64}$ is the minimum resolution needed to capture the
morphology accurately in a qualitative sense. Our investigation is
therefore consistent with the statement in \citet{Xu:1995} that
$R_{60}$ ``has captured the dominant dynamical effects present in the
evolution''.

\begin{figure*}
\resizebox{130mm}{!}{\includegraphics{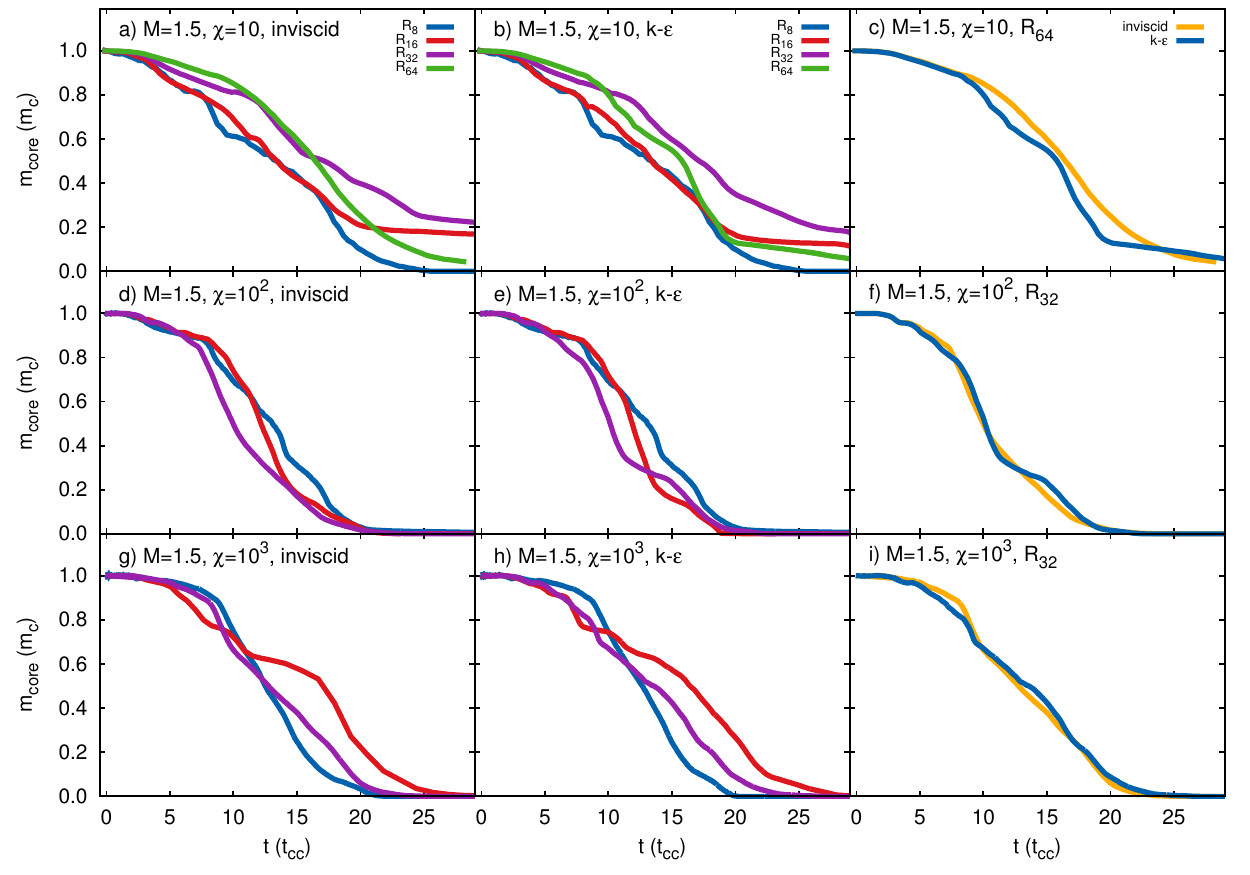}}
   \caption{Time evolution of the core mass, $m_{\rm core}$, for
     simulations with $M=1.5$. The top row has $\chi=10$, the middle
     row has $\chi=10^{2}$, and the bottom row has $\chi=10^{3}$. The
     calculations are made at various resolutions for inviscid (left
     column) and k-$\epsilon$ (middle column) simulations. A
     comparison at the indicated resolution is made between the inviscid
     and k-$\epsilon$ results in the right column.}
    \label{fig:mcore_restest_M1.5}
\end{figure*}

\begin{figure*}
\resizebox{130mm}{!}{\includegraphics{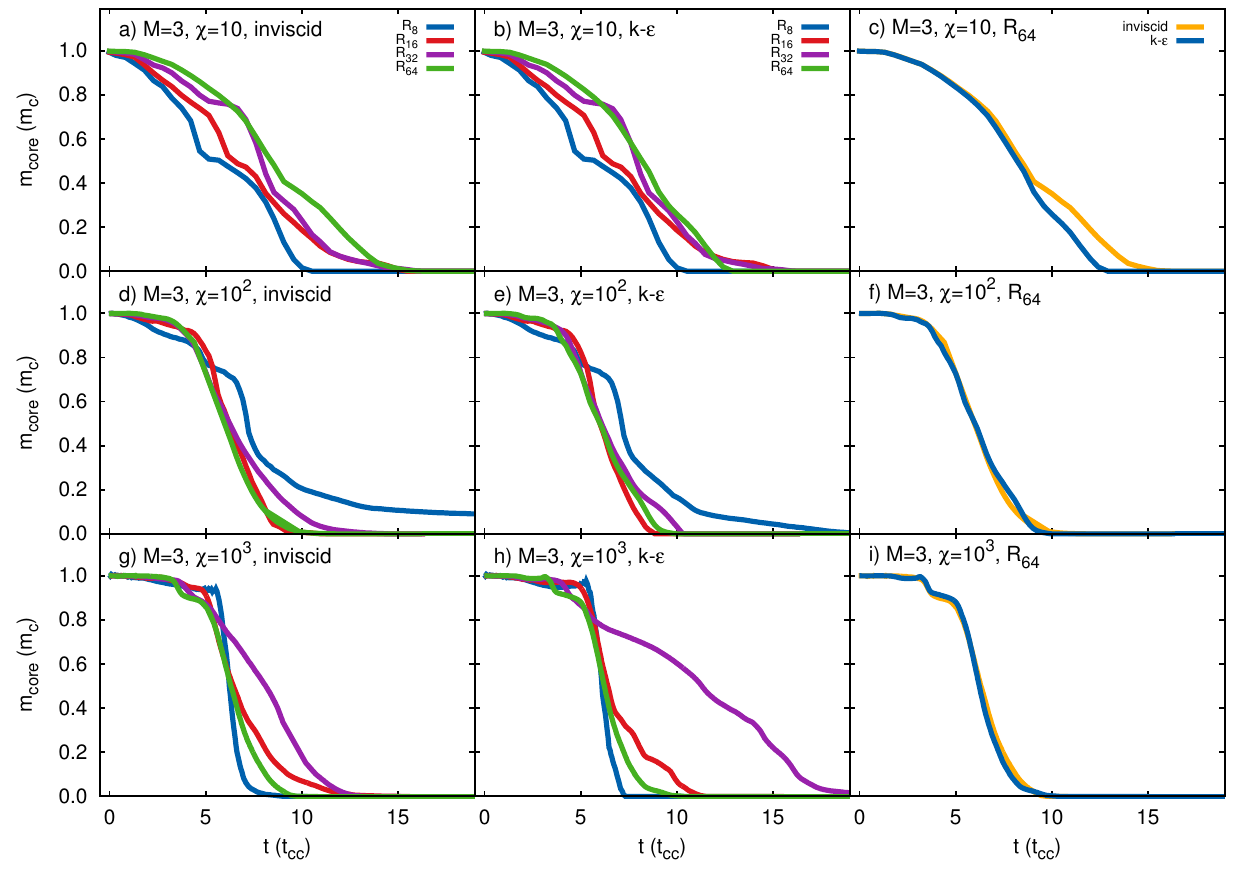}}
   \caption{As Fig.~\ref{fig:mcore_restest_M1.5} but for $M=3$.}
    \label{fig:mcore_restest_M3}
\end{figure*}

\begin{figure*}
\resizebox{130mm}{!}{\includegraphics{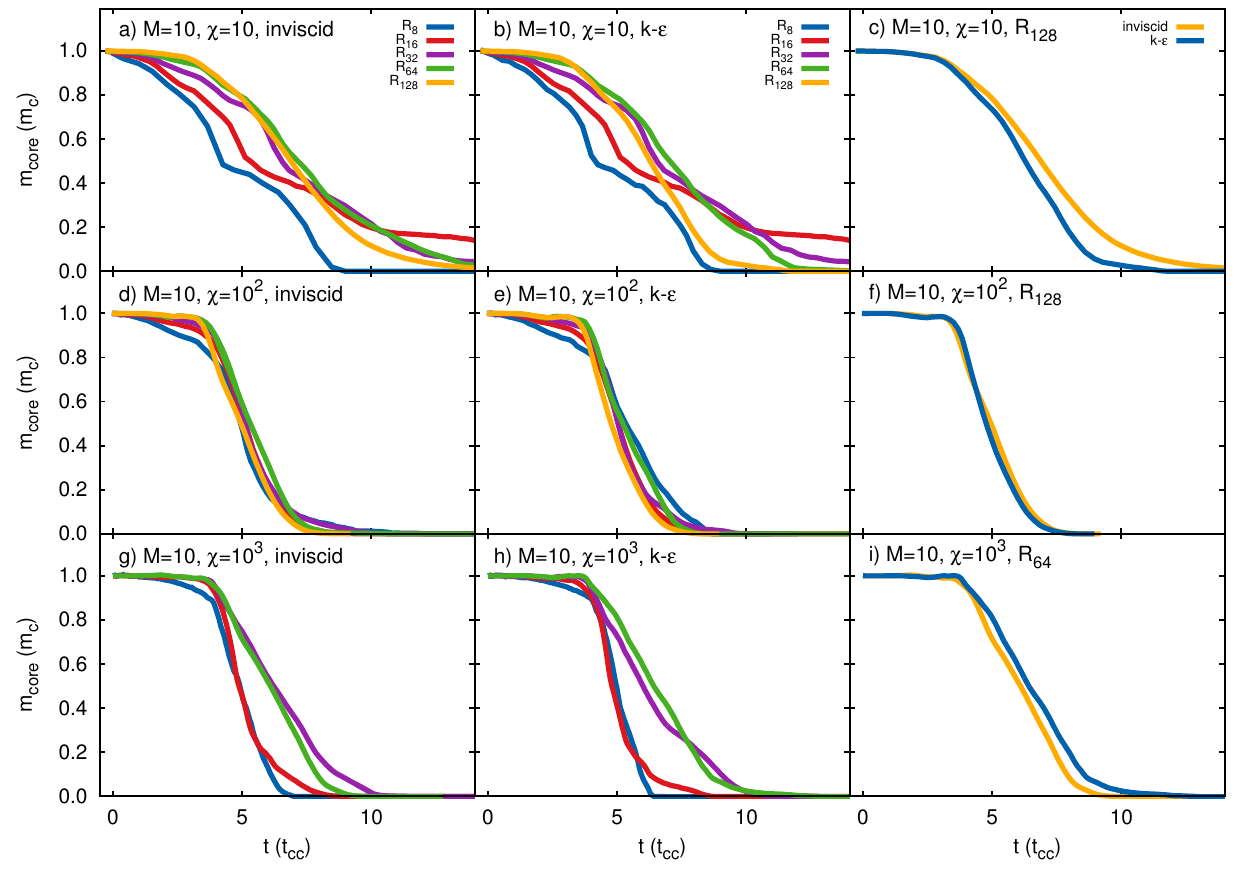}}
   \caption{As Fig.~\ref{fig:mcore_restest_M1.5} but for $M=10$.}
    \label{fig:mcore_restest_M10}
\end{figure*}

\begin{figure*}
\resizebox{130mm}{!}{\includegraphics{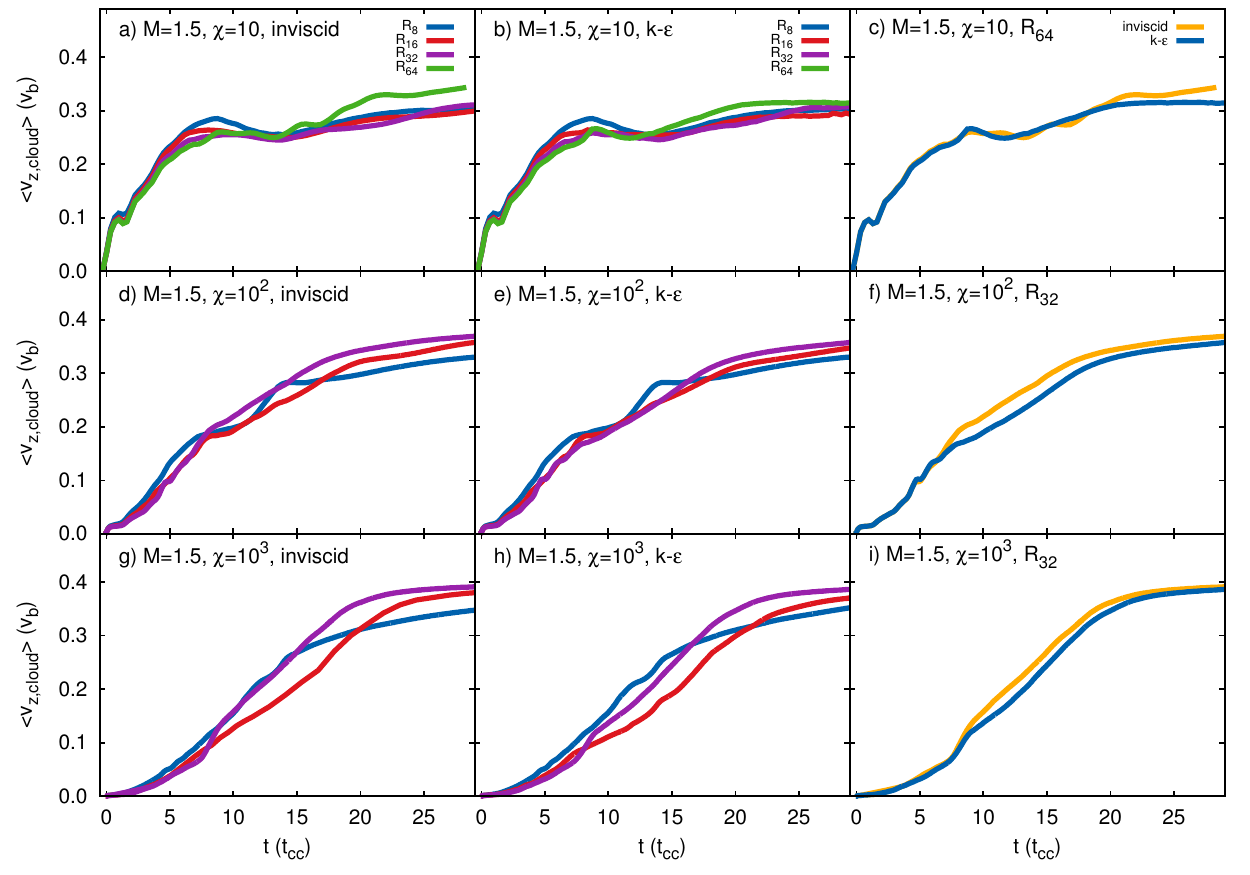}}
   \caption{Time evolution of the cloud velocity, $<v_{\rm z,cloud}>$, for
     simulations with $M=1.5$. The top row has $\chi=10$, the middle
     row has $\chi=10^{2}$, and the bottom row has $\chi=10^{3}$. The
     calculations are made at various resolutions for inviscid (left
     column) and k-$\epsilon$ (middle column) simulations. A
     comparison at the indicated resolution is made between the inviscid
     and k-$\epsilon$ results in the right column.}
    \label{fig:vzcloud_restest_M1.5}
\end{figure*}

\begin{figure*}
\resizebox{130mm}{!}{\includegraphics{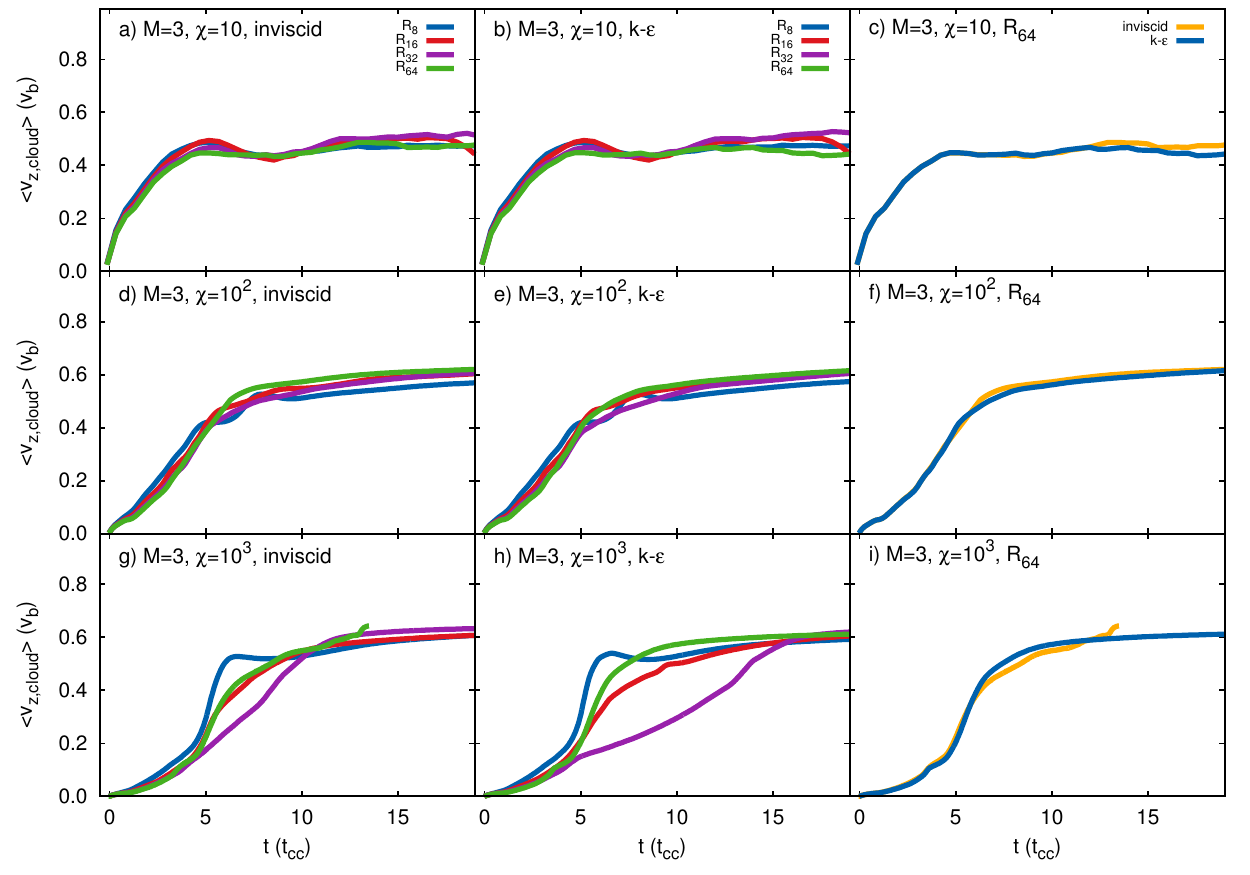}}
   \caption{As Fig.~\ref{fig:vzcloud_restest_M1.5} but for $M=3$.}
    \label{fig:vzcloud_restest_M3}
\end{figure*}

\begin{figure*}
\resizebox{130mm}{!}{\includegraphics{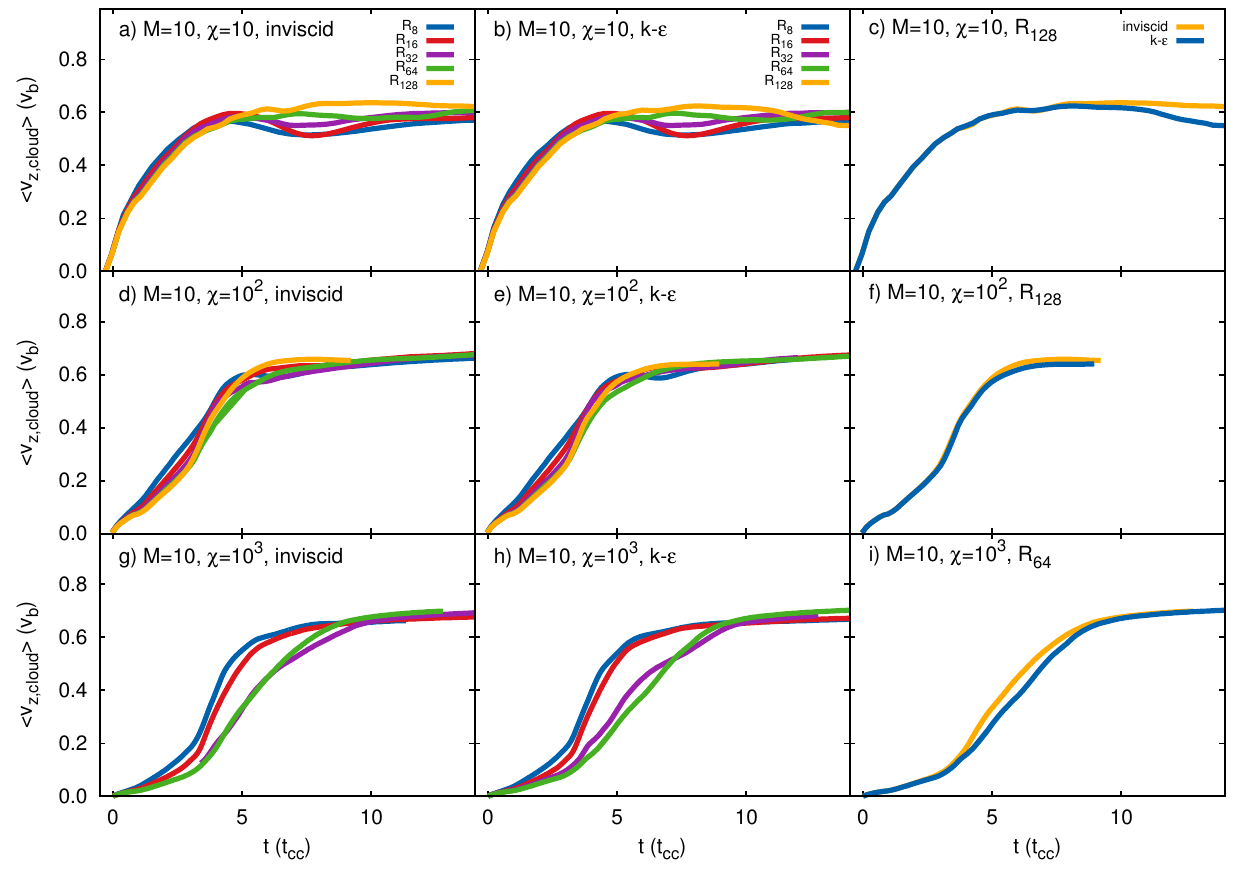}}
   \caption{As Fig.~\ref{fig:vzcloud_restest_M1.5} but for $M=10$.}
    \label{fig:vzcloud_restest_M10}
\end{figure*}

\subsection{Time Evolution}
\label{sec:restest_evolution}
Figs.~\ref{fig:mcore_restest_M1.5}-\ref{fig:mcore_restest_M10} show
the time evolution of the core mass, $m_{\rm core}$, for 3D
simulations with $\chi = 10, 10^{2}$ and $10^{3}$ and $M=1.5, 3$ and
10. In each figure, the left panels (a, d and g) show results at
different resolutions from the ``inviscid'' calculations, while the
centre panels (b, e and h) show corresponding results from the
$k$-$\epsilon$ calculations. The right panels (c, f and i) compare the
highest resolution simulations from each of these models.  In a
similar fashion,
Figs.~\ref{fig:vzcloud_restest_M1.5}-\ref{fig:vzcloud_restest_M10}
show the time evolution of the mean cloud velocity, $<v_{\rm
  z,cloud}>$, for the same runs. Since the displayed profiles are
generally less disparate and show a tighter correlation for the cloud
velocity than for the core mass we will concentrate on the latter in
the following discussion.

Consider first the $M=10$ results (see
Fig.~\ref{fig:mcore_restest_M10}). For $\chi=10$ and $\chi=10^{3}$ it
is clear that there are very large differences in the time dependent
behaviour of $m_{\rm core}$ between simulations at resolutions below
$R_{32}$. However, for both values of $\chi$ the $R_{32}$ and $R_{64}$
simulations are reasonably matched. This is true for both the
``inviscid'' and $k$-$\epsilon$ cases. Surprisingly, the simulations
for $\chi=10^{2}$ are much less dependent on resolution. In all cases,
the clouds in simulations with higher resolutions lose core mass
initially more slowly, but then show more rapid core mass loss at
later times (this is particularly true for the $\chi=10$
simulations). The former is due to the higher numerical viscosity and
thickness of the shear layer, while the latter is caused by the larger
dynamic range of instabilities which eventually develop. The important
point is that simulations with a resolution of $R_{32}$ appear to show
that the time evolution of $m_{\rm core}$ and $<v_{\rm z,cloud}>$ are
reasonably converged.

In contrast, 2D axisymmetric simulations of an adiabatic shock
striking a spherical cloud have $m_{\rm core}$ and $<v_{\rm z,cloud}>$
profiles which do not converge until $\sim R_{128}$ for ``inviscid''
calculations \citep[see, e.g., Fig.~2 in][]{Pittard:2009}. That the
resolution requirement for 3D calculations is lower than for 2D
calculations is likely due to the fundamentally different behaviour of
2D versus 3D turbulence (e.g., the absence of vortex-stretching in 2D,
and the transport of kinetic energy from large to small scales in 3D
(and vice-versa in 2D)). This behaviour indicates that fully 3D
simulations better represent the actual flow, and can ``outperform''
2D simulations of higher resolution (at least for the behaviour of
$m_{\rm core}$ and $<v_{\rm z,cloud}>$ - further work is necessary to
see if this is true for other integrated quantities and for the flow
field in general). On the other hand, it remains the case that higher
resolution simulations will develop smaller scale instabilities, and
that one can expect that resolutions in excess of $R_{32}$ will be
needed for the convergence of other properties, such as minimum and
maximum quantities \citep[cf.][]{Niederhaus:2007}.
 
Another surprise is that we see little difference between the inviscid
and $k$-$\epsilon$ 3D simulations\footnote{The inviscid and
  $k$-$\epsilon$ simulations are identical at low resolution, and only
  begin to differ at higher resolution (see, e.g., the $M=1.5$, 3 and
  10 results for $\chi=10$).
  Hence the right columns of
  Figs.~\ref{fig:mcore_restest_M1.5}-\ref{fig:vzcloud_restest_M10}
  indicate the {\em maximum} differences found between the inviscid
  and $k$-$\epsilon$ runs. Where there is a difference, the clouds in
  the $k$-$\epsilon$ simulations appear to lose mass slightly faster
  when $\chi=10$. However, this is not generally true for
  $\chi=10^{2}$ and $10^{3}$.}.  In particular, there is {\em no
  indication} that the $k$-$\epsilon$ simulations converge at a lower
resolution than their inviscid counterparts (see
Fig.~\ref{fig:mcore_restest_M10}).  This is in contrast to the 2D
simulation results presented in \citet{Pittard:2009}, where a major
finding was that the $k$-$\epsilon$ simulations converged at
significantly lower resolutions (roughly $R_{32}$) than their inviscid
counterparts (roughly $R_{128}$). It seems that the ability of
instabilities to grow in any direction means that 3D calculations more
accurately capture the real behaviour of such systems whether or not a
$k$-$\epsilon$ model is used. Thus there seems to be little benefit in
employing the $k$-$\epsilon$ model in 3D shock-cloud simulations
(though its lower resolution requirements for convergence mean that it
remains useful for 2D simulations).

For completeness we consider the resolution-dependent behaviour of 3D
simulations with a shock Mach number $M=3$
(Figs.~\ref{fig:mcore_restest_M3} and~\ref{fig:vzcloud_restest_M3})
and $M=1.5$ (Figs.~\ref{fig:mcore_restest_M1.5}
and~\ref{fig:vzcloud_restest_M1.5}). The same general behaviour is
seen in the $M=3$ simulations as in the $M=10$ simulations and thus
the same broad conclusions can be drawn (e.g., $R_{32}$ is roughly the
minimum needed for $m_{\rm core}$ and $<v_{\rm z,cloud}>$ to be
reasonably converged).

\begin{figure*}
\resizebox{175mm}{!}{\includegraphics{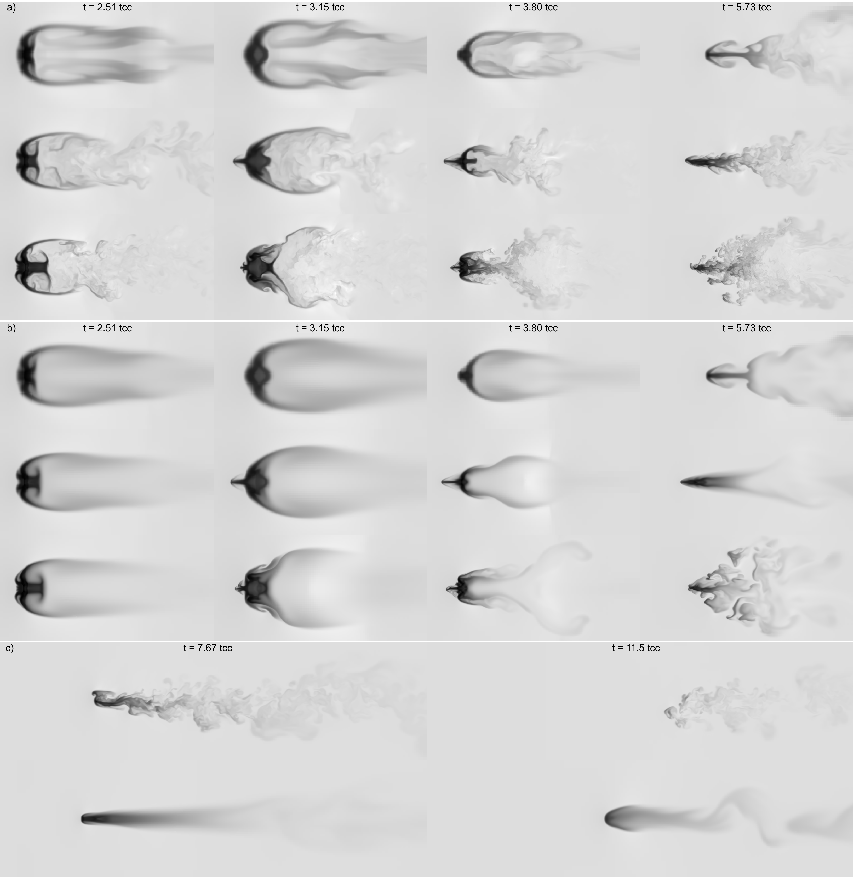}}
\caption{Comparison of the 3D $M=3$, $\chi=10^{3}$ simulations with
  time and resolution in the $Z=0$ plane. A grayscale of the
  logarithmic density is shown between $\rho_{\rm amb}$ (white) and
  $4\times \rho_{\rm c}$ (black). a) Inviscid models. b)
  $k$-$\epsilon$ models. In both a) and b), results at resolutions
  $R_{16}$ (top), $R_{32}$ (middle), and $R_{64}$ (bottom) are
  shown. c) At later times the $R_{32}$ inviscid (top) and
  $k$-$\epsilon$ (bottom) simulations are compared.  The frames show
  the region ($0 < X < 12$, $-3 < Y < 3$) at $t=2.51$ and
  $3.15\,t_{\rm cc}$, ($0 < X < 20$, $-5 < Y < 5$) at $t=3.80\,t_{\rm
    cc}$, ($0 < X < 26$, $-6.5 < Y < 6.5$) at $t=5.73\,t_{\rm cc}$,
  and ($0 < X < 50$, $-7 < Y < 7$) at $t=7.67$ and $11.5\,t_{\rm cc}$
  (all in units of $r_{\rm c}$). Note that in this figure the
  $X$-axis is plotted horizontally.}
    \label{fig:m10chi1e3r32dis}
\end{figure*}

The only significant discrepancy between the inviscid and
$k$-$\epsilon$ simulations occurs when $M=3$ and $\chi=10^{3}$ - the
$R_{32}$ $k$-$\epsilon$ simulation appears at odds with the others.
Fig.~\ref{fig:m10chi1e3r32dis} compares the morphology of the inviscid
and $k$-$\epsilon$ as a function of resolution at a number of
different times. Panels a) and b) show that the nature of the
interaction and the way the cloud is destroyed depends strongly on the
resolution. However, apart from a ``smoother'' wake in the
$k$-$\epsilon$ models, the cloud morphology is otherwise almost
identical between panels a) and b). The major difference is that the
cloud has a much more compact cross-section in the $R_{32}$
calculations at $t=5.73\,t_{\rm cc}$. It is not obvious from the
preceding panels why this occurs. It is at about this time that the
core mass and mean cloud velocity in the $R_{32}$ $k$-$\epsilon$
simulation start to diverge from the other models (see
Figs.~\ref{fig:mcore_restest_M3} and~\ref{fig:vzcloud_restest_M3}). In
the inviscid simulation, the cloud is already showing some asymmetry
at $t=5.73\,t_{\rm cc}$. This becomes more pronounced at later times,
as shown in Fig.~\ref{fig:m10chi1e3r32dis}c), and the cloud develops
significant transverse motions which speeds up its mixing and
acceleration.  In contrast, in the $k$-$\epsilon$ simulation the cloud
remains compact and symmetrical to very late times. As such, its core
mass drops more slowly as it suffers less ablation, and its
acceleration is much slower. This behaviour accounts for the
differences seen in Figs.~\ref{fig:mcore_restest_M3}
and~\ref{fig:vzcloud_restest_M3}. It highlights the fact that
instabilities develop differently at different resolutions, and
supports our earlier statement that this can eventually influence the
global mixing and acceleration of cloud material.


High resolutions, particularly at high cloud density contrasts, are
very computationally demanding when $M=1.5$ because of the gentler
nature of the interaction and the longer run times which ensue.  For
this reason we were unable to perform calculations above $R_{64}$ for
$\chi=10^{2}$, and above $R_{32}$ for $\chi=10^{3}$. We note that
$m_{\rm core}$ and $<v_{\rm z,cloud}>$ appear reasonably converged for
$R_{32}$ and $R_{64}$ when $M=1.5$ and $\chi=10$. However, further
work is needed to determine whether $R_{32}$ is an adequate resolution
for $m_{\rm core}$ and $<v_{\rm z,cloud}>$ $M=1.5$ when $\chi \gtsimm
100$.

The only published resolution test for 3D hydrodynamic simulations
that we are aware of in the astrophysics literature is shown in Fig.~5
of \citet{Xu:1995}.  They plot the time evolution of a number of cloud
properties, including $m_{\rm core}$ and $<v_{\rm z,cloud}>$, for
simulations with $M=10$ and $\chi=10$, with 11, 25 and 53 cells per
cloud radius. They find smaller differences between the $R_{25}$ and
$R_{53}$ simulations, than between the $R_{11}$ and $R_{25}$
simulations, and thus claim that their highest resolution simulation
captures the dominant dynamical effects. Our work is consistent with
these claims.

\subsection{Convergence Tests}
\label{sec:restest_convergence}
To gain further insight into the effect of the grid resolution on our
simulations we examine the variation of some integral quantities
computed from the datasets. One method for examining the degree of
convergence between simulations at different resolution is to study
the relative error, which is defined as the fractional difference
between the value measured at resolution $N$ and the value at the
finest resolution, $f$:
\begin{equation}
\label{eq:relerror}
\Delta Q_{N} = \frac{|Q_{N} - Q_{f}|}{|Q_{f}|}.
\end{equation}
If $\Delta Q_{N}$ shows a monotonic decrease with increasing
resolution, and is small, then self-convergence is occurring
\citep[e.g.,][]{Yirak:2010}. This behaviour translates into an
asymptotic levelling off with increasing resolution of a particular
quantity.

However, there is also the danger of ``false convergence'', whereby
further increases in resolution show that the models actually have not
converged \citep[see, e.g., Fig.~10a in][]{Niederhaus:2007}. This can
happen when an important flow feature is resolved for the first time
(e.g., the standoff of the bowshock, or the cooling zone of radiative
shocks). Furthermore, simulations which demonstrate convergence at a
particular simulation time may well not be converged at a later
time. This is expected for simulations of the ``inviscid'' Euler
equations, where RT and KH instabilities in simulations at different
resolution will break up the cloud differently, thus eventually
affecting the convergence of integral quantities. Therefore, any
statement that such simulations are ``converged'' must be qualified by
a time, and by the caveat that this does not imply that convergence
exists at later times.

\begin{figure*}
\includegraphics{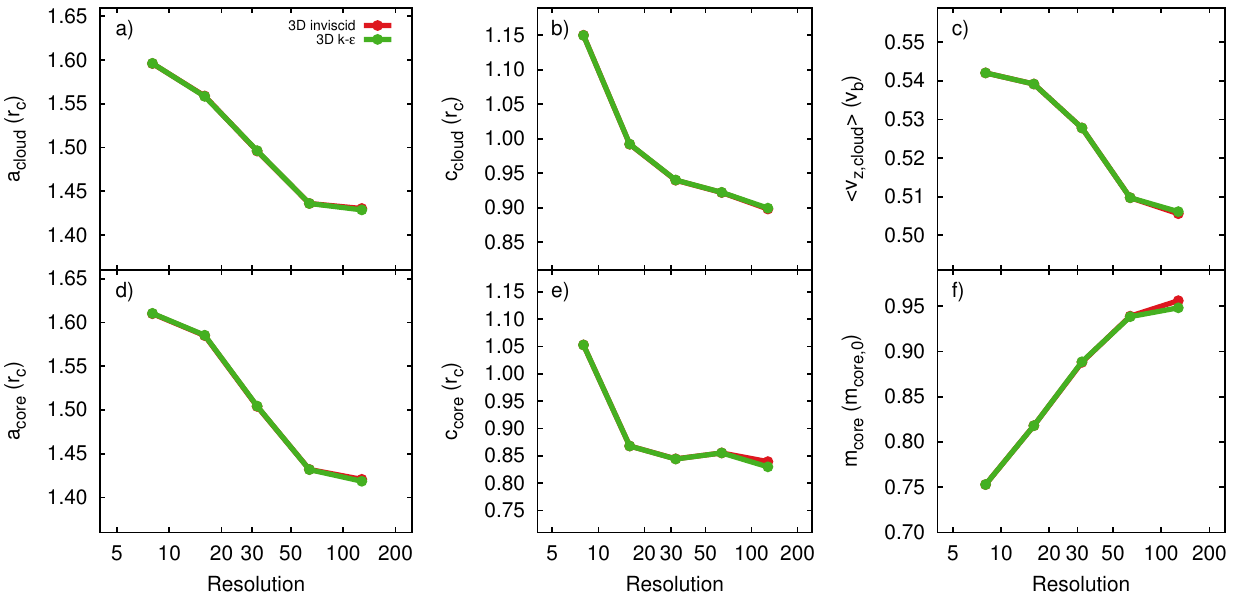}
   \caption{Integral quantities from the $M=10$, $\chi=10$,
     simulations at $t=3.14\,t_{\rm cc}$, plotted as a function of the
     grid resolution.}
    \label{fig:m10_chi1e1_convergence_error_2Dvs3D_evol}
\end{figure*}

The variation in $a_{\rm cloud}$, $c_{\rm cloud}$, $<v_{\rm
  z,cloud}>$, $a_{\rm core}$, $c_{\rm core}$ and $m_{\rm core}$ with
the spatial resolution for the $M=10$, $\chi=10$ 3D inviscid and
$k$-$\epsilon$ simulations is shown in
Fig.~\ref{fig:m10_chi1e1_convergence_error_2Dvs3D_evol}. We again
notice the very good agreement between the inviscid and $k$-$\epsilon$
simulations. We also see that some quantites appear to be converged
($c_{\rm core}$), some appear to show signs of convergence ($a_{\rm
  cloud}$, $a_{\rm core}$ and $m_{\rm core}$), while some are clearly
not converged ($c_{\rm cloud}$ and $<v_{\rm z,cloud}>$). Therefore,
our simulations are not formally converged at this time. However, they
are at sufficient resolution that \emph{some} global quantities
are. Clearly, it would be useful to extend this convergence study to
still higher resolutions.

\begin{figure*}
\includegraphics{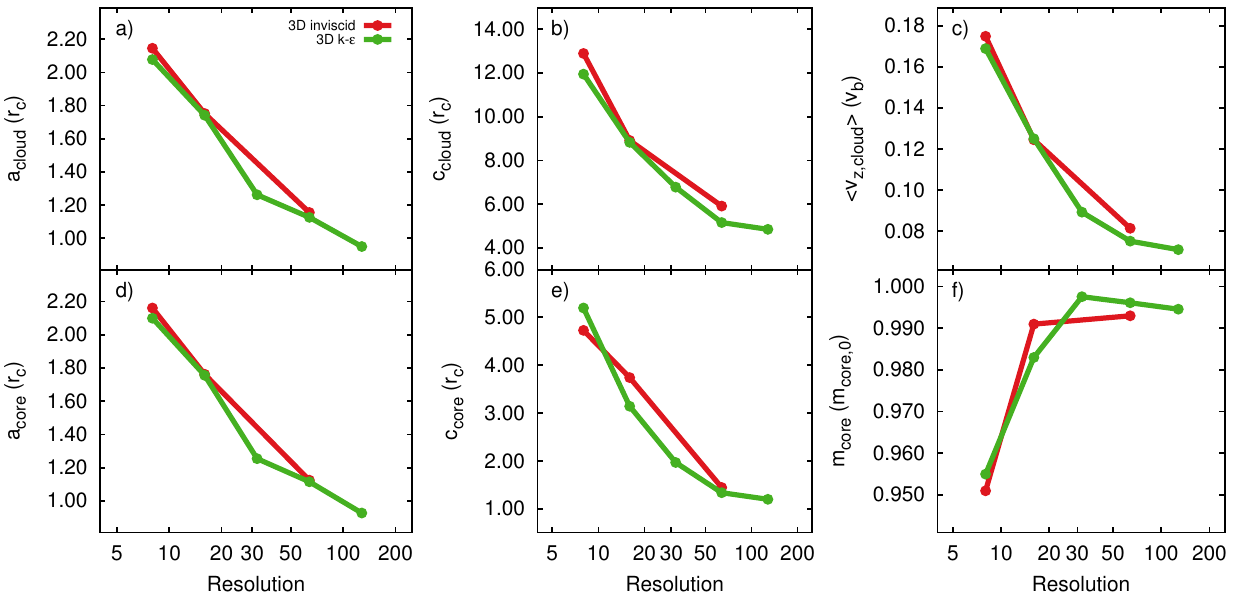}
   \caption{Integral quantities from the $M=10$, $\chi=10^{3}$,
     simulations at $t=2.9\,t_{\rm cc}$, plotted as a function of the
     grid resolution.}
    \label{fig:m10_chi1e3_convergence_error_2Dvs3D_3tcc}
\end{figure*}

We are also interested in the variation of these integral quantities
with resolution from simulations with a higher density contrast.
Fig.~\ref{fig:m10_chi1e3_convergence_error_2Dvs3D_3tcc} shows this
behaviour for models with $M=10$ and $\chi=10^{3}$. We see that there
is clear asymptotic levelling off of $c_{\rm cloud}$, $<v_{\rm
  z,cloud}>$, and $c_{\rm core}$ ($k$-$\epsilon$ only), indicating
that the solutions are converging for these quantities. However, there
is no levelling off for $a_{\rm core}$ and $a_{\rm cloud}$, indicating
clear non-convergence. $m_{\rm core}$ may be showing signs of
convergence, but more data is needed. We conclude that our simulations
are again not formally converged, but it appears that the highest
resolution simulations have sufficient resolution that some of the
integral quantities are showing signs of convergence at this time.

\begin{figure*}
\includegraphics{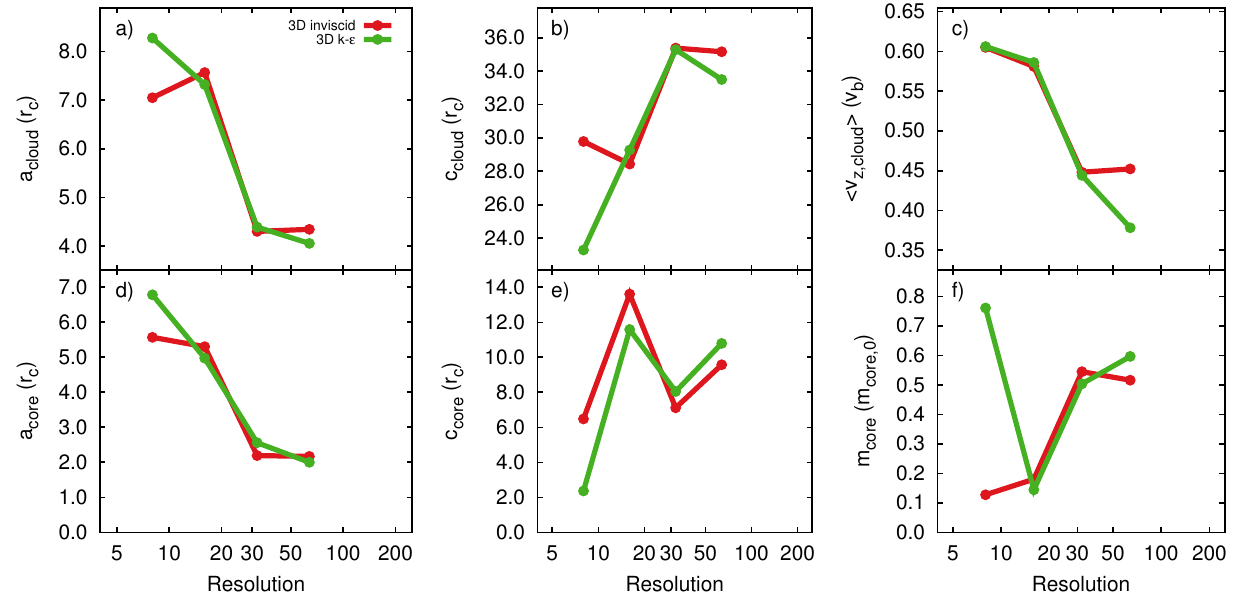}
   \caption{Integral quantities from the $M=10$, $\chi=10^{3}$,
     simulations at $t=6.0\,t_{\rm cc}$, plotted as a function of the
     grid resolution.}
    \label{fig:m10_chi1e3_convergence_error_2Dvs3D_6tcc}
\end{figure*}

Moving to later times ($t=6\,t_{\rm cc}$),
Fig.~\ref{fig:m10_chi1e3_convergence_error_2Dvs3D_6tcc} shows that
none of the quantities (except perhaps $a_{\rm core}$) display any
signs of convergence. This demonstrates that as the simulations
advance in time they move from showing \emph{some} convergence to
showing non-convergence. Thus we can be totally clear that still
higher resolution is necessary in order to obtain formal convergence
(at $t\approx3\,t_{\rm cc}$, let alone at later times), if indeed this
is even possible given the nature of the inviscid Euler equations.

\subsection{Timescales}

\begin{figure*}
\includegraphics{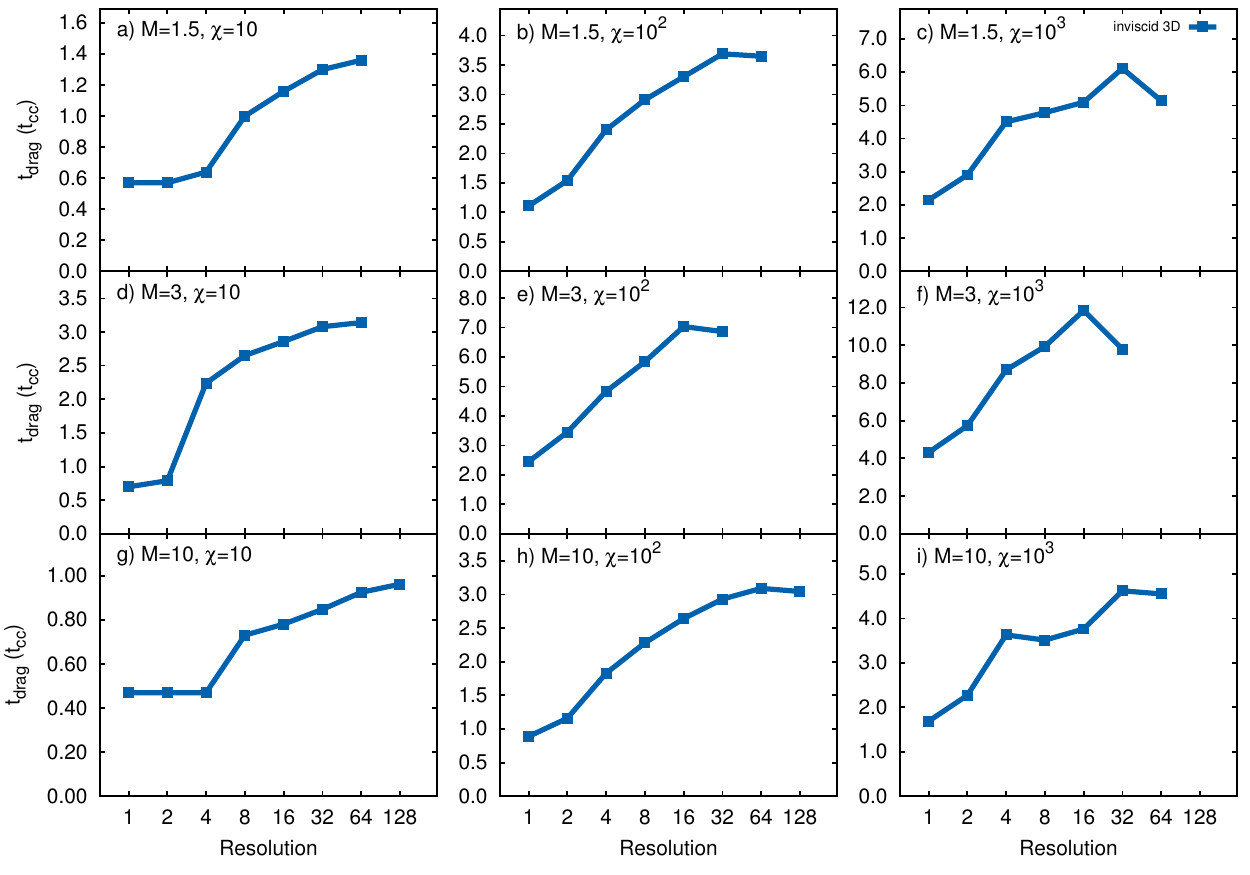}
   \caption{Resolution dependence of $t_{\rm drag}$ (for the
      cloud) as functions of the Mach number $M$ and
     cloud density contrast $\chi$.}
    \label{fig:tdragrestest}
\end{figure*}

\begin{figure*}
\includegraphics{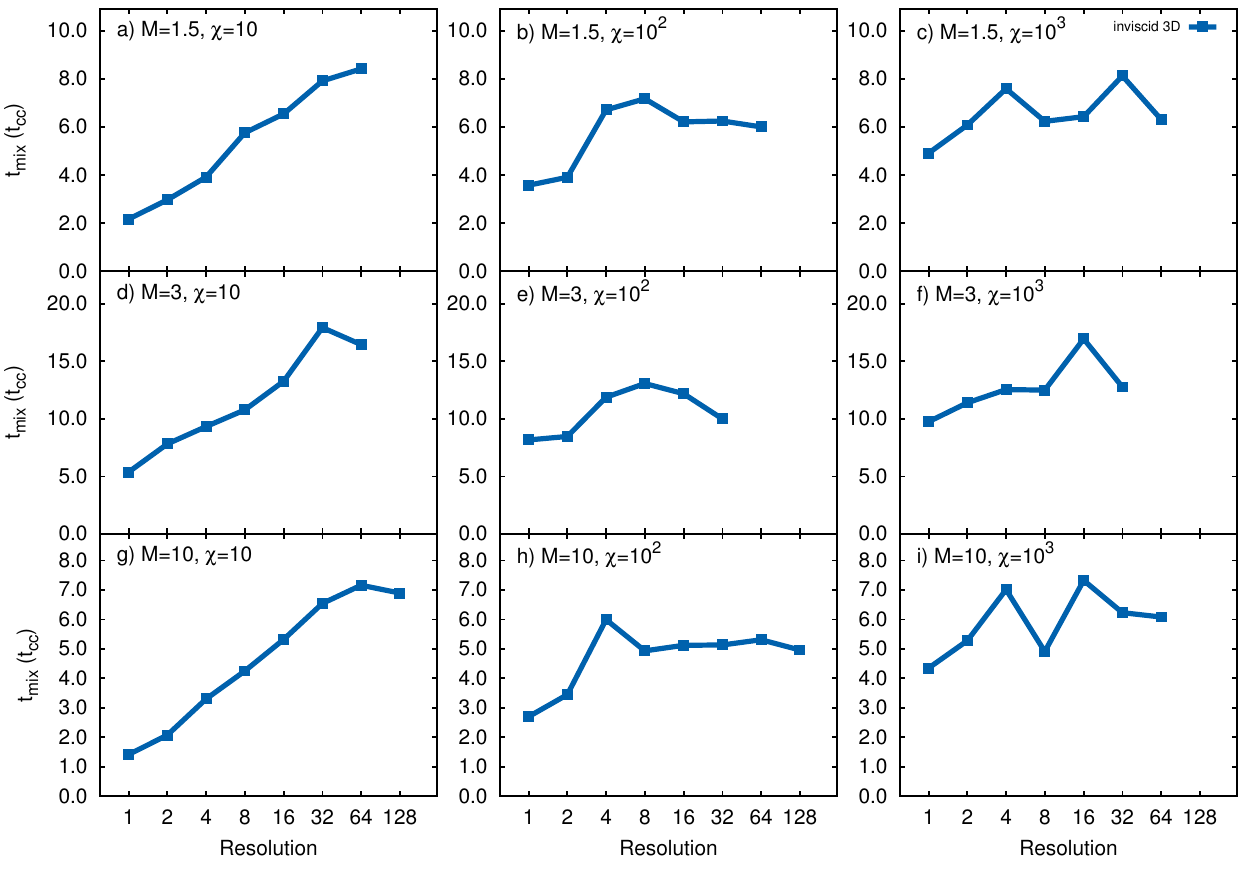}
   \caption{Resolution dependence of $t_{\rm mix}$ as functions of the Mach number $M$ and
      cloud density contrast $\chi$.}
    \label{fig:tmixrestest}
\end{figure*}

Figs.~\ref{fig:tdragrestest} and~\ref{fig:tmixrestest} examine the
resolution dependence of $t_{\rm drag}$ and $t_{\rm mix}$. In general,
$t_{\rm drag}$ increases with resolution. However, there are few signs
of convergence towards an asymptote, so further resolution tests are
needed to determine $t_{\rm drag}$ accurately. Since, $t_{\rm mix} >
t_{\rm drag}$ in all cases, it is not surprising that we do not see
formal convergence for $t_{\rm mix}$ either.

While formal convergence is not seen, in many cases $t_{\rm drag}$ and
$t_{\rm mix}$ level-off at higher resolutions, indicating that their
values may be reasonably close to the ``true'' value.  It is also of
interest to note by how much $t_{\rm drag}$ and $t_{\rm mix}$ are
underestimated in lower resolution calculations. By averaging the
values in Fig.~\ref{fig:tdragrestest} we find that $t_{\rm drag}$ is
only 40\% of its maximum resolution value at $R_{1}$, climbing to 83\%
at $R_{8}$ and 93\% at $R_{16}$, and is on average within 1\% of its
maximum resolution value for $R_{32}$ and higher resolutions.

There is significantly more scatter in the resolution dependence of
$t_{\rm mix}$, as shown in the bottom panel of
Fig.~\ref{fig:tmixrestest}.  However, like $t_{\rm drag}$, there is a
clear trend that $t_{\rm mix}$ is underestimated in lower resolution
calculations. At $R_{1}$, $t_{\rm mix}$ is on average 56\% of its
value from our highest resolution simulations, and is on average 10\%
lower at $R_{8}$. In some simulations it can be only 20\% of its true
value, while in others it can be nearly 30\% longer. Note that there
is a tight correlation in the trend of $t_{\rm mix}$ when $\chi=10$,
for the 3 Mach numbers investigated: at $R_{1}$ and $R_{8}$, $t_{\rm
  mix}$ is 26\% and 64\% of the true value, respectively.

Figs.~\ref{fig:tdragrestest} and~\ref{fig:tmixrestest} indicate that
clouds will be accelerated and destroyed more rapidly than they should
be when they are poorly resolved in numerical simulations. This has
implications for simulations of a wide-range of multiphase flows,
including the collision of stellar winds
\citep{Stevens:1992,Walder:2002,Pittard:2007a,Pittard:2009,Parkin:2011},
the interaction of stellar winds, jets and SNe with their local
environment
\citep[e.g.][]{Jun:1996,Steffen:2004,Tenorio-Tagle:2006,Yirak:2008,Rogers:2013,Dale:2014},
the SN-regulated ISM
\citep[e.g.,][]{deAvillez:2005,Joung:2006,Joung:2009,Hill:2012,Kim:2013,Creasey:2013,Hennebelle:2014,Walch:2015,Girichidis:2015},
galactic outflows and superwinds
\citep[e.g.,][]{Strickland:2000,Cooper:2008,Dubois:2008,Ceverino:2009,Fujita:2009,Hopkins:2012,Marinacci:2014,Schaye:2015,Vorobyov:2015,Kimm:2015},
ram-pressure stripping of the ISM from galaxies
\citep[e.g.,][]{Tonnesen:2009,Roediger:2014,Vijayaraghavan:2015}, and
AGN feedback \citep[e.g.,][]{Sutherland:2007,Wagner:2012}. It also
affects studies of the interaction of a shock with multiple clouds
\citep[e.g.,][]{Poludnenko:2002,Melioli:2005,Aluzas:2012}, and
mass-loaded flows in general \citep[see the review
by][]{Pittard:2007b}.

\bsp	
\label{lastpage}
\end{document}